\documentclass{article}
\usepackage{microtype}
\usepackage{graphicx}
\usepackage{booktabs}
\usepackage{float}
\usepackage{caption}
\usepackage{placeins}
\usepackage{longtable}
\usepackage{array}
\usepackage{amsmath}
\usepackage{amssymb}
\usepackage{mathtools}
\usepackage{xurl}
\usepackage{hyperref}
\usepackage[accepted]{icml2024}
\providecommand{\tightlist}{\setlength{\itemsep}{0pt}\setlength{\parskip}{0pt}}

\icmltitlerunning{GenAI Models Capture Urban Science}

\begin{document}
\twocolumn[
\icmltitle{GenAI Models Capture Urban Science but Oversimplify Complexity}
\begin{icmlauthorlist}
\icmlauthor{Yecheng Zhang}{thu}
\icmlauthor{Rong Zhao}{thu}
\icmlauthor{Zimu Huang}{thu}
\icmlauthor{Xinyu Wang}{thu}
\icmlauthor{Yue Ma}{thu}
\icmlauthor{Ying Long}{thu}
\end{icmlauthorlist}
\icmlaffiliation{thu}{School of Architecture, Tsinghua University, Beijing, China}
\icmlcorrespondingauthor{Ying Long}{ylong@tsinghua.edu.cn}
\icmlkeywords{urban science, generative artificial intelligence, urban data synthesis, virtual urban laboratory}
\vskip 0.3in
]
\printAffiliationsAndNotice{}
\chead{\small\bf GenAI Models Capture Urban Science}

\begin{abstract}
Generative artificial intelligence (GenAI) models are increasingly used for scientific data generation, yet their alignment with empirical knowledge in urban science remains unclear. We therefore ask whether generated urban data reproduce empirical regularities and support repeatable experiments. We introduce AI4US, a framework for evaluating data synthesis and conditional intervention across text and image modalities. Four cases spanning urban systems, within-city structure, neighbourhood vitality and streetscape perception are evaluated against published parameters, observed urban data and human judgements. Generated outputs recovered recognizable scaling and distance-decay patterns and yielded measurable associations between neighbourhood morphology indicators and pedestrian activity, while model judgements showed positive agreement with sampled human choices. Controlled changes to urban conditions produced repeatable output responses across all four cases. However, generated data often compressed empirical numerical coverage, local variation and visual diversity. In an inspectable GenAI model, intermediate activations linearly distinguished some urban relationships, while activation edits changed only selected output scores. AI4US provides an empirically grounded approach for assessing GenAI as the virtual urban laboratory in urban data synthesis and controlled model experiments, while revealing its tendency to simplify empirical complexity.
\end{abstract}

\section{Introduction}\label{introduction}

Urban science has long sought to decipher the complex behaviours of cities, from early demographic surveys to computational modelling{[}1--3{]}. Remote sensing, sensor networks and digital platforms have greatly expanded the scope of urban observation{[}3{]}, while making representation, governance and accountability part of urban research design{[}4{]}. Systematic comparative studies now examine many cities within common analytical designs{[}5{]}. Yet a city cannot be replicated in the manner of a laboratory specimen, and spatial, institutional, social and technological conditions often change together. Urban science therefore needs complementary settings in which urban conditions can be generated, varied repeatedly and evaluated against observations of real places.

Historically, urban experimentation has evolved through several paradigms. Early approaches treated the city as a living laboratory and used observation, policy trials and natural experiments to compare urban conditions{[}2,6{]}. Physical and analogue models offered controllable surrogates, and the subsequent rise of cellular automata, land-use and transport models, and agent-based simulations introduced rule-based digital experimentation{[}7{]}. Such models make their assumptions inspectable, but researchers must specify the variables, interactions and parameters in advance. Models trained on large corpora may offer another experimental medium alongside observation and rule-based simulation because they can generate urban data and respond to altered conditions without a task-specific urban simulator.

Advances in generative artificial intelligence (GenAI) are opening new approaches to data-driven scientific discovery{[}8--12{]}. Large language models learn statistical regularities from text and structured data, while multimodal models extend the same interface to images. In biology and materials science, related systems have supported protein-structure prediction and the search for novel materials{[}10,11{]}. Urban science presents a different but complementary test. GenAI can generate numerical city records, spatial profiles and streetscape scenes, and it can respond when features of those outputs are changed. Plausible output alone is not sufficient for scientific use: generated relationships, parameter values and data variation must be compared with empirical observations, and responses to changed conditions must be measured under repeatable designs.

Initial studies have shown promise, using GenAI models to generate synthetic mobility data, simulate residents' daily behaviours, or assess urban systems{[}13--15{]}. A recent dataset of 1,260 standardized urban-experience narratives across 21 cities, four actor profiles and three language models enables controlled comparisons of contextual consistency, model variation and representational bias{[}16{]}. Geographic evaluations also reveal systematic bias and inconsistent factual accuracy across locations and models{[}17,18{]}. This work demonstrates the reach of large models but leaves a broader evaluation problem: whether generated urban data reproduce established relationships, whether their parameters and variation resemble empirical observations, and whether the generated data can support controlled experiments.

Our case selection began with a systematic review of data-driven urban research in 57 journals, which identified recurring work on transport, spatial quality, vitality, urban form and perception{[}3{]}. We selected four established cases that each provide a measurable relationship or outcome, an empirical reference, an object that can be generated and a condition that can be varied. Urban Scaling tests power-law relationships between Population and Road or GDP across cities, together with the rank--size distribution of Population{[}19--21{]}. Distance Decay tests the inverse-S change in land density with distance from a city centre{[}22--25{]}. Jacobs-informed Vitality tests associations between five neighbourhood morphology indicators and observed Pedestrian Activity{[}26,27{]}. Place Pulse Perception uses the six human-judgement dimensions defined in Place Pulse{[}28{]}. Together, the cases span urban systems, within-city structure, neighbourhoods and streetscapes. The three numerical cases use text-only prompts to generate urban tables or profiles, whereas the Perception case uses streetscape images together with text instructions to evaluate visual judgements.

A reproduced pattern may reflect familiar city facts, a relation stated in the prompt or computations performed for the current task. Recent work has shown that spatial, temporal and geographic attributes can be decoded from language-model states and that activation edits can alter selected outputs{[}29--32{]}. We therefore designed a focused open-model analysis to test whether selected relation distinctions are readable from intermediate activations and whether targeted changes to those activations shift the corresponding output.

Here, we introduce AI4US as a common design for Data Synthesis and Conditional Intervention. Data Synthesis generates urban data, estimates their relationships and variation, and compares them with published parameters, city data, neighbourhood observations or human judgements. Baseline prompts define the requested variables, units and output format; Blueprint prompts add empirically informed value coverage, spatial organization or heterogeneity. Conditional Intervention varies an urban condition within a fixed setting and measures the model response. GPT-4o performs the numerical tasks and image scoring, gpt-image-2 generates and edits streetscapes, and Qwen2.5-VL-72B provides the open-model follow-up. AI4US separates three questions that plausible output alone cannot answer: whether a recognizable relationship is present, whether its parameter values and variation resemble empirical observations, and whether the model responds consistently when an urban condition is changed.

\section{Results}\label{results}

\subsection{GenAI-based data synthesis supports controlled experiments in virtual urban settings}\label{genai-based-data-synthesis-supports-controlled-experiments-in-virtual-urban-settings}

AI4US evaluates GenAI at the scales of urban systems, within-city structure, neighbourhoods and streetscapes (Fig. 1). The three text-modality cases generate numerical tables or profiles to examine cross-city scaling between Population and Road/GDP, within-city land-density profiles across distance from the city centre, and relationships between five neighbourhood-morphology indicators and Pedestrian Activity. The image-modality case uses streetscape images and text instructions to examine six Place Pulse judgements. Each case has an external empirical reference in the form of published parameters, urban data, neighbourhood observations or human labels.

The study first synthesizes urban data and estimates relationships, then compares generated data with empirical observations, and finally changes urban conditions within fixed settings. In the three numerical cases, the changed condition is specified as a value in the text input; in Perception, matched image edits alter the same base streetscape before scoring. The Baseline prompt supplies the definitions required to perform each task. The Blueprint prompt adds case-specific information about numerical coverage, spatial organization or heterogeneity. The open-model analysis follows these experiments: it compares response patterns with GPT-4o, tests whether selected relation types can be read from hidden states and measures whether targeted activation edits shift relative answer scores.

Across these four scales, the shared sequence treats recognizable generated structure as the initial observation, empirical magnitude and variation as the calibration test, and the response to one controlled change as the experimental test. Together, these steps distinguish plausible pattern reproduction from empirical correspondence and repeatable model behaviour.

\begin{figure*}[!t]
\centering
\includegraphics[width=\textwidth,height=0.62\textheight,keepaspectratio]{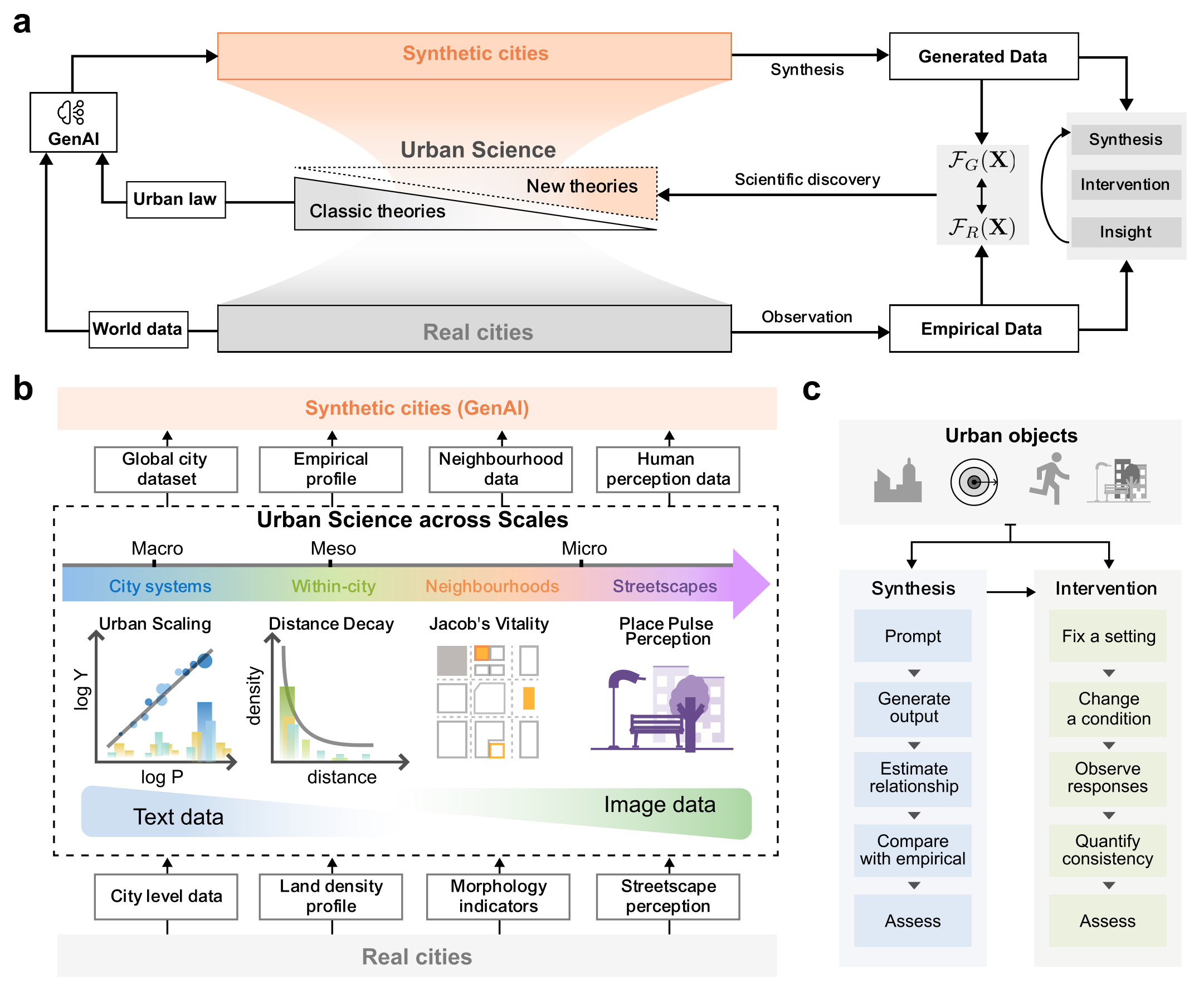}
\caption{\textbf{AI4US links generated and empirical urban data across four urban-science cases.} \textbf{a,} Conceptual view of GenAI within urban science. GenAI outputs and observations of real cities provide generated and empirical data. For urban objects (X), (F\_G(X)) and (F\_R(X)) denote relationships estimated from generated and real-city data, respectively. Comparing these relationships supports data synthesis, controlled intervention and the evaluation of urban-science hypotheses; the path from classic to new theories represents a prospective cycle of hypothesis development. \textbf{b,} The four cases span city systems, within-city structure, neighbourhoods and streetscapes: Urban Scaling, Distance Decay, Jacob's Vitality and Place Pulse Perception. The first three use text data, whereas Place Pulse Perception uses image data. The boxes above the case sequence name the corresponding GenAI-generated data products, and those below name the real-city data objects used as empirical references. \textbf{c,} Data Synthesis prompts the model to generate an urban output, estimates its relationship, compares it with an empirical reference and assesses the result. Conditional Intervention holds a setting fixed, changes one condition, observes the model responses and assesses their consistency. The open-model analysis uses Qwen2.5-VL-72B as the primary example to compare matched responses with GPT-4o, probe selected relationship distinctions in intermediate activations and test targeted state interventions.}
\label{fig:1}
\end{figure*}

\subsection{GenAI reproduces recognizable urban regularities}\label{genai-reproduces-recognizable-urban-regularities}

The four cases produced measurable urban relationships and perceptual judgements (Fig. 2). Both Urban Scaling prompts produced consistently estimable power-law relationships. Under Baseline and Blueprint, the mean Road scaling exponents were 0.887 {[}0.862, 0.910{]} and 0.850 {[}0.820, 0.881{]}, close to the empirical estimate of 0.849 for US metropolitan areas. The corresponding GDP exponents were 1.044 {[}1.015, 1.072{]} and 1.013 {[}0.987, 1.039{]}, below the empirical estimate of 1.126{[}19{]}. The rank--size slope shifted from \(-0.477\) {[}\(-0.499\), \(-0.455\){]} to \(-1.531\) {[}\(-1.591\), \(-1.470\){]}, closer to our estimate of \(-1.115\) from FHWA data. Blueprint improved the rank--size slope, whereas the Road and GDP exponents did not move closer to their empirical estimates.

The median fitted land-density profile declined with distance from the city centre under both Distance Decay prompt conditions. Median inverse-S \(R^2\) was 0.997 under Baseline and 0.995 under Blueprint. Median steepness \(\alpha\) increased from 1.430 to 1.691, peripheral density \(c\) from 0.010 to 0.016, and spatial scale \(D\) from 22.892 to 37.114 km. Across 150 empirical city-period curves from 50 cities, the corresponding medians were \(\alpha=3.210\), \(c=0.070\) and \(D=15.750\,\mathrm{km}\){[}25{]}. Median fitted steepness was lower in both generated conditions than in the empirical profiles.

Jacobs-informed Vitality tables had positive mean coefficients for Residential Density, Functional Mix and Short Blocks: 0.599, 0.171 and 0.136 under Baseline, and 0.699, 0.081 and 0.157 under Blueprint. Mean coefficients for Aged Buildings and Tall Buildings were weakly negative. At the 52 Melbourne sites, Residential Density, Functional Mix and Aged Buildings were positive, Short Blocks was negative and Tall Buildings was close to zero. Short Blocks and Aged Buildings therefore had opposite signs in the generated and empirical models. Using only the five morphology indicators, GPT-4o ranked 0.603 {[}0.515, 0.681{]} of empirical site pairs in the observed Pedestrian Activity order.

GPT-4o showed positive agreement with sampled human choices across all six Place Pulse dimensions. Each dimension used 1,000 pairwise comparison records, and each task retained one sampled Place Pulse vote rather than a pairwise majority label. Across all 6,000 tasks, three-class accuracy was 0.550 {[}0.537, 0.562{]} and Cohen's \(\kappa\) was 0.210 {[}0.190, 0.230{]}. Dimension-specific \(\kappa\) was higher for Beautiful, Wealthy and Safe than for Depressing, Lively and Boring. This comparison provides an independent human-judgement reference for the subsequent scoring of generated streetscapes{[}28{]}.

Taken together, the models reproduced recognizable relationship forms and showed positive but modest agreement with human choices. Comparison with the empirical references gave a more specific result: Road scaling was close to its reference, GDP scaling was shallower, land-density decline was more gradual, and two Vitality coefficients reversed sign relative to observations. Thus, recognizing urban structure in generated outputs is distinct from matching its observed magnitude and direction.

Variable definitions, empirical-data construction and sensitivity analyses are described in Methods 4.3--4.4 and the Supplementary Methods and Results.

\begin{figure*}[!t]
\centering
\includegraphics[width=\textwidth,height=0.62\textheight,keepaspectratio]{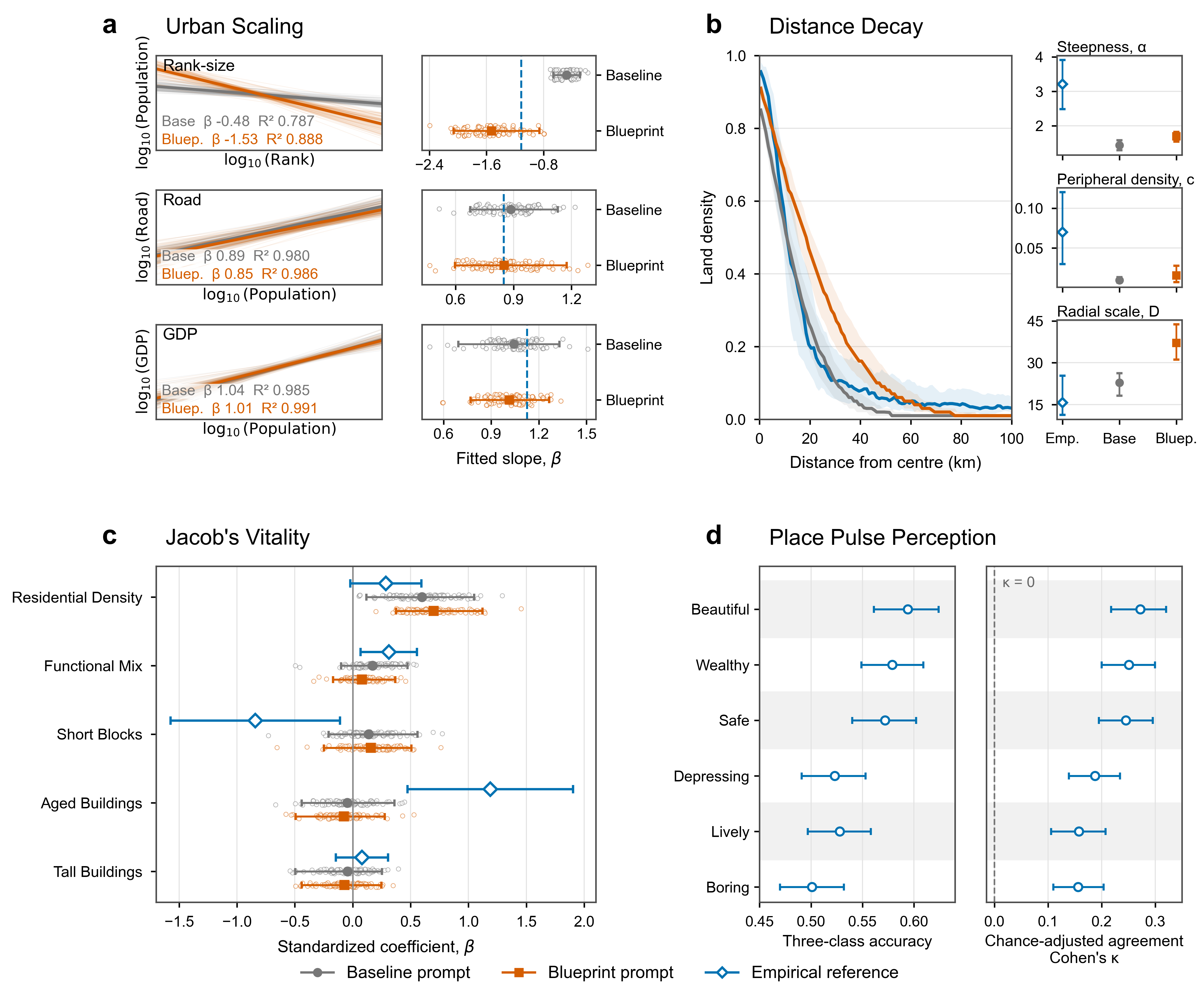}
\caption{\textbf{Generated urban relationships and an external human-judgement reference.} \textbf{a,} Urban Scaling. Thin lines show table-level base-10 log--log fits and thick lines show the mean fit. Points summarize condition means and horizontal ranges span the 2.5th--97.5th percentiles across fitted tables; blue dashed lines mark the empirical slopes. The Rank-size, Road and GDP slope distributions use independent x-axis ranges, with row-specific tick values shown on all three axes. \textbf{b,} Distance Decay. Curves show median annular land density and shading shows interquartile ranges. The adjacent panels report medians and interquartile ranges for inverse-S steepness \(\alpha\), peripheral density \(c\), and radial scale \(D\) in kilometres; for \(f(r)=c+(1-c)/[1+\exp\{\alpha(2r/D-1)\}]\), the transition midpoint is \(r=D/2\). The empirical parameter distribution contains 150 city-period curves from 50 cities. \textbf{c,} Jacob's Vitality. Points show mean standardized coefficients from five-variable regressions. Generated ranges span the 2.5th--97.5th percentiles across tables, whereas empirical bars are HC3 95\% confidence intervals across 52 sites. \textbf{d,} Place Pulse Perception evaluator. The aligned axes show three-class accuracy and Cohen's \(\kappa\) between GPT-4o choices and one sampled human vote for 1,000 image pairs per dimension; bars are task-bootstrap 95\% confidence intervals. Accuracy is the raw share of identical LEFT, RIGHT or EQUAL labels; \(\kappa\) adjusts for agreement expected from the human and model marginal class distributions. Across all 6,000 pairs, accuracy was 0.550 {[}0.537, 0.562{]} and \(\kappa\) was 0.210 {[}0.190, 0.230{]}. Grey, orange and blue denote Baseline, Blueprint and empirical references in \textbf{a--c}. Road denotes public-road lane-miles and GDP denotes Gross Metropolitan Product.}
\label{fig:2}
\end{figure*}

\subsection{GenAI-generated data show less variation than empirical observations}\label{genai-generated-data-show-less-variation-than-empirical-observations}

Generated data preserved readily estimable patterns while compressing or redistributing empirical variation (Fig. 3). In a separate generic-city Urban Scaling cohort, Blueprint reduced the Wasserstein distances from generated Population, total Road length and city GDP to an empirical reference of 3,525 Chinese physical-city polygons from 0.626, 0.626 and 0.804 to 0.193, 0.338 and 0.388. The corresponding within-table magnitude ranges were 2.665, 2.008 and 2.512 orders of magnitude, below the empirical ranges of 3.858, 3.806 and 4.316. Blueprint expanded numerical coverage and brought the three marginal distributions closer to this empirical reference in that cohort.

For Distance Decay, the main differences concerned local variation. Empirical profiles had a median unique-value fraction of 0.32, a normalized second difference of 0.0177 and a median of 19 outward density reversals. Median reversals were zero under both Baseline and Blueprint. Blueprint increased value resolution and local curvature, while a minority of profiles still contained outward reversals.

For Vitality, Blueprint standard deviations for Residential Density, Short Blocks and Pedestrian Activity approached those across the 52 sites. Functional Mix was 1.79 times the empirical standard deviation, and Aged Buildings and Tall Buildings were 0.87 and 1.22 times the empirical values. Dispersion therefore changed differently across variables.

Blueprint increased CLIP{[}33{]} distances among generated streetscapes, while reducing nearest-neighbour similarity to Place Pulse images in both directions and increasing the centroid distance between the two sets. Sensitivity analyses used DINOv2{[}34{]} and ResNet-50{[}35{]}; DINOv2 showed the same directional change in nearest-neighbour similarity, whereas the ResNet-50 change was near zero. Under the primary CLIP analysis, Blueprint therefore combined greater dispersion within the generated set with lower proximity to the Place Pulse distribution.

Blueprint broadened several generated distributions, while numerical range, local spatial irregularity and proximity to the empirical image distribution changed by different amounts and sometimes in different directions. The relevant test of generated-data quality therefore depends on the form of variation needed for the urban question, not on overall dispersion alone.

The comparison metrics and empirical sampling procedures are defined in Methods 4.5 and the Supplementary Methods and Results.

\begin{figure*}[!t]
\centering
\includegraphics[width=\textwidth,height=0.62\textheight,keepaspectratio]{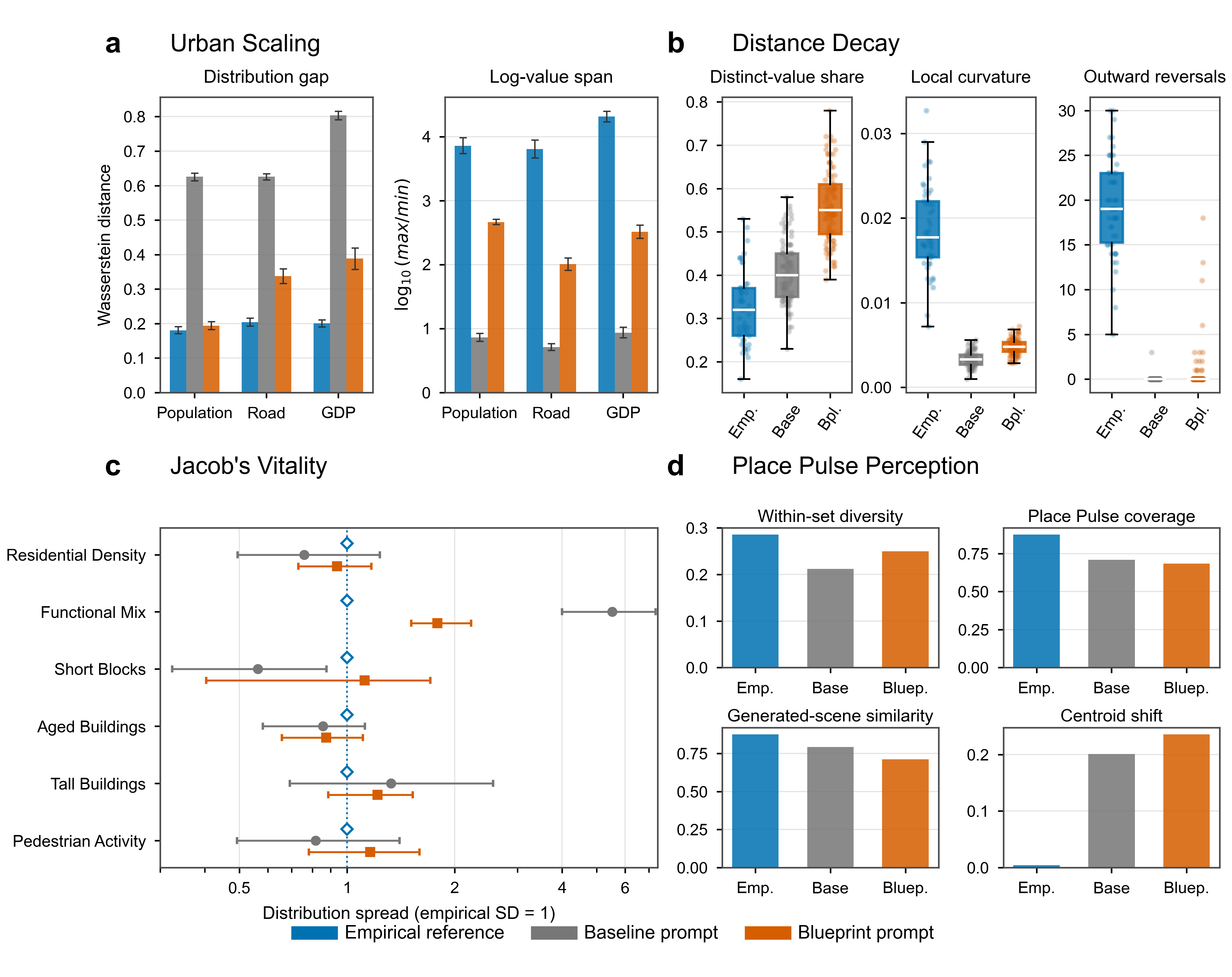}
\caption{\textbf{Variation in generated data relative to empirical observations.} \textbf{a,} Urban Scaling distribution cohort. Distribution gap is the one-dimensional Wasserstein distance between generated and empirical \(\log_{10}(x/\sum x)\) within-table share distributions for Population, total Road length and GDP, using an empirical reference of 3,525 Chinese physical-city polygons; lower values indicate closer distributions. Baseline and Blueprint error bars are percentile 95\% confidence intervals from 10,000 resamples of complete 100-city tables, and empirical bars span the 2.5th--97.5th percentiles across 100 independent 100-city resamples. Log-value span is \(\log_{10}(\max x/\min x)\), so values of 1 and 2 correspond to 10-fold and 100-fold spans. This cohort is separate from the source-matched coefficient experiment in Fig. 2a. \textbf{b,} Distance Decay. After every annular-density profile was rounded to two decimal places, distinct-value share was the number of distinct values divided by the number of distance bands; local curvature was the mean absolute second difference divided by the profile range (denominator 1 for a constant profile); and outward reversals counted adjacent outward steps in which density increased. Local curvature is therefore a normalized second-difference measure, not fitted geometric curvature. Boxplots summarize profile distributions and points show individual profiles. \textbf{c,} Jacob's Vitality. Points show each variable's generated within-table standard deviation relative to the corresponding standard deviation across 52 empirical sites after the analysis transformation; the empirical reference is one and bars span the 2.5th--97.5th percentiles. \textbf{d,} Place Pulse Perception. Within-set diversity is mean pairwise CLIP cosine distance among generated images. Place Pulse coverage averages, over empirical images, the maximum cosine similarity to the generated set; generated-scene similarity averages the reverse nearest-neighbour direction over generated images; centroid shift is the cosine distance between the two set centroids. Empirical references use split halves of a 3,840-image Place Pulse set from 16 cities. Grey, orange and blue denote Baseline, Blueprint and empirical references, respectively.}
\label{fig:3}
\end{figure*}

\subsection{GenAI shows repeatable responses to controlled changes in urban conditions}\label{genai-shows-repeatable-responses-to-controlled-changes-in-urban-conditions}

Across all four cases, changing an urban condition within a fixed setting produced directional and repeatable GenAI responses (Fig. 4). When Population was increased across five levels in 100 fictional metropolitan settings, the mean GDP scaling exponent was 1.034 {[}1.008, 1.061{]} and the Road exponent was 0.682 {[}0.655, 0.707{]}. Their mean difference was 0.352 {[}0.314, 0.392{]}, and GDP responses were stronger in 99 of 100 settings. GDP therefore increased superlinearly and Road sublinearly with Population.

Predicted land density generally declined as Distance from centre increased. Of 1,000 prespecified near--far comparisons, 0.816 {[}0.772, 0.858{]} had the expected ordering. Among the 67 complete five-band profiles, the adjacent-band decline rate was 0.899. Forty-one of the 100 city settings produced strictly decreasing profiles.

For each of 38 neighbourhood settings, we kept the other four morphology indicators fixed and set Functional Mix to locally plausible low, middle and high values, defined as the \(q_{10}\), \(q_{50}\) and \(q_{90}\) among the eight most similar observed sites. Thirty-one settings produced the complete High--Mid--Low ordering in Pedestrian Activity, for a fixed-denominator rate of 0.816 {[}0.666, 0.908{]}. The three pairwise ordering rates ranged from 0.816 to 0.842.

Streetscape editing produced the complete \(3\times6\) perception-effect matrix. Natural increased Beautiful, Lively, Safe and Wealthy and reduced Depressing and Boring. Traffic increased Lively and reduced Beautiful. Built primarily increased Wealthy and reduced Depressing. Each theme effect was measured against a same-scene editing control. To create this control, the same baseline scene passed through the same image-editing process with an instruction to make no intentional visible change. The inferential unit was the scene (\(n=12\)).

Across the numerical and image tasks, changing one urban condition while holding the rest fixed yielded an explicit directional comparison. These comparisons turn qualitative urban propositions into repeatable tests of GenAI responses that can be evaluated against empirical data or human judgements.

The intervention estimands, fixed denominators and uncertainty procedures are defined in Methods 4.6 and the Supplementary Methods and Results.

\begin{figure*}[!t]
\centering
\includegraphics[width=\textwidth,height=0.62\textheight,keepaspectratio]{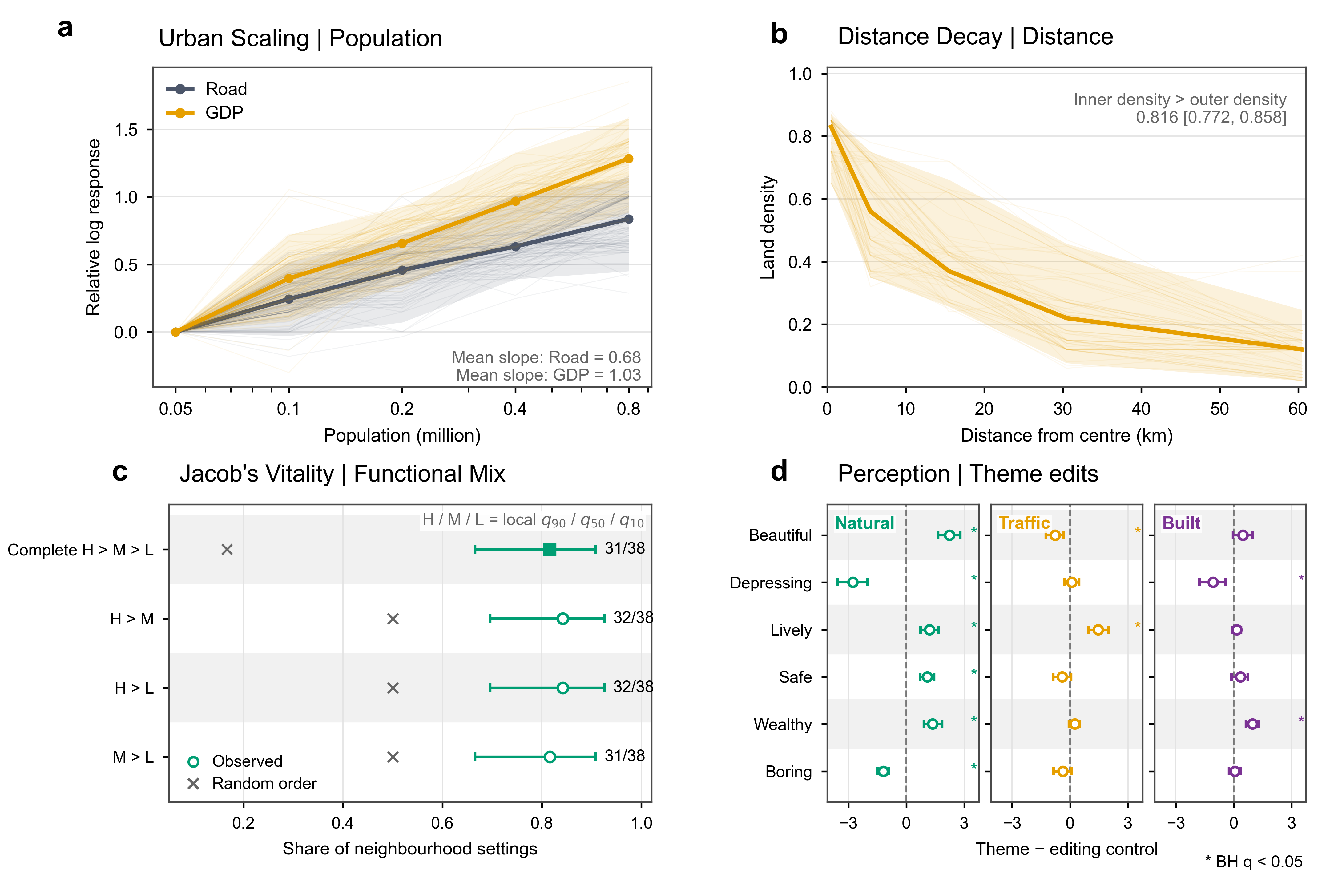}
\caption{\textbf{GenAI responses to controlled changes in urban conditions.} \textbf{a,} Population was varied from 0.05 to 0.8 million in 100 fictional metropolitan settings. ``Relative log response'' is \(\log_{10}[Y_s(p)/Y_s(0.05\ \mathrm{million})]\), calculated separately for Road and GDP. Thin lines show setting-level responses; thick lines are pointwise medians and shading spans the 2.5th--97.5th percentiles across settings; these are descriptive ranges, not confidence intervals. Annotations give the mean setting-level log--log slopes. \textbf{b,} Model-reported annular land-density fraction was estimated at five 1-km distance bands in 100 fictional city settings. The annotation is the share of all 1,000 prespecified near--far comparisons in which inner density exceeded outer density; missing or invalid comparisons contribute zero. Thin curves show 67 complete profiles; the thick curve and shading are defined as in \textbf{a}. Brackets give the 95\% setting-bootstrap confidence interval. \textbf{c,} For Jacob's Vitality, Functional Mix was set to local low (L), middle (M) and high (H) values in 38 neighbourhood settings while the other four morphology indicators were fixed. H, M and L are the local \(q_{90}\), \(q_{50}\) and \(q_{10}\) among eight similar observed sites. The square shows the complete \(H>M>L\) Pedestrian Activity ordering and circles its three pairwise components; bars are Wilson 95\% confidence intervals and labels report hits/38. Crosses show random-order probabilities (1/6 complete; 1/2 pairwise). \textbf{d,} Natural, Traffic and Built edits were compared with same-scene editing controls across 12 street scenes. Points show mean theme-minus-control differences on six 0--10 perception scales; positive values favour the theme edit. Bars are scene-bootstrap 95\% confidence intervals; asterisks denote Benjamini--Hochberg \(q<0.05\) from exact paired sign-flip tests across the \(3\times6\) matrix.}
\label{fig:4}
\end{figure*}

\subsection{Response profiling, linear probing and activation intervention test selected state--output links}\label{response-profiling-linear-probing-and-activation-intervention-test-selected-stateoutput-links}

Because GPT-4o does not expose internal activations, we used Qwen2.5-VL-72B as the primary inspectable model. The analysis had three stages: response profiling, linear probing and activation intervention (Profile · Probe · Intervene). Response profiling compared matching Road, GDP, land-density, activity and Perception responses from GPT-4o and Qwen2.5-VL-72B under the same task descriptions and output templates. Linear probing tested whether intermediate states distinguished each expected relationship from its reverse. Activation intervention then changed candidate states while holding the prompt fixed and measured shifts in relative answer scores. This final stage differs from Conditional Intervention, which changes an urban condition in the input rather than the model state.

In the entry-by-entry comparison of the paired tables, GPT-4o and Qwen2.5-VL-72B were correlated at 0.932 {[}0.900, 0.963{]} for Urban Scaling, 0.977 {[}0.970, 0.991{]} for Distance Decay and 0.998 {[}0.998, 0.998{]} for Vitality. Figure 5 shows the corresponding mean response curves, obtained by averaging matching entries at each tested Population value, distance or Functional Mix value across the metropolitan, city or neighbourhood examples; their correlations were 0.975, 0.997 and 0.998. The complete \(3\times6\) Perception pattern was correlated at 0.929 {[}0.862, 0.940{]}. These comparisons provide the behavioural basis for the Qwen internal-state analyses below.

We summarized the difference between expected and reversed prompts as one direction in the model's intermediate states. AUC measures how often the expected prompt ranks above its reversed counterpart, with 0.5 indicating chance and 1 perfect separation. On independent confirmation prompts, AUC was 0.944 for Urban Scaling and 1.000 for Vitality, exceeding the prespecified 0.90 criterion, and 0.750 for Distance Decay. Because Urban Scaling and Vitality exceeded this threshold, their directions were also tested during generation of new numerical tables. In all six held-out tables for each task, the prespecified summary statistic was unchanged. Separate prespecified short-answer experiments used directions selected specifically for those experiments. Adding or subtracting the Scaling direction, which we call activation editing, shifted the Road and GDP answer logits; logits are the model's pre-softmax scores for the answers. Dose and erasure edits of the Vitality direction shifted the Pedestrian Activity answer logits without changing the final answer category. The separately selected short-answer directions therefore shifted answer scores, whereas the directions tested during table generation did not change the prespecified table statistic.

For Perception, the 18 GPT-4o differences between theme images and their editing controls defined the comparison pattern. We then copied the image-token states produced by each same-scene editing control into the corresponding positions for its theme-edited image. To summarize the change, each of the 18 Qwen effects was weighted by the size of the corresponding GPT-4o difference, and its sign set the expected direction. Replacing all theme-image token states produced a reduction of 3.712 {[}3.084, 4.312{]} logit units, exceeding an equal-norm random perturbation at the same positions. Replacing the quarter of image-token positions with the largest mean absolute RGB differences produced a reduction of 1.473 {[}1.203, 1.745{]}, exceeding random and low-change image-position controls. An equal-total-norm text-token perturbation caused a larger change; this comparison therefore does not establish modality specificity.

Together, response profiling, linear probing and activation intervention separate three levels of evidence: whether GPT-4o and Qwen produce similar response patterns, whether Qwen intermediate states distinguish the task relation, and whether changing those Qwen states changes an answer. These stages did not always yield the same result: selected state changes shifted short-answer scores and Perception effects, but the state changes tested during table generation did not alter the prespecified table statistic. This separation identifies more precisely when task-related information is detectable in model states and when changing those states affects an output.

Probe selection, held-out confirmation and activation-edit estimands are described in Methods 4.7 and Supplementary Results, ``Open-model analyses of selected output regularities''.

\begin{figure*}[!t]
\centering
\includegraphics[width=\textwidth,height=0.55\textheight,keepaspectratio]{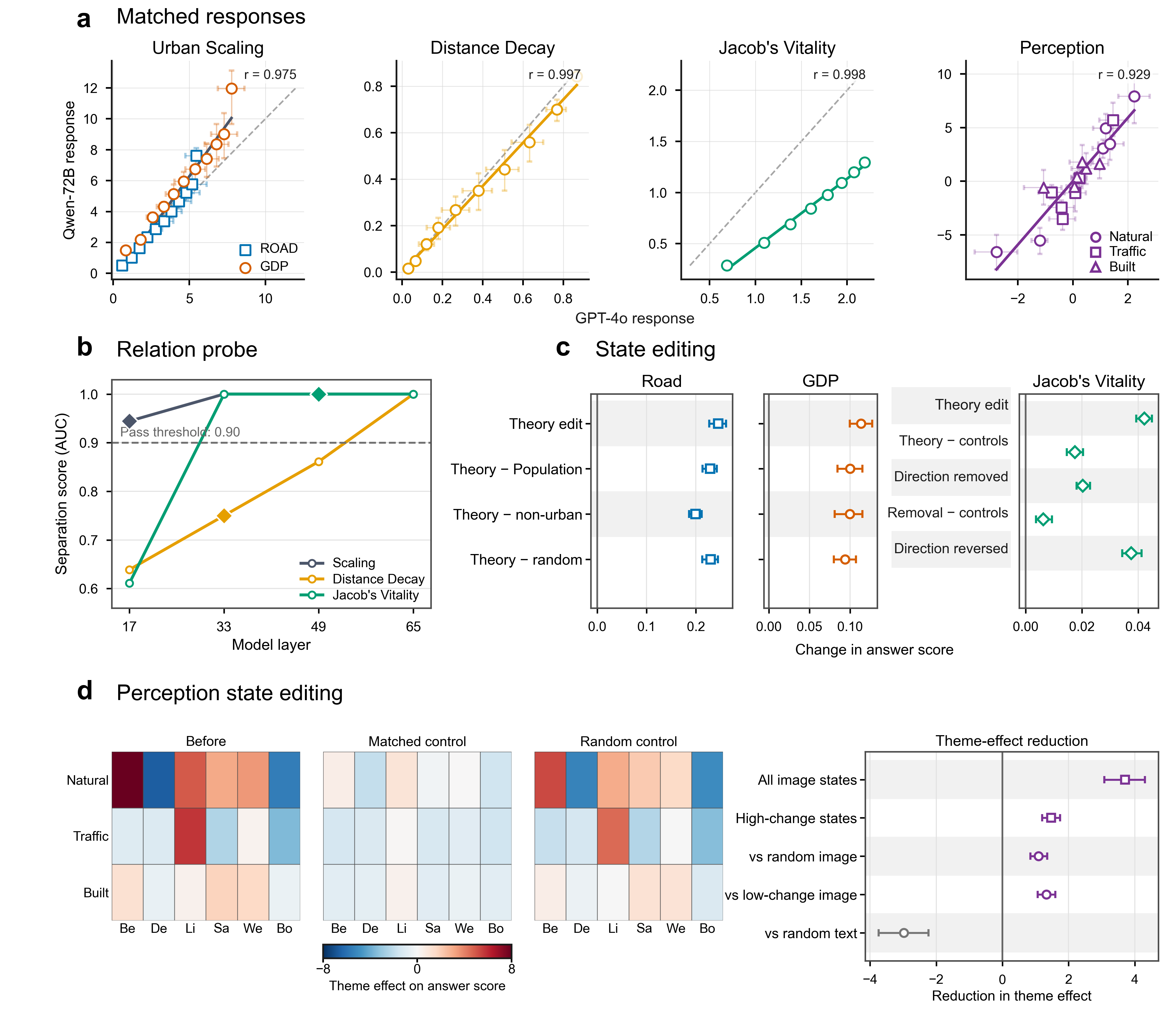}
\caption{\textbf{Matched open-model responses, relation probes and state editing.} \textbf{a,} Mean GPT-4o and Qwen2.5-VL-72B response profiles for Urban Scaling, Distance Decay, Jacob's Vitality and the \(3\times6\) Perception pattern. Numerical points average matching table positions across examples; \(r\) is the Pearson correlation and error bars are bootstrap 95\% confidence intervals. \textbf{b,} Confirmation-set AUC for separating expected from reversed relation prompts across model layers (0.5, chance; 1, perfect separation). Filled diamonds mark layers selected on development data, open circles show other candidates and the dashed line is the prespecified 0.90 threshold. \textbf{c,} Short-answer state-editing effects for Scaling (Road and GDP) and Jacob's Vitality (Pedestrian Activity), expressed as changes in restricted pre-softmax answer scores in logit units. \textbf{Theory edit} is the dose--response slope along the selected task-relation direction. Scaling contrasts subtract the absolute slopes of the named Population, non-urban or random controls. For Jacob's Vitality, \textbf{Theory \(-\) controls} subtracts the mean absolute slope across the Functional Mix input, irrelevant-number and three norm-matched random directions. \textbf{Direction removed} neutralizes the selected direction at the development-set midpoint; \textbf{Direction reversed} reflects it across that midpoint; \textbf{Removal \(-\) controls} subtracts matched control changes. Points are means and bars are bootstrap 95\% confidence intervals across held-out examples. \textbf{d,} Perception answer-score patterns before state editing, after replacing theme-image-token states with matched same-scene control states and after an equal-norm random perturbation. Heatmap cells average scene-level theme-minus-control answer scores; positive values favour the theme image. Be, De, Li, Sa, We and Bo denote Beautiful, Depressing, Lively, Safe, Wealthy and Boring. The forest plot reports the GPT-4o-reference-weighted clean-minus-edited reduction in the 18-cell Qwen pattern. \textbf{All image states} uses every image-token position; \textbf{High-change states} uses the top quartile ranked by outcome-blind mean absolute RGB difference. The three ``vs'' rows subtract matched random-image, low-change-image and random-text control effects. Bars are scene-bootstrap 95\% confidence intervals; positive values favour targeted image-state replacement. The negative random-text contrast means that this comparison does not establish modality specificity. Panels \textbf{b--d} use Qwen2.5-VL-72B.}
\label{fig:5}
\end{figure*}

\section{Discussion}\label{discussion}

AI4US connects text-based numerical tasks and image-based perception tasks within a common experimental design. Generated outputs recovered recognizable urban regularities and responded repeatably to controlled changes. Comparing them with empirical parameters, neighbourhood observations, human judgements and observed variation then showed which properties were recovered, compressed or redistributed. This connects two questions that are often examined separately: what structure generated urban data contain and how the model responds when a condition in those data is varied. Their scientific value depends on both the relationships they contain and the extent to which their parameters and variation match the phenomenon under study.

The numerical tasks often produced strong, readily estimable trends while compressing numerical ranges, local fluctuations or differences between settings. The Vitality comparison further showed that a regular generated coefficient pattern may differ in sign and strength from neighbourhood observations. In the image task, Blueprint increased dispersion within the generated set but reduced its proximity to Place Pulse. Blueprint therefore changed different properties of the outputs in different directions. Prompt design should be evaluated against the property required by a study: relationship magnitude, numerical coverage, local spatial variation or coverage of an empirical image distribution.

The empirical comparisons also clarify what it means to reproduce an urban regularity. Scaling exponents depend on city definitions, sample composition and fitting choices{[}20,21{]}. Radial density parameters depend on the definition of the centre and the spatial extent examined{[}22--25{]}. The neighbourhood activity reference is specific to the 52 Melbourne sites. Place Pulse records the judgements of a particular participant population and image collection{[}28{]}. The values used here are empirical reference points rather than universal constants. Reporting generated parameter distributions beside these references separates recovery of a relationship form, agreement with its observed magnitude and recovery of the underlying data variation.

Conditional Intervention gives the generated data an experimental role. Changing an urban condition within an otherwise fixed synthetic setting produced directional model responses in all four cases. These response profiles quantify model sensitivity to alternative urban inputs and provide hypotheses for subsequent tests with field data, policy evaluations or human participants. In this role, AI4US complements mechanistic urban simulations such as agent-based models or cellular automata, which represent urban dynamics through explicit rules.

The two human-scale cases show how empirical comparison changes interpretation. Pedestrian Activity was measured independently at 52 Melbourne sites: the external coefficient pattern differed from the generated tables, while morphology-only ranking achieved a pairwise accuracy of 0.603. Place Pulse choices provide an external reference for the image evaluator, while the theme interventions estimate GPT-4o-predicted changes in controlled synthetic scenes. Natural edits shifted the complete six-dimensional profile in a favourable direction, Traffic increased predicted liveliness while reducing predicted beauty, and Built edits produced more selective changes. These model-predicted changes define a complete set of hypotheses for subsequent human evaluation.

The open-model experiments linked selected responses to task-related intermediate states. Following related representation and model-editing studies{[}29--32{]}, we used linear readouts and activation edits to examine selected model responses. Similar responses for named and anonymous cities, together with adaptation to deliberately reversed rules, show that current task instructions and relation cues contribute beyond direct city-name retrieval or repetition of one invariant formula. Training-example memory and interpolation among learned statistical patterns may also contribute, and the present controls do not separate them. In Qwen2.5-VL-72B, selected relation types were linearly decodable from intermediate activations. Editing those activations shifted short-answer logits for Scaling and Vitality, and replacing theme-image states weakened the Perception effect profile. The prespecified statistics calculated from newly generated tables were unchanged under the numerical state edits, separating short-answer sensitivity from table-generation outcomes. In Qwen2.5-VL-7B, the external Perception profile recurred, but targeted state replacement did not outperform an equal-norm random perturbation; the 72B replacement result therefore did not extend to the tested smaller checkpoint (Supplementary Results).

AI4US is most useful for urban questions with a measurable empirical reference and a clearly defined condition that can be varied. It can compare models and prompting strategies, identify where generated data compress numerical ranges, local variation or image distributions, and formulate hypotheses for subsequent field data, policy evaluation or human studies. Extending the framework to additional model families, geographic contexts and longitudinal observations will test the stability of these patterns. In this way, AI4US connects urban-data generation with empirically referenced model experiments while keeping final tests of urban hypotheses anchored in urban observations and human experience.

\section{Methods}\label{methods}

\subsection{Study design and case selection}\label{study-design-and-case-selection}

Data Synthesis estimates relationships in generated data and compares their parameters and variation with empirical references. Conditional Intervention changes a case-specific urban condition within a fixed setting and measures the corresponding model outputs. The open-model analysis tests whether selected relation distinctions are linearly decodable from hidden states and whether activation edits shift selected output logits.

A systematic review of urban spatial big-data research in 57 representative journals provided the field-level literature landscape for case selection{[}3{]}. The four cases examined here each have a measurable relationship or outcome, an external empirical reference, a generatable object and a condition that can be changed within a fixed setting. Urban Scaling, Distance Decay and Jacobs-informed Vitality generate numerical data through text prompts. Place Pulse Perception generates and edits images, with human judgements providing the empirical reference.

\subsection{Models and prompt conditions}\label{models-and-prompt-conditions}

GPT-4o was used for numerical data synthesis, numerical interventions, empirical-site ranking and streetscape perception scoring. Streetscape generation and editing used the image-generation interface identified as \texttt{gpt-image-2} at the time of use. Hidden-state experiments primarily used Qwen2.5-VL-72B-Instruct-AWQ{[}36{]}. A within-family Perception follow-up used Qwen2.5-VL-7B-Instruct{[}37{]}; full methods and results are reported in the Supplementary Information. The three numerical tasks provided text input only, whereas Perception used image--text input. GPT-4o temperature was 0 for Scaling, Decay, intervention and perception scoring, and 0.7 for Vitality data synthesis. Model identifiers, access dates and other decoding settings are reported in Supplementary Methods; Supplementary Data 1 identifies the open-model checkpoints.

Each synthesis task compared Baseline and Blueprint prompts. Baseline specified the object, variables, units and output format. Blueprint additionally specified case-specific value coverage, spatial or system organization, or heterogeneity. All reported relationship parameters were estimated from the returned data. Some conditions, most explicitly Distance Decay, used empirical value envelopes and directional spatial organization. The Supplementary Information lists these components, identifies which quantities came from empirical sources, and reports factorial ablations and format-sensitivity analyses.

The primary numerical synthesis tasks requested all tested values and their outputs together in one table. A matched format-sensitivity analysis found that requesting one row at a time preserved the Scaling and Vitality relationships less reliably (Supplementary Methods, ``Joint and separate generation'').

\subsection{Empirical references}\label{empirical-references}

Urban Scaling parameters use Bettencourt's estimates for 2006 US Metropolitan Statistical Areas{[}19{]}. Road denotes public-road lane-miles and GDP denotes Gross Metropolitan Product. The empirical exponents are \(0.849\pm0.038\) for Road and \(1.126\pm0.023\) for GDP. We estimated a rank--size slope of \(-1.115\) from Population values for 448 urbanized areas in FHWA \emph{Highway Statistics 2006}{[}38{]}; metropolitan GDP provenance follows BEA{[}39{]}. Distribution comparisons use 3,525 Chinese physical-city polygons with complete positive Population, Road and GDP values. GDP was aggregated from a 2019 1-km revised real-GDP grid{[}40{]}, Population from PopSE 2020{[}41{]}, and Road from a 2019 Amap layer. Construction of this 3,525-city reference and its inclusion criteria are detailed in Supplementary Results, ``Urban Scaling'', and Supplementary Data 2.

Distance Decay combines land-density profiles calculated in 1-km concentric distance bands for 50 cities in China, the United States and Europe. Three periods per city provide 150 city-period parameter estimates{[}25{]}, whereas profile comparisons use 2019 GISA 1.0 impervious-surface data{[}42{]}. The inverse-S function follows published urban-density work{[}22--25{]}. City selection, source rasters and processing are reported in Supplementary Results, ``Distance Decay''.

The Vitality reference links City of Melbourne pedestrian counts and sensor locations{[}43,44{]} with CLUE blocks and associated land-use and building attributes{[}45{]}. Five morphology indicators were summarized within 400-m buffers around the sensors. Functional Mix was normalized land-use entropy, \(M=-\sum_{g=1}^{K}p_g\log p_g/\log K\), over \(K=9\) land-use groups. An eligibility procedure that did not use Pedestrian Activity outcomes retained 52 sites; variable construction is reported in Supplementary Results, ``Jacobs-informed Vitality''.

Place Pulse 2.0 contains 110,988 streetscape images from 56 cities and 1,169,078 human pairwise comparisons across six dimensions{[}28{]}. The model--human comparison sampled 1,000 comparison records per dimension. The representation-distribution comparison used 3,840 images from 16 cities that did not appear in the generation prompts; sampling and city composition are reported in Supplementary Results, ``Place Pulse Perception''{[}46{]}.

\subsection{Data synthesis and relationship estimation}\label{data-synthesis-and-relationship-estimation}

\textbf{Urban Scaling.} Each request jointly generated Population, Road and GDP for 100 fictional metropolitan areas, which were ranked by Population. The rank--size relation was

\[
\log_{10}(\mathrm{Population}_i)
=\log_{10}a_Z+\beta_Z\log_{10}r_i+\varepsilon_{Z,i} .
\tag{1}
\]

Road and GDP were fitted as

\[
\begin{aligned}
\log_{10}y_i={}&\log_{10}a_y
+\beta_y\log_{10}(\mathrm{Population}_i)\\
&+\varepsilon_{y,i},\\
&y\in\{\mathrm{Road},\mathrm{GDP}\}.
\end{aligned}
\tag{2}
\]

Here \(i\) indexes a metropolitan area after descending Population order, and \(r_i\) is its dimensionless rank, with rank 1 assigned to the largest Population. The positive quantities \(a_Z\) and \(a_y\) are scale constants, \(\beta_Z\) and \(\beta_y\) are dimensionless log--log slopes, and \(\varepsilon_{Z,i}\) and \(\varepsilon_{y,i}\) are residuals on the base-10 log scale. Before transformation, Population is measured in persons, Road in public-road lane-miles and GDP as annual Gross Metropolitan Product in billions of 2006 US dollars; the numerical values of the scale constants therefore depend on these units.

Equations (1) and (2) were estimated by ordinary least squares on the log-transformed values, matching the slope definition used for the metropolitan reference{[}19{]}. Heteroskedasticity-consistent standard errors were used within tables{[}47{]}. Each prompt condition was queried 100 times. One response per condition could not be fitted, leaving 99 fitted tables and 98 matched Baseline--Blueprint pairs. Two Baseline and one Blueprint response contained exactly 100 rows; the fitted responses contained means of 90.5 and 90.3 valid rows, respectively, all of which entered the analysis. Percentile 95\% confidence intervals were obtained from 10,000 resamples of the fitted tables. As fitting sensitivities, equation (2) was re-estimated with the Theil--Sen median-slope estimator{[}48{]} and after removing the single largest generated metropolitan area; both retained the mean Road-sublinear and GDP-superlinear directions (Supplementary Table 7).

\textbf{Distance Decay.} Each fictional city contained 100 concentric 1-km bands. The proportion of impervious urban land at distance \(r\) was fitted by

\[
f(r)=\frac{1-c}{1+\exp\left[\alpha\left(\frac{2r}{D}-1\right)\right]}+c .
\tag{3}
\]

Here \(f(r)\) is the dimensionless proportion of a distance band occupied by impervious urban land, \(r\) is radial distance from the city centre in kilometres, \(\alpha\) is a dimensionless steepness parameter, \(c\) is the dimensionless peripheral background density, and \(D\) is the characteristic radial distance in kilometres. The inflection point occurs at \(r=D/2\). Baseline and Blueprint yielded 94 and 95 estimable curves, including 90 matched pairs. Profile-level summaries used 10,000 resamples of complete city profiles.

\textbf{Jacobs-informed Vitality.} Each request jointly generated 100 neighbourhoods. Let \(Y_j\) denote Pedestrian Activity, \(P_j\) Residential Density, \(M_j\) Functional Mix, \(B_j\) Mean Block Area, \(A_j\) Aged Buildings and \(T_j\) Tall Buildings. Short Blocks was represented as \(-\log B_j\). Each table fitted

\[
\begin{aligned}
z[\log(1+Y_j)]={}&\beta_0
+\beta_Pz[\log(1+P_j)]\\
&+\beta_Mz(M_j)+\beta_Sz[-\log B_j]\\
&+\beta_Az(A_j)+\beta_Tz(T_j)+e_j .
\end{aligned}
\tag{4}
\]

Here \(j\) indexes neighbourhoods, \(z[\,\cdot\,]\) denotes standardization within each table, and \(\log\) denotes the natural logarithm. \(Y_j\) is the outcome, whereas \(P_j\), \(M_j\), \(B_j\), \(A_j\) and \(T_j\) are the five morphology predictors. \(\beta_0\) is the intercept; \(\beta_P\), \(\beta_M\), \(\beta_S\), \(\beta_A\) and \(\beta_T\) are the partial standardized coefficients for Residential Density, Functional Mix, Short Blocks, Aged Buildings and Tall Buildings, respectively; and \(e_j\) is the residual on the standardized outcome scale. All transformed predictors and the outcome were standardized within each table, so the coefficients in equation (4) are dimensionless. Each prompt condition completed 100 independent tables. HC3 heteroskedasticity-consistent standard errors were calculated within tables{[}47{]}, and table-level summaries used 10,000 resamples of complete generated tables. Equation (4) was also fitted to the 52-site empirical table using the same transformations and standardization; HC3 intervals describe the empirical coefficients, with an eight-spatial-group bootstrap retained as a dependence-sensitive uncertainty check. Separately, GPT-4o ranked anonymous pairs of sites from the five morphology indicators alone. Observed Pedestrian Activity supplied the independent reference order used to evaluate those rankings.

\textbf{Place Pulse Perception.} Image synthesis used a \(2\times2\times2\) factorial design comprising explicit scene components, spatial coordination, and regional or scene heterogeneity. Eight conditions each planned 100 images; 797 images were available for analysis. GPT-4o independently judged 1,000 Place Pulse pairs for each perception dimension. Rows were sampled without replacement with random state 42, reinitialized within each dimension. Each task compared the model choice with the \texttt{winner} label from one sampled human vote, with all three labels \(\mathcal L=\{\mathrm{LEFT},\mathrm{RIGHT},\mathrm{EQUAL}\}\) retained. LEFT means that the left image was judged higher on the named dimension, RIGHT means the same for the right image, and EQUAL denotes a tie. If \(n_{kl}\) is the number of tasks assigned human label \(k\) and model label \(l\), then \(N=\sum_{k\in\mathcal L}\sum_{l\in\mathcal L}n_{kl}\) is the number of tasks, \(p_k^{H}=\sum_{l\in\mathcal L}n_{kl}/N\) is the human marginal probability of label \(k\), and \(p_k^{M}=\sum_{h\in\mathcal L}n_{hk}/N\) is the model marginal probability of the same label. Chance-corrected agreement was

\[
\begin{aligned}
P_o&=\frac{1}{N}\sum_{k\in\mathcal L}n_{kk},\\
P_e&=\sum_{k\in\mathcal L}p_k^{H}p_k^{M},\\
\kappa&=\frac{P_o-P_e}{1-P_e}.
\end{aligned}
\tag{5}
\]

Here \(P_o\) is the observed proportion of exact human--model label matches, \(P_e\) is the agreement expected from the human and model marginal label distributions, and \(\kappa\) is Cohen's chance-corrected agreement coefficient. \(P_o\), \(P_e\) and \(\kappa\) are dimensionless.

The task pair was the analysis unit within each perception dimension, and percentile 95\% confidence intervals used 2,000 task-pair resamples. The image-synthesis factorial analysis likewise used 2,000 image resamples.

\subsection{Generated--empirical variation}\label{generatedempirical-variation}

For Urban Scaling variable \(x\), values in each 100-city table were converted to shares \(q_i=x_i/\sum_jx_j\) and log shares \(\ell_i=\log_{10}q_i\). If \(F_A\) and \(F_B\) are the empirical distribution functions of two log-share samples, distribution distance and numerical coverage were

\[
\begin{aligned}
W_1(A,B)&=\int_{-\infty}^{\infty}|F_A(u)-F_B(u)|\,\mathrm du,\\
R_{10}(x)&=\log_{10}\!\left(\frac{\max_i x_i}{\min_i x_i}\right).
\end{aligned}
\tag{6}
\]

Each generated table's \(W_1\) was averaged across 100 independent 100-city resamples from the 3,525-city empirical dataset. The empirical reference compared each empirical resample with the other empirical resamples, excluding self-comparisons. This distributional comparison used a separate generic-city prompt-component experiment, in which the variables were Population, total Road length and city GDP rather than the source-matched lane-mile and Gross Metropolitan Product definitions used for the coefficient experiment in Section 4.4. It contained 91 complete Baseline tables and 58 complete Blueprint tables. The complete 100-city table was the analysis unit, and percentile 95\% confidence intervals used 10,000 table resamples. Its results describe marginal coverage and numerical range within that prompt family; they are not used to estimate the coefficients in Fig. 2a.

For a Distance Decay profile \(y_1,\ldots,y_n\) ordered outwards from the centre, value resolution, normalized local curvature and outward reversals were

\[
\begin{aligned}
U(y)&=\frac{|\{y_i\}_{i=1}^{n}|}{n},\\
C_2(y)&=\frac{1}{(n-2)s_y}
\sum_{i=1}^{n-2}|y_{i+2}-2y_{i+1}+y_i|,\\
R_+(y)&=\sum_{i=1}^{n-1}\mathbb I(y_{i+1}>y_i).
\end{aligned}
\tag{7}
\]

where \(s_y=\max_i y_i-\min_i y_i\) and was set to one for a constant profile. Empirical and generated profiles were rounded to two decimal places before these three local-detail statistics were calculated. The city profile, not an individual distance-band value, was the analysis unit; interval estimates used 10,000 profile resamples.

For Vitality variable \(v\), let \(g_v(\cdot)\) denote the same transformation used in the relationship model: \(\log(1+x)\) for Residential Density and Pedestrian Activity, \(\log B\) for Mean Block Area, and the identity transformation for the remaining variables. The dispersion ratio for generated table \(b\) was

\[
\rho_{b,v}=\frac{\operatorname{sd}_{j}\{g_v(x_{b,j,v})\}}
{\operatorname{sd}_{e}\{g_v(x_{e,v})\}},
\tag{8}
\]

where the denominator was calculated across the 52 empirical sites. Each generated table was one analysis unit, and interval estimates used 10,000 table resamples.

Image comparisons used 3,840 Place Pulse images. Embeddings were normalized to unit vectors \(\mathbf z_i\). Mean within-set cosine distance was

\[
D_{\mathrm{within}}(G)
=\frac{2}{|G|(|G|-1)}
\sum_{i<j}\left(1-\mathbf z_i^{\mathsf T}\mathbf z_j\right).
\tag{9}
\]

For generated set \(G\) and empirical set \(E\), empirical coverage by generated images, generated-image proximity to the empirical support, and centroid displacement were

\[
\begin{aligned}
S_{E\rightarrow G}
&=\frac{1}{|E|}\sum_{e\in E}\max_{g\in G}
\mathbf z_e^{\mathsf T}\mathbf z_g,\\
S_{G\rightarrow E}
&=\frac{1}{|G|}\sum_{g\in G}\max_{e\in E}
\mathbf z_g^{\mathsf T}\mathbf z_e.
\end{aligned}
\tag{10}
\]

\[
D_{\mathrm{centroid}}(G,E)
=1-\frac{\bar{\mathbf z}_G^{\mathsf T}\bar{\mathbf z}_E}
{\|\bar{\mathbf z}_G\|_2\|\bar{\mathbf z}_E\|_2}.
\tag{11}
\]

The two nearest-neighbour directions answer different questions: \(S_{E\rightarrow G}\) asks whether empirical images are covered by the generated set, whereas \(S_{G\rightarrow E}\) asks whether generated images remain near the empirical support. Each image set was the analysis unit. Empirical references used repeated size-matched, disjoint splits of the Place Pulse set, and percentile 95\% confidence intervals used 2,000 image resamples. The primary encoder was OpenAI CLIP ViT-L/14{[}33{]}, with RGB conversion, bicubic resizing and centre cropping to 224 pixels, and the published CLIP channel normalization. Sensitivity analyses used DINOv2 ViT-L/14{[}34{]}, with bicubic resizing to 256 pixels followed by a 224-pixel centre crop, and ResNet-50 ImageNet1K\_V2{[}35{]}, with bilinear resizing to 232 pixels followed by a 224-pixel centre crop. The two sensitivity encoders used ImageNet channel normalization, and all encoder outputs were \(\ell_2\)-normalized before cosine calculations. Supplementary Methods and Supplementary Data 1 identify the encoder checkpoints and their preprocessing settings.

\subsection{Conditional interventions}\label{conditional-interventions}

Each numerical intervention began with a fictional metropolitan-system, city or neighbourhood description. Within that description, one urban condition changed while every other stated detail stayed fixed. We refer to one such description as a setting and denote it by \(b\); \(c\) denotes the changed condition and \(\mathbf{x}_{b,-c}\) the remaining details. The response between two tested values was

\[
\Delta Y_b(c_1,c_0)
=Y(c_1,\mathbf{x}_{b,-c})-Y(c_0,\mathbf{x}_{b,-c}).
\tag{12}
\]

Urban Scaling varied Population from 50,000 to 800,000 across five levels in 100 fictional metropolitan settings and recorded Road and GDP. The plotted response was \(\widetilde y_{b,k}=\log_{10}(y_{b,k}/y_{b,1})\) against \(\widetilde P_k=\log_{10}(P_k/P_1)\). Equation (2) was fitted separately within each setting; the setting-level exponents \(\hat\beta_{b,\mathrm{Road}}\) and \(\hat\beta_{b,\mathrm{GDP}}\) were the estimands. Percentile 95\% confidence intervals used 10,000 resamples of the fictional metropolitan settings.

Distance Decay queried five annular bands \(r_1<\cdots<r_5\): 0--1, 5--6, 15--16, 30--31 and 60--61 km in each of \(B=100\) fictional city settings. The fixed-denominator near--far score was

\[
\begin{aligned}
H_b^{D}&=\frac{1}{\binom{5}{2}}
\sum_{i<j}\mathbb I\{Y_b(r_i)>Y_b(r_j)\},\\
\bar H^{D}&=\frac{1}{B}\sum_{b=1}^{B}H_b^{D}.
\end{aligned}
\tag{13}
\]

A missing or invalid comparison contributed zero. The adjacent-band score \(A_b^{D}=\frac14\sum_{i=1}^{4}\mathbb I\{Y_b(r_i)>Y_b(r_{i+1})\}\) and strict-decline indicator \(C_b^{D}=\prod_{i=1}^{4}\mathbb I\{Y_b(r_i)>Y_b(r_{i+1})\}\) were auxiliary summaries; \(A_b^{D}\) was reported only for profiles with all five valid bands. The city setting was the inferential unit, and percentile 95\% confidence intervals used 10,000 setting resamples.

Vitality varied Functional Mix in \(B=38\) neighbourhood settings while holding the other four morphology variables fixed. For each setting, low, intermediate and high were the local \(q_{10}\), \(q_{50}\) and \(q_{90}\) among its eight nearest observed neighbourhoods in the fixed morphology space. Settings were retained only when all three counterfactual profiles fell within the prespecified leave-one-out empirical support threshold. The pairwise and complete-order scores were

\[
\begin{aligned}
H_b^{V}={}&\frac{1}{3}\bigl[
\mathbb I(Y_{b,H}>Y_{b,M})+\mathbb I(Y_{b,H}>Y_{b,L})\\
&\quad+\mathbb I(Y_{b,M}>Y_{b,L})\bigr],\\
C_b^{V}={}&\mathbb I(Y_{b,H}>Y_{b,M}\ \land\ Y_{b,M}>Y_{b,L}).
\end{aligned}
\tag{14}
\]

Each pairwise rate and the complete-order rate used all 38 settings as the denominator; a missing output counted as a failure. Wilson 95\% confidence intervals summarized binomial rates at the setting level.

Perception used 12 baseline scenes. For each scene, the image-editing process produced Natural, Traffic and Built theme edits and one editing control that requested no intentional visible change. GPT-4o independently scored the 12 baseline images, 36 theme-edited images and 12 editing controls on six 0--10 scales. The control-adjusted effect for perception dimension \(p\) and theme \(c\) was

\[
\Delta S_{p,c}
=\frac{1}{N_c}\sum_{i=1}^{N_c}
\left(S_{p,i,c}-S_{p,i,\mathrm{control}}\right),
\qquad N_c=12 .
\tag{15}
\]

Here \(S_{p,i,\mathrm{control}}\) is the score of the same-scene editing control. Because the theme image and its control originate from the same baseline scene, subtracting their scores is equivalent to subtracting the control's baseline-relative change from the theme image's baseline-relative change. The scene was the inferential unit. Percentile 95\% confidence intervals used 10,000 scene resamples; exact paired sign-flip tests were adjusted by Holm's method for three prespecified relationships and by the Benjamini--Hochberg procedure for the complete \(3\times6\) matrix.

\subsection{Open-model hidden-state analysis}\label{open-model-hidden-state-analysis}

Because GPT-4o does not expose internal activations, the selected tasks were repeated with Qwen2.5-VL-72B, whose intermediate states can be inspected and edited. We first established that the two models gave similarly shaped outputs from the same task descriptions. In Urban Scaling, each model filled a table showing Road and GDP as Population changed. In Distance Decay, each filled a table showing land density as distance from the city centre changed. In Vitality, each filled a table showing Pedestrian Activity as Functional Mix changed. Each numerical task used several fictional metropolitan, city or neighbourhood examples. Within one example, the surrounding details stayed fixed and one table contained all tested Population, distance or Functional Mix values.

GPT-4o and Qwen received the same description and table template for each example. We paired corresponding Road, GDP, land-density or activity values in the two returned tables. For the curves in Fig. 5a, matching values at the same Population, distance or Functional Mix position were averaged across the fictional examples. In the notation below, \(b\) identifies one pair of tables, \(k\) identifies one corresponding analysed value in those tables, \(t\) identifies the task and \(m\) identifies the model. For Scaling, \(k\) spans both Road and GDP changes. Thus \(y^{(m)}_{t,b,k}\) is the value extracted from model \(m\) at one corresponding table position, and \(\bar y^{(m)}_{t,k}\) is that value averaged across the paired tables. The entry-by-entry correlation and mean-response-curve correlation were

\[
\begin{aligned}
r_t^{\mathrm{entry}}
&=\operatorname{corr}_{b,k}\!\left(
y^{(\mathrm{GPT\! -\!4o})}_{t,b,k},y^{(\mathrm{Qwen})}_{t,b,k}\right),\\
r_t^{\mathrm{curve}}
&=\operatorname{corr}_{k}\!\left(
\bar y^{(\mathrm{GPT\! -\!4o})}_{t,k},\bar y^{(\mathrm{Qwen})}_{t,k}\right).
\end{aligned}
\tag{16}
\]

For Perception, each model produced 18 mean effects from three themes and six perception dimensions across 12 scenes. Correlation intervals used 30,000 resamples of the fictional examples or scenes, recalculating the corresponding table values or mean response curves in each resample.

We next asked whether intermediate activations distinguished the expected numerical relationship from its reverse. Each pair of prompts described the same task with opposite relationship directions. Hidden states were extracted from separate development, confirmation and held-out fictional examples. For task \(t\), \(\bar{\mathbf h}_{t,+}\) and \(\bar{\mathbf h}_{t,-}\) are the mean hidden states for the expected and reversed prompts. Their normalized difference defines the task direction \(\mathbf d_t\). Projecting a hidden state \(\mathbf h_i\) onto this direction gives its scalar readout \(s_{i,t}\):

\[
\mathbf d_t=
\frac{\bar{\mathbf h}_{t,+}-\bar{\mathbf h}_{t,-}}
{\lVert\bar{\mathbf h}_{t,+}-\bar{\mathbf h}_{t,-}\rVert_2},
\qquad s_{i,t}=\mathbf h_i^{\mathsf T}\mathbf d_t .
\tag{17}
\]

This readout follows related work that decodes spatial, temporal and geographic information from model activations{[}29,32{]}.

Decoder blocks were selected on development prompts using prespecified task-specific rules; confirmation prompts played no role in selection. On the independent confirmation split, AUC is the probability that an expected-relationship prompt receives a higher readout score than its reversed counterpart, with ties receiving half weight. AUC is 0.5 at chance and 1 under perfect separation:

\[
\operatorname{AUC}_t=\Pr(s_{+,t}>s_{-,t})+\tfrac12\Pr(s_{+,t}=s_{-,t}).
\tag{18}
\]

For table-generation edits, a confirmation AUC of at least 0.90 was required. Scaling and Vitality met this criterion, whereas Distance Decay did not. The short-answer tests in Fig. 5c were separate theory-specific experiments with predefined prompt pairs, decoder blocks and confirmation rules. The Supplementary Information gives the selection rules and selected hidden-state indices.

In the separate short-answer experiments, activation editing added or subtracted the selected task direction from the hidden state while leaving the prompt unchanged. Scaling and Vitality used doses \(\lambda\in\{-1,-0.5,0,0.5,1\}\). Let \(m_{b,t}(\lambda)\) be the pre-softmax score for the expected answer minus that for the reversed answer, reoriented so that a positive value supports the expected relationship. The dose effect for each fictional example was the ordinary-least-squares slope

\[
\begin{aligned}
\gamma_{b,t}&=\operatorname{slope}_{\lambda}\{m_{b,t}(\lambda)-m_{b,t}(0)\},\\
\operatorname{Spec}^{\mathrm{slope}}_{b,t}(\mathcal C)
&=\gamma_{b,t}-|\mathcal C|^{-1}\sum_{c\in\mathcal C}|\gamma_{b,c}|,\\
E^{\mathrm{erase}}_{b,t}&=m_{b,t}(0)-m^{\mathrm{erase}}_{b,t},\\
\operatorname{Spec}^{\mathrm{erase}}_{b,t}
&=E^{\mathrm{erase}}_{b,t}-|\mathcal C|^{-1}\sum_{c\in\mathcal C}|m_{b,c}-m_{b,t}(0)|,\\
E^{\mathrm{reflect}}_{b,t}&=m_{b,t}(0)-m^{\mathrm{reflect}}_{b,t}.
\end{aligned}
\tag{19}
\]

For Scaling, \(\mathcal C\) was reported separately for the Population and non-urban directions and as the mean over five equal-norm random directions. For Vitality, slope specificity averaged the predefined controls; erasure specificity used matched equal-norm control changes. Reflection moved the edited hidden state's projection across the development-set midpoint. The table-generation experiment instead used the task-relation directions selected by the common three-task screen together with matched controls. Table-generation editing was evaluated for Scaling and Vitality, which met both the 0.90 confirmation-AUC criterion and the baseline-table criteria; Distance Decay was excluded because its confirmation AUC was 0.750. For each qualifying task, one dose unit \(u_t\) was the median positive projection gap between paired expected and reversed development prompts at the selected hidden-state index. During greedy generation with key--value caching, the edit \(\lambda u_t\mathbf d_t\) was added at the selected decoder block to the last active sequence position on every autoregressive step: the final prompt position on the first pass and the current generated position thereafter. The task-relation direction and matched focal-input, non-urban and seeded equal-norm random directions were each tested at \(\lambda=-1\) and \(+1\) for every held-out fictional example. For the task-relation direction, let \(Q_{b,t}(\lambda)\) be the statistic in the generated table: \(\beta_{\mathrm{GDP}}-\beta_{\mathrm{Road}}\) for Scaling and \(\log(1+Y_H)-\log(1+Y_L)\) for Vitality, where \(Y_H\) and \(Y_L\) are the generated activities at the high and low Functional Mix endpoints. The symmetric task-relation effect was

\[
E_{b,t}=\frac{Q_{b,t}(+1)-Q_{b,t}(-1)}{2}.
\tag{20}
\]

The fictional example was the analysis unit for both short-answer and table-generation analyses. Percentile 95\% confidence intervals used 30,000 resamples of these examples. Adding, removing or replacing candidate states follows related function-vector and activation-editing methods{[}30,31,49{]}.

Finally, Perception tested whether the Qwen response used information carried by the edited image. The model received a theme image and its same-scene editing control and selected LEFT, RIGHT or EQUAL for a named perception dimension. If the theme appeared on the left, its signed score was \(s=\ell_{\mathrm{LEFT}}-\ell_{\mathrm{RIGHT}}\); if it appeared on the right, \(s=\ell_{\mathrm{RIGHT}}-\ell_{\mathrm{LEFT}}\). Here \(\ell\) is the model's pre-softmax score for an answer. The EQUAL score was retained but did not enter this signed contrast. Averaging across scenes produced 18 signed effects, one for each combination of three themes and six dimensions. These effects were compared with the corresponding GPT-4o comparison pattern using equation (16).

We then copied the image-token states produced by the same-scene editing control into the corresponding theme-image positions at the selected decoder layer. At that decoder layer, no other state positions were replaced. Let \(r_k\) be the prespecified GPT-4o comparison effect for theme--dimension combination \(k\), and let \(s^{\mathrm{before}}_{b,k}\) and \(s^{q}_{b,k}\) be the Qwen score before and after state replacement \(q\) for scene \(b\). The reduction in the 18-effect pattern, weighted by the GPT-4o comparison effects, was

\[
\begin{aligned}
A_b(q)&=\frac{\sum_{k=1}^{18}r_k
\{s^{\mathrm{before}}_{b,k}-s^{q}_{b,k}\}}
{\sum_{k=1}^{18}|r_k|},\\
\operatorname{Spec}_b(q,q_0)&=A_b(q)-|A_b(q_0)|.
\end{aligned}
\tag{21}
\]

Controls used equal-norm random image perturbations and equal-total-norm text-token perturbations. For spatial localization, the theme image and editing control were resized to the same arrangement as the image tokens. We then compared token positions with high, low and random mean absolute RGB differences. The scene was the analysis unit, and \(A_b\) is measured in model-logit units. Percentile 95\% confidence intervals used 30,000 scene resamples.

The follow-up Qwen2.5-VL-7B Perception check used restricted next-token scores over LEFT, RIGHT and EQUAL and compared editing-control state replacement with an equal-norm random replacement at the same positions. The 7B follow-up protocol was specified after the 72B result but before the 7B run. Supplementary Methods and Results report the checkpoint, hardware, image-processing, denominator and inference details; Supplementary Data 1 identifies the checkpoint.

\section*{Data availability}\label{data-availability}
\addcontentsline{toc}{section}{Data availability}

Published empirical sources are cited in Methods and the Supplementary Information. Numerical data underlying Figs. 2--5 and Supplementary Figs. 1--12 are provided in Source Data. Prompt templates and inference settings are provided in Supplementary Data 1, empirical-source, inclusion and city-centre information in Supplementary Data 2, and the parsed model outputs used in the analyses, including the deidentified Place Pulse task index and 7B Perception follow-up, in Supplementary Data 3. Third-party rasters, street-view images, raw Place Pulse votes and City of Melbourne records remain available from their cited sources and are not redistributed.

\section*{Code availability}\label{code-availability}
\addcontentsline{toc}{section}{Code availability}

Code and source tables for the reported analyses are available in the accompanying code archive and in the public repository at \url{https://github.com/24kchengYe/AI4UrbanScience}. Third-party model weights and restricted or non-redistributable raw inputs are available from their cited providers or sources and are not redistributed.

\section*{References}\label{references}
\addcontentsline{toc}{section}{References}

\begin{enumerate}
\def\labelenumi{\arabic{enumi}.}
\item
  Geddes, P. \emph{Cities in Evolution: An Introduction to the Town Planning Movement and to the Study of Civics}. Williams \& Norgate (1915).
\item
  Long, Y. \& Zhang, E. City laboratory: embracing new data, new elements, and new pathways to invent new cities. \emph{Environ. Plan. B Urban Anal. City Sci.} \textbf{51}, 1068--1072 (2024). \url{https://doi.org/10.1177/23998083241246630}
\item
  Tu, T., Zhang, E. \& Long, Y. Profile and theoretical advances in urban big data studies: a systematic review of 57 representative journals (2013--2023). \emph{Environ. Plan. B Urban Anal. City Sci.} \textbf{53}, 673--688 (2026). \url{https://doi.org/10.1177/23998083251346582}
\item
  Kitchin, R. The ethics of smart cities and urban science. \emph{Phil. Trans. R. Soc. A} \textbf{374}, 20160115 (2016). \url{https://doi.org/10.1098/rsta.2016.0115}
\item
  Sudmant, A., Creutzig, F. \& Mi, Z. Replicate and generalize to make urban research coherent. \emph{Int. J. Urban Sci.} \textbf{29}, 582--603 (2025). \url{https://doi.org/10.1080/12265934.2024.2382706}
\item
  Sun, G., Choe, E. Y. \& Webster, C. Natural experiments in healthy cities research: how can urban planning and design knowledge reinforce the causal inference? \emph{Town Plan. Rev.} \textbf{94}, 87--108 (2023). \url{https://doi.org/10.3828/tpr.2022.14}
\item
  Batty, M. A generic framework for computational spatial modelling. In \emph{Agent-Based Models of Geographical Systems} 19--50 (Springer, 2012). \url{https://doi.org/10.1007/978-90-481-8927-4_2}
\item
  Birhane, A., Kasirzadeh, A., Leslie, D. \& Wachter, S. Science in the age of large language models. \emph{Nat. Rev.~Phys.} \textbf{5}, 277--280 (2023). \url{https://doi.org/10.1038/s42254-023-00581-4}
\item
  Binz, M. et al.~How should the advancement of large language models affect the practice of science? \emph{Proc. Natl Acad. Sci. USA} \textbf{122}, e2401227121 (2025). \url{https://doi.org/10.1073/pnas.2401227121}
\item
  Jumper, J. et al.~Highly accurate protein structure prediction with AlphaFold. \emph{Nature} \textbf{596}, 583--589 (2021). \url{https://doi.org/10.1038/s41586-021-03819-2}
\item
  Zeni, C. et al.~A generative model for inorganic materials design. \emph{Nature} \textbf{639}, 624--632 (2025). \url{https://doi.org/10.1038/s41586-025-08628-5}
\item
  Gottweis, J. et al.~Accelerating scientific discovery with Co-Scientist. \emph{Nature} \textbf{655}, 487--496 (2026). \url{https://doi.org/10.1038/s41586-026-10644-y}
\item
  Balsa-Barreiro, J., Cebrián, M., Menéndez, M. \& Axhausen, K. Leveraging generative AI models in urban science. In \emph{Principles and Advances in Population Neuroscience}, \emph{Current Topics in Behavioral Neurosciences} \textbf{68}, 239--275 (Springer Nature Switzerland, 2024). \url{https://doi.org/10.1007/7854_2024_482}
\item
  Ye, X. et al.~Artificial intelligence in urban science: why does it matter? \emph{Ann. GIS} \textbf{31}, 181--189 (2025). \url{https://doi.org/10.1080/19475683.2025.2469110}
\item
  Wang, J. et al.~Large language models as urban residents: an LLM agent framework for personal mobility generation. \emph{Adv. Neural Inf. Process. Syst.} \textbf{37}, 124547--124574 (2024). \url{https://doi.org/10.52202/079017-3957}
\item
  Balsa-Barreiro, J., Rabbani, S., Difallah, D., Menendez, M. \& Cebrian, M. A large-scale LLM-generated dataset for exploring social interactions and urban experiences across 21 global cities. \emph{Discover Data} \textbf{4}, 5 (2026). \url{https://doi.org/10.1007/s44248-026-00105-2}
\item
  Manvi, R., Khanna, S., Burke, M., Lobell, D. B. \& Ermon, S. Large language models are geographically biased. In \emph{Proceedings of the 41st International Conference on Machine Learning}, \emph{PMLR} \textbf{235}, 34654--34669 (2024). \url{https://proceedings.mlr.press/v235/manvi24a.html}
\item
  Abbasi, O. R., Welscher, F., Weinberger, G. \& Scholz, J. The world as large language models see it: exploring the reliability of LLMs in representing geographical features. arXiv:2506.00203 (2025). \url{https://arxiv.org/abs/2506.00203}
\item
  Bettencourt, L. M. A. The origins of scaling in cities. \emph{Science} \textbf{340}, 1438--1441 (2013). \url{https://doi.org/10.1126/science.1235823}
\item
  Arcaute, E. et al.~Constructing cities, deconstructing scaling laws. \emph{J. R. Soc. Interface} \textbf{12}, 20140745 (2015). \url{https://doi.org/10.1098/rsif.2014.0745}
\item
  Cabrera-Arnau, C. \& Bishop, S. R. The effect of dragon-kings on the estimation of scaling law parameters. \emph{Sci. Rep.} \textbf{10}, 20226 (2020). \url{https://doi.org/10.1038/s41598-020-77232-6}
\item
  Jiao, L. Urban land density function: a new method to characterize urban expansion. \emph{Landsc. Urban Plan.} \textbf{139}, 26--39 (2015). \url{https://doi.org/10.1016/j.landurbplan.2015.02.017}
\item
  Pun-Cheng, L. S. C. Distance decay. In \emph{International Encyclopedia of Geography} 1--5 (Wiley, 2016). \url{https://doi.org/10.1002/9781118786352.wbieg0179}
\item
  Batty, M. \& Kim, K. S. Form follows function: reformulating urban population density functions. \emph{Urban Stud.} \textbf{29}, 1043--1069 (1992). \url{https://doi.org/10.1080/00420989220081041}
\item
  Jiao, L. \& Dong, T. Inverse S-shape rule of urban land density distribution and its applications. \emph{J. Geomatics} \textbf{43}, 8--16 (2018). \url{https://doi.org/10.14188/j.2095-6045.2018217}
\item
  Jacobs, J. \emph{The Death and Life of Great American Cities}. Random House (1961).
\item
  Huang, J. et al.~Re-examining Jane Jacobs' doctrine using new urban data in Hong Kong. \emph{Environ. Plan. B Urban Anal. City Sci.} \textbf{50}, 76--93 (2023). \url{https://doi.org/10.1177/23998083221106186}
\item
  Dubey, A., Naik, N., Parikh, D., Raskar, R. \& Hidalgo, C. A. Deep learning the city: quantifying urban perception at a global scale. In \emph{European Conference on Computer Vision 2016} 196--212 (Springer, 2016). \url{https://doi.org/10.1007/978-3-319-46448-0_12}
\item
  Gurnee, W. \& Tegmark, M. Language models represent space and time. In \emph{The Twelfth International Conference on Learning Representations} (2024). \url{https://openreview.net/forum?id=jE8xbmvFin}
\item
  Todd, E., Li, M. L., Sharma, A. S., Mueller, A., Wallace, B. C. \& Bau, D. Function vectors in large language models. In \emph{The Twelfth International Conference on Learning Representations} (2024). \url{https://openreview.net/forum?id=AwyxtyMwaG}
\item
  Meng, K., Bau, D., Andonian, A. \& Belinkov, Y. Locating and editing factual associations in GPT. \emph{Adv. Neural Inf. Process. Syst.} \textbf{35}, 17359--17372 (2022). \url{https://doi.org/10.52202/068431-1262}
\item
  De Sabbata, S., Mizzaro, S. \& Roitero, K. Geospatial mechanistic interpretability of large language models. In \emph{Geography According to Foundation Models}, \emph{Frontiers in Artificial Intelligence and Applications} \textbf{422}, 135--150 (IOS Press, 2026). \url{https://doi.org/10.3233/FAIA260476}
\item
  Radford, A. et al.~Learning transferable visual models from natural language supervision. In \emph{Proceedings of the 38th International Conference on Machine Learning}, \emph{PMLR} \textbf{139}, 8748--8763 (2021). \url{https://proceedings.mlr.press/v139/radford21a.html}
\item
  Oquab, M. et al.~DINOv2: learning robust visual features without supervision. \emph{Trans. Mach. Learn. Res.} (2024). \url{https://openreview.net/forum?id=a68SUt6zFt}
\item
  He, K., Zhang, X., Ren, S. \& Sun, J. Deep residual learning for image recognition. In \emph{Proceedings of the IEEE Conference on Computer Vision and Pattern Recognition} 770--778 (2016). \url{https://doi.org/10.1109/CVPR.2016.90}
\item
  Qwen Team. Qwen2.5-VL-72B-Instruct-AWQ model card. Hugging Face (2025). \url{https://huggingface.co/Qwen/Qwen2.5-VL-72B-Instruct-AWQ} (accessed 14 August 2026).
\item
  Qwen Team. Qwen2.5-VL-7B-Instruct model card. Hugging Face (2025). \url{https://huggingface.co/Qwen/Qwen2.5-VL-7B-Instruct} (accessed 25 August 2026).
\item
  Federal Highway Administration. \emph{Highway Statistics 2006, Table HM-72: Urbanized Areas} (2007). \url{https://www.fhwa.dot.gov/policy/ohim/hs06/pdf/hm72.pdf} (accessed 15 August 2026).
\item
  U.S. Bureau of Economic Analysis. \emph{Gross Domestic Product by Metropolitan Area, 2006 and Revised 2004--2005} (2008). \url{https://www.bea.gov/news/2008/gross-domestic-product-metropolitan-area-2006-and-revised-2004-2005} (accessed 15 August 2026).
\item
  Chen, J. et al.~Global 1 km × 1 km gridded revised real gross domestic product and electricity consumption during 1992--2019 based on calibrated nighttime light data. \emph{Sci. Data} \textbf{9}, 202 (2022). \url{https://doi.org/10.1038/s41597-022-01322-5}
\item
  Chen, Y., Xu, C., Ge, Y., Zhang, X. \& Zhou, Y. A 100 m gridded population dataset of China's seventh census using ensemble learning and big geospatial data. \emph{Earth Syst. Sci. Data} \textbf{16}, 3705--3718 (2024). \url{https://doi.org/10.5194/essd-16-3705-2024}
\item
  Huang, X., Li, J., Yang, J., Zhang, Z., Li, D. \& Liu, X. 30 m global impervious surface area dynamics and urban expansion pattern observed by Landsat satellites: from 1972 to 2019. \emph{Sci. China Earth Sci.} \textbf{64}, 1922--1933 (2021). \url{https://doi.org/10.1007/s11430-020-9797-9}
\item
  City of Melbourne. \emph{Pedestrian Counting System (counts per hour)}. \url{https://data.melbourne.vic.gov.au/explore/dataset/pedestrian-counting-system-monthly-counts-per-hour/} (accessed 15 August 2026).
\item
  City of Melbourne. \emph{Pedestrian Counting System - Sensor Locations}. \url{https://data.melbourne.vic.gov.au/explore/dataset/pedestrian-counting-system-sensor-locations/} (accessed 15 August 2026).
\item
  City of Melbourne. \emph{Blocks for Census of Land Use and Employment (CLUE)}. \url{https://data.melbourne.vic.gov.au/explore/dataset/blocks-for-census-of-land-use-and-employment-clue/} (accessed 15 August 2026).
\item
  Salesses, P. \& Hidalgo, C. A. \emph{Place Pulse}. figshare Dataset (2020). \url{https://doi.org/10.6084/m9.figshare.11859993.v1}
\item
  White, H. A heteroskedasticity-consistent covariance matrix estimator and a direct test for heteroskedasticity. \emph{Econometrica} \textbf{48}, 817--838 (1980). \url{https://doi.org/10.2307/1912934}
\item
  Sen, P. K. Estimates of the regression coefficient based on Kendall's tau. \emph{J. Am. Stat. Assoc.} \textbf{63}, 1379--1389 (1968). \url{https://doi.org/10.1080/01621459.1968.10480934}
\item
  Marinescu, R., Gruber, V.-E. \& Fajardo Vargas, D. Medical interpretability and knowledge maps of large language models. In \emph{The Fourteenth International Conference on Learning Representations} (2026). \url{https://openreview.net/forum?id=BhqFWlYKUi}
\end{enumerate}

\section*{Acknowledgements}\label{acknowledgements}
\addcontentsline{toc}{section}{Acknowledgements}

This work was supported by the National Natural Science Foundation of China (grant nos. 62394331 and 62394335).

\section*{Author contributions}\label{author-contributions}
\addcontentsline{toc}{section}{Author contributions}

Y.Z. contributed to Conceptualization, Methodology and Writing -- original draft. R.Z. contributed to Data curation and Investigation. Y.Z. and R.Z. contributed to Visualization. Z.H., X.W. and Y.M. contributed to Investigation. Y.L. contributed to Supervision, Project administration and Writing -- review \& editing. All authors reviewed and approved the manuscript.

\section*{Competing interests}\label{competing-interests}
\addcontentsline{toc}{section}{Competing interests}

The authors declare no competing interests.

\clearpage
\onecolumn
\setcounter{secnumdepth}{0}
\setcounter{table}{0}
\renewcommand{\thetable}{S\arabic{table}}
\setcounter{figure}{0}
\renewcommand{\thefigure}{S\arabic{figure}}
\setlength{\tabcolsep}{3pt}
\small
\section{Supplementary Information}\label{supplementary-information}

\emph{GenAI Models Capture Urban Science but Oversimplify Complexity}

\subsection{Supplementary Methods}\label{supplementary-methods}

\subsubsection{Cases and empirical references}\label{cases-and-empirical-references}

AI4US evaluates two experimental tasks across four cases: synthesizing urban datasets and testing how model outputs respond when an urban condition changes. Urban Scaling examines cross-metropolitan relationships among Population, public-road lane-miles (Road) and Gross Metropolitan Product (GDP). Distance Decay examines land density across distances from the city centre. Jacobs-informed Vitality examines five neighbourhood-morphology indicators and Pedestrian Activity. Place Pulse Perception examines six human-labelled streetscape judgements. The first three cases use text-only prompts to generate numerical tables; the fourth generates and edits streetscape images and evaluates them through image--text inputs, with Place Pulse judgements as the empirical reference.

Each case has an external empirical reference. Scaling parameters are compared with published estimates for United States metropolitan systems and empirical road-network data. Distance Decay parameters are compared with 150 city-period curves from 50 cities. Vitality is compared with morphology and pedestrian-count observations at 52 Melbourne sites. Perception judgements are compared with Place Pulse 2.0 human votes. The archived Chinese physical-city polygon data are used only for the distributional comparison of Population, Road and GDP.

An earlier systematic review of 1,425 articles from 57 journals on urban spatial big-data research identified recurring themes in urban transport, spatial quality, vitality, form and perception{[}1{]}. The four selected cases each have a measurable relationship or outcome, an external empirical reference, an object that can be generated and an urban condition that can be varied experimentally (Supplementary Table 1).

Data Synthesis uses two prompt conditions. The Baseline prompt defines the requested variables, physical units and output format. The Blueprint prompt adds case-specific information about plausible numerical coverage, spatial organization and heterogeneity. Intervention experiments change one urban condition while keeping the remaining details of the same fictional urban description fixed. Scaling, Decay and Vitality specify that condition numerically in the text input; Perception uses matched edits to change the visual content of the same base streetscape before scoring. Open-model analyses then examine intermediate activations in Qwen2.5-VL-72B-Instruct-AWQ; Perception additionally includes a Qwen2.5-VL-7B-Instruct within-family follow-up whose plan was fixed after the 72B result and before the 7B run.

\textbf{Supplementary Table 1 \textbar{} Cases, generated objects and empirical references.}

{\def\LTcaptype{none} 
\begin{longtable}[]{@{}
  >{\raggedright\arraybackslash}p{(\linewidth - 8\tabcolsep) * \real{0.2000}}
  >{\raggedright\arraybackslash}p{(\linewidth - 8\tabcolsep) * \real{0.2000}}
  >{\raggedright\arraybackslash}p{(\linewidth - 8\tabcolsep) * \real{0.2000}}
  >{\raggedright\arraybackslash}p{(\linewidth - 8\tabcolsep) * \real{0.2000}}
  >{\raggedright\arraybackslash}p{(\linewidth - 8\tabcolsep) * \real{0.2000}}@{}}
\toprule\noalign{}
\begin{minipage}[b]{\linewidth}\raggedright
Case
\end{minipage} & \begin{minipage}[b]{\linewidth}\raggedright
Generated object
\end{minipage} & \begin{minipage}[b]{\linewidth}\raggedright
Focal relationship
\end{minipage} & \begin{minipage}[b]{\linewidth}\raggedright
Empirical reference
\end{minipage} & \begin{minipage}[b]{\linewidth}\raggedright
Main intervention
\end{minipage} \\
\midrule\noalign{}
\endhead
\bottomrule\noalign{}
\endlastfoot
Urban Scaling & Metropolitan tables & Population--Road and Population--GDP power laws & Published US metropolitan exponents and FHWA data & Population \\
Distance Decay & Radial density profiles & Land density--Distance from centre & 150 curves from 50 cities & Distance from centre \\
Jacobs-informed Vitality & Neighbourhood tables & Five morphology indicators--Pedestrian Activity & 52 Melbourne sites & Functional Mix \\
Place Pulse Perception & Streetscape images and six scores & Streetscape image--perception judgement & Place Pulse 2.0 human votes and images & Natural, Traffic and Built themes \\
\end{longtable}
}

\subsubsection{Theoretical quantities and empirical targets}\label{theoretical-quantities-and-empirical-targets}

Scaling laws are a fundamental concept in complex systems, describing how certain quantities change with system size, typically following a power-law relationship. In the context of urban systems, urban scaling laws serve as functional models for understanding cities as complex systems.

Urban Scaling uses three relationships{[}2--4{]}. After metropolitan areas are ordered from largest to smallest Population, \(i\) indexes an area and \(r_i\) is its dimensionless rank. The rank--size relationship is

\[
\log_{10}(\mathrm{Population}_i)
=\log_{10}a_Z+\beta_Z\log_{10}r_i+\varepsilon_{Z,i}.
\]

Here \(a_Z\) is a positive scale constant, \(\beta_Z\) is the dimensionless rank--size slope and \(\varepsilon_{Z,i}\) is a residual on the base-10 log scale. Road and GDP follow \(y_i=a_y\mathrm{Population}_i^{\beta_y}\) for \(y\in\{\mathrm{Road},\mathrm{GDP}\}\), estimated as a linear relationship in base-10 log space. The published 2006 US metropolitan references are \(\beta_{\mathrm{Road}}=0.849\pm0.038\) and \(\beta_{\mathrm{GDP}}=1.126\pm0.023\); the empirical rank--size slope used here is \(\beta_Z=-1.115\). Population is measured in persons, Road in public-road lane-miles and GDP as annual Gross Metropolitan Product in billions of 2006 US dollars, although the shorter labels are used in figures.

Distance decay describes a decline in a measured quantity as distance increases. Here we examine the inverse-S decline of land density with distance from the city centre.

Distance Decay uses Jiao's inverse-S land-density function{[}5--6{]}

\[
f(r)=\frac{1-c}{1+\exp\left[\alpha\left(\frac{2r}{D}-1\right)\right]}+c .
\]

where \(f(r)\) is the dimensionless proportion of a distance band occupied by impervious urban surface, \(r\) is radial distance from the city centre in kilometres, \(\alpha\) is a dimensionless steepness parameter, \(c\) is the dimensionless peripheral background density, and \(D\) is the characteristic radial distance in kilometres. The inflection point occurs at \(r=D/2\). Exponential and power-law functions are fitted to the same curves as alternatives.

Jacobs-informed Vitality operationalizes density, mixed use, short blocks and aged buildings from Jacobs' account of street life{[}7,8{]}. Tall Buildings was added as a locally measurable intensity indicator. Pedestrian Activity is the outcome.

GenAI perceptual judgements were benchmarked against sampled human choices from the MIT Place Pulse dataset. Cohen's kappa measures chance-corrected agreement between GPT-4o choices and one sampled Place Pulse human vote per task.

Place Pulse Perception uses the six dimensions Safe, Beautiful, Wealthy, Lively, Boring and Depressing defined by the human pairwise-comparison dataset{[}9{]}.

\subsubsection{Empirical-data construction}\label{empirical-data-construction}

\textbf{Supplementary Table 2 \textbar{} Empirical datasets and their roles in the study.}

{\def\LTcaptype{none} 
\begin{longtable}[]{@{}
  >{\raggedright\arraybackslash}p{(\linewidth - 6\tabcolsep) * \real{0.2308}}
  >{\raggedright\arraybackslash}p{(\linewidth - 6\tabcolsep) * \real{0.2308}}
  >{\raggedleft\arraybackslash}p{(\linewidth - 6\tabcolsep) * \real{0.3077}}
  >{\raggedright\arraybackslash}p{(\linewidth - 6\tabcolsep) * \real{0.2308}}@{}}
\toprule\noalign{}
\begin{minipage}[b]{\linewidth}\raggedright
Reference use
\end{minipage} & \begin{minipage}[b]{\linewidth}\raggedright
Source and spatial unit
\end{minipage} & \begin{minipage}[b]{\linewidth}\raggedleft
Final analysis sample
\end{minipage} & \begin{minipage}[b]{\linewidth}\raggedright
Role in the study
\end{minipage} \\
\midrule\noalign{}
\endhead
\bottomrule\noalign{}
\endlastfoot
Scaling parameters & 2006 US metropolitan and urbanized-area statistics; FHWA public-road lane-miles and BEA Gross Metropolitan Product & Published metropolitan estimates; 448 urbanized-area populations for the rank--size reference & Scaling-parameter comparison \\
Scaling distributions & 2019 1-km GDP{[}10{]}, 2020 PopSE population{[}11{]} and a locally archived 2019 Amap road layer aggregated to the Chinese physical-city polygon dataset & 3,675 entities; 3,525 positive complete records & Population, Road and GDP distribution comparison \\
Scaling boundary sensitivity & Chinese municipal administrative units & 681 units; 676 positive complete records & Sensitivity to the urban spatial unit \\
Distance parameters & 28 Chinese and 22 US/European cities, three periods each{[}6{]} & 150 city-period inverse-S parameter sets & Empirical range of \(\alpha\), \(c\) and \(D\) \\
Distance local detail & GISA 1.0 2019 impervious-surface raster{[}12{]}, 1-km annuli around uniform GeoNames centres{[}13{]} & 50 radial profiles & \(R^2\), value resolution, local curvature and outward reversals \\
Vitality & City of Melbourne CLUE blocks/\allowbreak buildings/\allowbreak dwellings/\allowbreak land use and Pedestrian Counting System{[}14--16{]} & 52 eligible sensor sites with 400-m morphology buffers & Coefficients, spatial holdout and empirical-site ranking \\
Perception judgements & Place Pulse 2.0 pairwise human choices{[}9,17{]} & 1,000 sampled comparison records per dimension & Human-judgement agreement \\
Perception image distribution & Place Pulse images from 16 held-out cities{[}17{]} & 3,840 distinct images & Image-representation distribution comparison \\
\end{longtable}
}

The physical-city dataset contained 3,675 polygons, of which 3,525 had finite positive Population, Road and GDP values and entered the distribution analysis. Road is the length of the locally archived 2019 Amap road layer summarized within each polygon. Supplementary Data 2 describes the data sources and inclusion criteria and identifies third-party or locally archived raw inputs that are not redistributed.

For each Place Pulse dimension, 1,000 rows were sampled without replacement from the archived \texttt{final\_data.csv} table using \texttt{pandas.DataFrame.sample(n=1000,\ random\_state=42)}, with the random state reinitialized for each dimension. The six samples were concatenated in the order Safe, Beautiful, Lively, Wealthy, Boring and Depressing. The \texttt{winner} field in each sampled row supplied the primary reference label; a same-pair plurality label was evaluated as a sensitivity analysis. A deidentified index of the 6,000 analysed tasks is supplied in Supplementary Data 3 under the Place Pulse CC BY 4.0 licence{[}17{]}.

\subsubsection{Generation design}\label{generation-design}

\paragraph{Models and inference conditions}\label{models-and-inference-conditions}

Numerical synthesis, intervention and empirical-site ranking used GPT-4o through the MindCraft OpenAI-compatible HTTPS chat-completions service. Calls analysed here ran from 25 July to 14 August 2026, and both requests and responses identified the model as \texttt{gpt-4o}. Intervention, ranking and Perception scoring used temperature 0, whereas Vitality synthesis used 0.7. Generated and edited streetscapes were scored separately without condition labels. The Place Pulse comparison uses GPT-4o choices obtained on 17 September 2025 with the corresponding task specification; Supplementary Data 1 gives the available model label, date and settings for this analysis.

Baseline streetscapes and their theme or cue edits were produced from 27 July to 3 August 2026 through an OpenAI image-generation interface that identified itself as \texttt{gpt-image-2}. A contemporary-model sensitivity used the \texttt{gpt-5.6} gateway alias through the OpenAI-compatible \texttt{sub2api} HTTPS chat-completions service on 26 July 2026 from 09:57 to 13:59 UTC; requests and responses identified the same alias. Distance Decay used temperature 0.7 and the other replicated numerical tasks used temperature 0. The \texttt{gpt-5.6} and GPT-4o sensitivity analyses used the same task wording and requested values of Population, distance or Functional Mix.

Hidden-state analyses used Qwen2.5-VL-72B-Instruct-AWQ on three RTX 4090 GPUs. Numerical tasks used text-only inputs and Perception used image--text inputs. The within-family Perception follow-up used Qwen/Qwen2.5-VL-7B-Instruct, float16 with SDPA on one 24-GB NVIDIA GeForce RTX 4090, and restricted next-token logits over LEFT, RIGHT and EQUAL. Supplementary Data 1 identifies the open-model checkpoints.

Supplementary Data 1 provides the complete prompt text, model identifiers and inference settings, and Supplementary Data 3 provides parsed outputs.

\paragraph{Blueprint composition}\label{blueprint-composition}

Three prompt components were varied in a \(2\times2\times2\) design. Each numerical condition used 100 independent generations; each image condition planned 100 generated images.

\textbf{Supplementary Table 3 \textbar{} Blueprint components and principal readouts.}

{\def\LTcaptype{none} 
\begin{longtable}[]{@{}
  >{\raggedright\arraybackslash}p{(\linewidth - 6\tabcolsep) * \real{0.2500}}
  >{\raggedright\arraybackslash}p{(\linewidth - 6\tabcolsep) * \real{0.2500}}
  >{\raggedright\arraybackslash}p{(\linewidth - 6\tabcolsep) * \real{0.2500}}
  >{\raggedright\arraybackslash}p{(\linewidth - 6\tabcolsep) * \real{0.2500}}@{}}
\toprule\noalign{}
\begin{minipage}[b]{\linewidth}\raggedright
Case
\end{minipage} & \begin{minipage}[b]{\linewidth}\raggedright
Prompt components
\end{minipage} & \begin{minipage}[b]{\linewidth}\raggedright
Basis of added information
\end{minipage} & \begin{minipage}[b]{\linewidth}\raggedright
Principal readouts
\end{minipage} \\
\midrule\noalign{}
\endhead
\bottomrule\noalign{}
\endlastfoot
Urban Scaling & Metropolitan-system organization; Population coverage; differences in Road and GDP among similarly sized metropolitan areas & FHWA and BEA sources define plausible Population, Road and GDP coverage. The Blueprint also describes variation among similarly sized metropolitan areas; power-law exponents are estimated from the returned tables. & Power-law fits, parameter error and marginal distributions \\
Distance Decay & Centre--transition--periphery organization; wide density coverage; between-city and local variation & The published 50-city profiles{[}5--6{]} inform the density envelope and radial organization. Inverse-S parameters are fitted to the returned profiles. & Inverse-S parameters, value resolution and local variation \\
Vitality & Neighbourhood-type organization; ranges for morphology and activity; between-neighbourhood heterogeneity & The 52 Melbourne sites define the variable ranges. Neighbourhood organization and heterogeneity structure the generated tables, from which the five coefficient estimates are obtained. & Five regression coefficients, empirical activity ranking and variable coverage \\
Perception & Spatial scene components; component co-occurrence; regional and scene heterogeneity & The Blueprint specifies spatial components, their co-occurrence and variation across regions and scenes. Place Pulse images enter later as the external representation reference. & Within-set representation diversity and location relative to Place Pulse \\
\end{longtable}
}

The Baseline prompts define the generated object, variables, physical units and output schema. The complete Blueprint adds the three components in the table above, while relationship parameters are estimated from the generated outputs. Supplementary Data 1 provides the Baseline, component, complete Blueprint, intervention, editing-control and other control prompts. The editing control passes the same baseline scene through the same image-editing process with an instruction to make no intentional visible change. ``Blueprint'' in the main article refers to the complete three-component condition.

\paragraph{Generation scale}\label{generation-scale}

The main numerical synthesis tasks jointly generated 100 metropolitan areas, 100 distance bands or 100 neighbourhoods per response. Sensitivity analyses requested 20, 50, 100, 200 or 300 rows for Scaling and Vitality. Distance Decay varied the number and width of physical distance bands. Perception varied the number of image pairs judged in a single request.

Longer requests reduced complete-return rates. Scaling returned complete tables in 100\%, 90\%, 91\%, 56\% and 5\% of generations at 20, 50, 100, 200 and 300 cities, respectively. These Scaling size tests used the earlier generic-city prompt family used for the distribution analysis in Fig. 3a, not the source-matched metropolitan prompt used for the coefficient estimates in Fig. 2a. Vitality returned complete tables in 98\%, 98\%, 100\%, 88\% and 22\% of generations at the corresponding neighbourhood counts. The main relationship directions were retained across the tested sizes. Within these size-sensitivity cohorts, the 100-row condition yielded 91/100 complete Scaling tables and 100/100 complete Vitality tables; the source-matched Scaling coefficient experiment is reported separately in the main text.

\textbf{Supplementary Table 4 \textbar{} Generation-size sensitivity in the generic-city Scaling cohort and the Vitality synthesis.}

{\def\LTcaptype{none} 
\begin{longtable}[]{@{}
  >{\raggedleft\arraybackslash}p{(\linewidth - 10\tabcolsep) * \real{0.1667}}
  >{\raggedleft\arraybackslash}p{(\linewidth - 10\tabcolsep) * \real{0.1667}}
  >{\raggedleft\arraybackslash}p{(\linewidth - 10\tabcolsep) * \real{0.1667}}
  >{\raggedleft\arraybackslash}p{(\linewidth - 10\tabcolsep) * \real{0.1667}}
  >{\raggedleft\arraybackslash}p{(\linewidth - 10\tabcolsep) * \real{0.1667}}
  >{\raggedleft\arraybackslash}p{(\linewidth - 10\tabcolsep) * \real{0.1667}}@{}}
\toprule\noalign{}
\begin{minipage}[b]{\linewidth}\raggedleft
Requested rows
\end{minipage} & \begin{minipage}[b]{\linewidth}\raggedleft
Scaling complete tables
\end{minipage} & \begin{minipage}[b]{\linewidth}\raggedleft
Scaling \(\beta_{\mathrm{Road}}\)
\end{minipage} & \begin{minipage}[b]{\linewidth}\raggedleft
Scaling \(\beta_{\mathrm{GDP}}\)
\end{minipage} & \begin{minipage}[b]{\linewidth}\raggedleft
Vitality complete tables
\end{minipage} & \begin{minipage}[b]{\linewidth}\raggedleft
Vitality mean valid rows
\end{minipage} \\
\midrule\noalign{}
\endhead
\bottomrule\noalign{}
\endlastfoot
20 & 100/100 & 0.828 & 1.010 & 98/100 & 19.60 \\
50 & 90/100 & 0.851 & 1.025 & 98/100 & 49.00 \\
100 & 91/100 & 0.840 & 1.052 & 100/100 & 100.00 \\
200 & 56/100 & 0.793 & 0.967 & 88/100 & 188.21 \\
300 & 5/100 & 0.796 & 1.049 & 22/100 & 148.93 \\
\end{longtable}
}

This table summarizes the generic-city generation-size experiment; the source-matched Baseline--Blueprint estimates in Supplementary Table 5 provide the main Scaling coefficient comparison. At 200 and 300 rows, estimates use all valid returned rows because few responses met the requested length.

\paragraph{Joint and separate generation}\label{joint-and-separate-generation}

Joint generation asks the model to return all tested values and their outputs together in one table. Separate generation asks for the output at one tested value per request. The surrounding fictional description, the tested Population or Functional Mix values and the requested output variables were identical in the two formats. Across 12 Scaling examples, joint tables produced mean Road and GDP exponents of 0.920 and 1.231, and all 12 retained a sublinear Road and superlinear GDP relationship. Separate requests produced mean exponents of 3.008 and 2.430, with GDP exceeding Road in 4 of 12 examples. In Vitality, the joint-table endpoint contrast between Functional Mix 0.90 and 0.10 was 1.244 {[}1.194, 1.294{]} log units, compared with \(-0.006\) {[}\(-0.280\), 0.221{]} for separate requests. Joint generation was therefore retained for the main synthesis tasks.

\subsubsection{Statistical definitions}\label{statistical-definitions}

\paragraph{Scaling and Decay}\label{scaling-and-decay}

The Scaling relationship follows the main-text definition

\[
\log_{10}y_i
=\log_{10}a_y+\beta_y\log_{10}(\mathrm{Population}_i)+\varepsilon_{y,i},
\qquad y\in\{\mathrm{Road},\mathrm{GDP}\}.
\]

where \(i\) indexes metropolitan areas, \(a_y\) is the positive outcome-specific scale constant, \(\beta_y\) is the dimensionless log--log slope and \(\varepsilon_{y,i}\) is the residual on the base-10 log scale. The corresponding rank--size symbols are \(r_i\), \(a_Z\), \(\beta_Z\) and \(\varepsilon_{Z,i}\), as in Article equation (1). Distance Decay uses the inverse-S model defined in the main text. Competing exponential and power-law models were fitted to the same set of physical distance bands.

The Scaling slopes were estimated by ordinary least squares on log-transformed values. HC3 standard errors account for heteroskedastic residual variance within a generated table{[}18{]}. Fitting sensitivity replaced ordinary least squares with the Theil--Sen median of pairwise slopes{[}19{]}, or removed the single largest Population value before refitting. These alternatives changed neither the mean Road-sublinear nor the mean GDP-superlinear direction (Supplementary Table 7).

The alternatives were \(f(r)=A\exp(-kr)+c\) and \(f(r)=Ar^{-b}+c\), with parameters estimated by nonlinear least squares under the same bounds and retained annuli. Model selection used the small-sample corrected Akaike information criterion,

\[
\mathrm{AICc}=n\log(\mathrm{RSS}/n)+2k+\frac{2k(k+1)}{n-k-1},
\]

where \(n\) is the number of retained annuli and \(k\) the number of fitted parameters. Inverse-S \(R^2\) is reported independently of that selection outcome.

For a generated radial profile \(f=(f_1,\ldots,f_m)\) with \(m\) annuli, the Article's general local-detail definitions specialize as follows. Numerical resolution was

\[
U(f)=\frac{|\{f_i\}_{i=1}^{m}|}{m}.
\]

Normalized local curvature was

\[
C_2(f)=\frac{1}{(m-2)s_f}\sum_{i=1}^{m-2}
|f_{i+2}-2f_{i+1}+f_i|,
\]

where \(s_f=\max_i f_i-\min_i f_i\) and was set to one for a constant profile, matching equation (7) in the Article.

The number of outward reversals was

\[
R_+(f)=\sum_{i=1}^{m-1}\mathbb I(f_{i+1}>f_i).
\]

\paragraph{Vitality relationships and empirical ranking}\label{vitality-relationships-and-empirical-ranking}

For generated table \(g\), the standardized multivariable relationship was

\[
\begin{aligned}
z_g[\log(1+Y_{j,g})]={}&\beta_{0,g}
+\beta_{P,g}z_g[\log(1+P_{j,g})]
+\beta_{M,g}z_g(M_{j,g})+\beta_{S,g}z_g[-\log B_{j,g}]\\
&+\beta_{A,g}z_g(A_{j,g})+\beta_{T,g}z_g(T_{j,g})+e_{j,g}.
\end{aligned}
\]

This is Article equation (4) applied within generated table \(g\). Here \(g\) indexes generated tables, \(j\) indexes neighbourhoods, \(z_g[\,\cdot\,]\) denotes standardization within table \(g\), and \(\log\) denotes the natural logarithm. \(Y\) is Pedestrian Activity; \(P\), \(M\), \(B\), \(A\) and \(T\) denote Residential Density, Functional Mix, Mean Block Area, Aged Buildings and Tall Buildings, respectively, with Short Blocks represented as \(-\log B\). \(\beta_{0,g}\) is the table-specific intercept; \(\beta_{P,g}\), \(\beta_{M,g}\), \(\beta_{S,g}\), \(\beta_{A,g}\) and \(\beta_{T,g}\) are the five partial standardized coefficients; and \(e_{j,g}\) is the residual on the standardized outcome scale. Residential Density was transformed as \(z_g[\log(1+P)]\); the remaining predictors were standardized after their operational transformations. HC3 heteroskedasticity-consistent standard errors were used within generated tables, and five-fold cross-validation estimated held-out prediction. Primary empirical coefficient intervals also used HC3; eight spatial groups were used for the dependence-sensitive bootstrap and held-out prediction.

For an empirical site pair \((i,j)\) with observed activity order \(o_{ij}\in\{-1,1\}\) and model ranking \(\hat o_{ij}\), pairwise accuracy was

\[
A_{pair}=\frac{1}{|\mathcal P|}\sum_{(i,j)\in\mathcal P}\mathbb I(\hat o_{ij}=o_{ij}).
\]

Spatial-group bootstrap resampled the eight groups and retained all eligible cross-site comparisons represented in each resample.

\paragraph{Perception agreement and theme effects}\label{perception-agreement-and-theme-effects}

For Place Pulse pairwise judgements, observed agreement was \(P_o\) and chance agreement from the human and model marginal choice frequencies was \(P_e\), using the definitions in Article equation (5). Cohen's kappa was

\[
\kappa=\frac{P_o-P_e}{1-P_e}.
\]

For perception dimension \(p\) and theme \(c\) across \(N_c=12\) scenes, the control-adjusted effect was

\[
\Delta S_{p,c}
=\frac{1}{N_c}\sum_{i=1}^{N_c}
\left(S_{p,i,c}-S_{p,i,\mathrm{control}}\right),
\qquad N_c=12.
\]

Here \(S_{p,i,\mathrm{control}}\) is the same-scene editing-control score. Because the theme image and editing control share the same baseline scene, each scene term is algebraically identical to \(\{S_{p,i,c}-S_{p,i,\mathrm{base}}\}-\{S_{p,i,\mathrm{control}}-S_{p,i,\mathrm{base}}\}\). Percentile intervals use 10,000 scene-level bootstrap resamples. Exact two-sided paired sign-flip tests operate on the 12 scene effects; Holm correction applies to three prespecified relationships and Benjamini--Hochberg correction to all 18 theme--dimension combinations.

\paragraph{Intervention estimands}\label{intervention-estimands}

For fictional example \(s\) with ordered tested values \(x_1<\cdots<x_L\) and output \(y_{s\ell}\), numerical interventions report the within-example slope from \(y_{s\ell}\) on \(x_\ell\) and a directional-order score

\[
Q_s=\frac{2}{L(L-1)}\sum_{j<k}\mathbb I\{(x_k-x_j)(y_{sk}-y_{sj})>0\}.
\]

For Decay, the expected direction is reversed because density should decline with distance. Vitality additionally reports the complete-order rate \(\Pr(PA_H>PA_M>PA_L)\) and the three pairwise order rates. The fixed denominator is 38 settings; an unparseable or incomplete response contributes zero to the corresponding order rate.

The Article's task-specific pairwise-order statistics \(H_b^D\) and \(H_b^V\) are the Distance Decay and Vitality specializations of \(Q_s\), respectively; the different symbol is retained here because \(Q_s\) is the general intervention definition across tasks.

\paragraph{Distributional comparison}\label{distributional-comparison}

One-dimensional Wasserstein distance compares samples \(A\) and \(B\) through their empirical cumulative distribution functions \(F_A\) and \(F_B\):

\[
W_1(A,B)=\int_{-\infty}^{\infty}|F_A(u)-F_B(u)|\,\mathrm du.
\]

Jensen--Shannon divergence was computed after applying the same fixed bins to the two distributions:

\[
\operatorname{JSD}(P,Q)=\frac{1}{2}\operatorname{KL}(P\|M)+
\frac{1}{2}\operatorname{KL}(Q\|M),\qquad M=\frac{P+Q}{2}.
\]

JSD is zero when the two binned distributions are identical and increases as their probability mass separates; it compares distribution shape rather than only the observed value range.

Within-table magnitude coverage was \(R_{10}(x)=\log_{10}(\max_i x_i/\min_i x_i)\) for positive values, using the notation of Article equation (6).

Distribution overlap was the shared probability mass under common fixed bins, \(O(P,Q)=\sum_b\min(p_b,q_b)\). Interquartile range used the 75th minus the 25th percentile. Shannon entropy was \(H(P)=-\sum_b p_b\log p_b\) on the same bins. Gini coefficients and top-decile shares were calculated directly from the positive values. The quantities are reported separately because they describe distinct aspects of data variation.

\paragraph{Image-representation comparison}\label{image-representation-comparison}

Let \(\mathbf z_i\) be the unit-normalized embedding of image \(i\) and \(d(\mathbf z_i,\mathbf z_j)=1-\mathbf z_i^{\mathsf T}\mathbf z_j\) its cosine distance. Within-set diversity for image set \(A\) was

\[
D_{\mathrm{within}}(A)
=\frac{2}{|A|(|A|-1)}\sum_{i<j}\left(1-\mathbf z_i^{\mathsf T}\mathbf z_j\right).
\]

Nearest-neighbour similarity from set \(A\) to set \(B\) was

\[
S_{A\rightarrow B}=\frac{1}{|A|}\sum_{i\in A}\max_{j\in B}\mathbf z_i^{\mathsf T}\mathbf z_j.
\]

The two directions answer different questions: \(S_{G\rightarrow E}\) measures how close each generated image is to some empirical image, whereas \(S_{E\rightarrow G}\) measures how well the generated set covers empirical images. Centroid distance was

\[
D_{\mathrm{centroid}}(A,B)
=1-\frac{\bar{\mathbf z}_A^{\mathsf T}\bar{\mathbf z}_B}
{\|\bar{\mathbf z}_A\|_2\|\bar{\mathbf z}_B\|_2}.
\]

These are general set definitions. Article equations (9)--(11) use \(A=G\) for generated images and \(B=E\) for empirical images, giving \(D_{\mathrm{within}}(G)\), \(S_{G\rightarrow E}\), \(S_{E\rightarrow G}\) and \(D_{\mathrm{centroid}}(G,E)\) without changing the estimands.

The main Place Pulse comparison used a city-held-out reference design and split-half empirical references for within-dataset baselines. The primary implementation used the OpenAI CLIP ViT-L/14 encoder{[}20{]} and checkpoint \texttt{ViT-L-14.pt}. Images were converted to RGB, bicubically resized on the shorter side to 224 pixels, centre-cropped to \(224\times224\) pixels and normalized with channel means \((0.48145466,0.45782750,0.40821073)\) and standard deviations \((0.26862954,0.26130258,0.27577711)\). The first sensitivity encoder was DINOv2 ViT-L/14{[}21{]}, checkpoint \texttt{dinov2\_vitl14\_pretrain.pth}. Its inputs were converted to RGB, bicubically resized on the shorter side to 256 pixels and centre-cropped to \(224\times224\) pixels. The final class-token representation was retained. The second sensitivity encoder was ResNet-50 ImageNet1K\_V2{[}22{]}, checkpoint \texttt{resnet50-0676ba61.pth}. Its inputs were converted to RGB, bilinearly resized on the shorter side to 232 pixels and centre-cropped to \(224\times224\) pixels. The classification head was replaced by an identity mapping to retain the penultimate representation. DINOv2 and ResNet-50 used channel means \((0.485,0.456,0.406)\) and standard deviations \((0.229,0.224,0.225)\). Outputs from all three encoders were \(\ell_2\)-normalized before cosine calculations. Supplementary Data 1 provides checkpoint and preprocessing details.

\paragraph{Hidden-state readout and intervention}\label{hidden-state-readout-and-intervention}

Each matched prompt pair described the same task twice: once with the expected relationship and once with that relationship reversed. The symbols \(+\) and \(-\) label these two prompts. For task \(t\), let \(\bar{\mathbf h}_{t,+}\) and \(\bar{\mathbf h}_{t,-}\) be their mean hidden states. Their normalized difference gives the task direction, and projecting the hidden state of evaluated prompt \(i\) onto that direction gives the scalar score

\[
\mathbf d_t=
\frac{\bar{\mathbf h}_{t,+}-\bar{\mathbf h}_{t,-}}
{\lVert\bar{\mathbf h}_{t,+}-\bar{\mathbf h}_{t,-}\rVert_2},
\qquad s_{i,t}=\mathbf h_i^{\mathsf T}\mathbf d_t.
\]

Confirmation AUC was \(\operatorname{AUC}_t=\Pr(s_{+,t}>s_{-,t})+\tfrac12\Pr(s_{+,t}=s_{-,t})\). It is the probability that an expected-relationship prompt ranks above its reversed counterpart, with ties receiving half weight; 0.5 denotes chance and 1 perfect separation. Candidate layers and token positions were selected separately for each task on development prompts only; confirmation and held-out prompts did not influence selection. For the common three-task screen, the earliest candidate decoder block within 0.005 of the maximum development AUC was selected: hidden-state index 17 for Scaling, 33 for Distance Decay and 49 for Vitality. These common-screen directions were tested during table generation for tasks that met the 0.90 confirmation rule. Separate short-answer experiments made their own selections. The Scaling short-answer experiment fixed index 61 from matched direct-versus-inverse development prompts. The Vitality short-answer experiment selected the earliest candidate index within 0.005 of its maximum development AUC, then required the lower confidence bound of confirmation AUC to exceed 0.80 and the default-state projection to be positive; index 49 met both requirements. Thus \(t\) indexes the task, not a token position. In the separate short-answer experiments, a dose intervention added or subtracted the selected task direction, \(\mathbf h'_i=\mathbf h_i+\lambda\mathbf d_t\), and estimated the slope of a task-specific answer logit contrast on \(\lambda\). Population, irrelevant-number, non-urban-relation and equal-norm random directions were matched controls. Full-state replacement copies the corresponding activation from a paired prompt and estimates the paired change in the same answer contrast.

In the table-generation experiment, only Scaling and Vitality met both the 0.90 confirmation rule and the prespecified baseline-generation rule: all six baseline tables were complete, at least five statistics were positive and the bootstrap lower bound for the mean was above zero. For each task, one dose unit \(u_t\) was the median positive paired projection gap in the development split (\texttt{0.2958365} for Scaling and \texttt{2.9615555} for Vitality). During greedy generation with key--value caching, \(\lambda u_t\mathbf d_t\) was added at the selected decoder block on every autoregressive step, at the final prompt position on the first pass and the current generated position thereafter. For each of six held-out fictional examples per task, tables were generated at \(\lambda=-1\) and \(+1\) for the task-relation direction and for matched focal-input, non-urban and seeded equal-norm random directions. This produced 48 edited generations per task and 96 in total. For the task-relation direction, let \(Q_{b,t}(\lambda)\) be the table statistic: \(\beta_{\mathrm{GDP}}-\beta_{\mathrm{Road}}\) for Scaling and \(\log(1+Y_H)-\log(1+Y_L)\) for Vitality, where \(Y_H\) and \(Y_L\) are the generated endpoint activities under high and low Functional Mix. The symmetric task-relation effect was

\[
E_{b,t}=\frac{Q_{b,t}(+1)-Q_{b,t}(-1)}{2}.
\]

Here \(b\) indexes one fictional metropolitan or neighbourhood example and \(t\) indexes the task. This is the same estimand as Article equation (20).

For Perception, image-state replacement used the same 18 theme-by-dimension effects as the behavioural task. At the selected decoder layer, the states at the theme-image token positions were replaced by states from the same-scene editing control. Each Qwen change was weighted by the direction and magnitude of the corresponding GPT-4o theme effect; specificity subtracts the absolute effect of an equal-norm random perturbation at the same positions. In the 72B localization analysis, the primary 0.01\% post-write equal-norm tolerance was met for 57/216 pairs because FP16 rounding affected the rescaled low-change condition. A 0.05\% sensitivity threshold chosen without examining the outcome effects was met for 216/216 pairs. The text-stream control remained larger under both tolerances.

The 7B within-family follow-up used two images plus text, processor bounds of 3,136 and 1,003,520 pixels, and single-GPU placement without CPU or disk offload. Phase A contained 216/216 usable image-pair evaluations before state replacement. Phase B replaced all theme-image-token states at zero-based decoder layer 16 with matched same-scene editing-control states and contained 216/216 usable editing-control state-replacement evaluations; each was paired with an equal-norm seeded-random replacement at the same positions. All 216 matched comparisons met the equal-norm criterion. Statistical inference used 12 scenes, 30,000 scene-bootstrap resamples and exact two-sided paired sign-flip tests.

\subsection{Supplementary Results}\label{supplementary-results}

\subsubsection{Urban Scaling}\label{urban-scaling}

\paragraph{Main synthesis results and empirical parameters}\label{main-synthesis-results-and-empirical-parameters}

Each prompt condition comprised 100 GPT-4o requests for 100 fictional metropolitan areas. One request in each condition ended without a usable output. Coefficient summaries use the 99 fit-valid tables per condition and 98 matched Baseline--Blueprint pairs. Exact 100-row compliance was 2/100 under Baseline and 1/100 under Blueprint; the fit-valid responses contained a mean of 90.5 and 90.3 valid rows, respectively. ``Complete'' requires exactly 100 valid metropolitan rows.

\textbf{Supplementary Table 5 \textbar{} Urban Scaling estimates under Baseline and Blueprint prompts and their empirical references.}

{\def\LTcaptype{none} 
\begin{longtable}[]{@{}
  >{\raggedright\arraybackslash}p{(\linewidth - 6\tabcolsep) * \real{0.2000}}
  >{\raggedleft\arraybackslash}p{(\linewidth - 6\tabcolsep) * \real{0.2667}}
  >{\raggedleft\arraybackslash}p{(\linewidth - 6\tabcolsep) * \real{0.2667}}
  >{\raggedleft\arraybackslash}p{(\linewidth - 6\tabcolsep) * \real{0.2667}}@{}}
\toprule\noalign{}
\begin{minipage}[b]{\linewidth}\raggedright
Quantity
\end{minipage} & \begin{minipage}[b]{\linewidth}\raggedleft
Baseline
\end{minipage} & \begin{minipage}[b]{\linewidth}\raggedleft
Complete Blueprint
\end{minipage} & \begin{minipage}[b]{\linewidth}\raggedleft
Empirical reference
\end{minipage} \\
\midrule\noalign{}
\endhead
\bottomrule\noalign{}
\endlastfoot
Rank--size slope \(\beta_Z\) & \(-0.477\) {[}\(-0.499\), \(-0.455\){]} & \(-1.531\) {[}\(-1.591\), \(-1.470\){]} & \(-1.115\) \\
Road exponent \(\beta_{\mathrm{Road}}\) & 0.887 {[}0.862, 0.910{]} & 0.850 {[}0.820, 0.881{]} & \(0.849\pm0.038\) \\
GDP exponent \(\beta_{\mathrm{GDP}}\) & 1.044 {[}1.015, 1.072{]} & 1.013 {[}0.987, 1.039{]} & \(1.126\pm0.023\) \\
Mean absolute Road-parameter error & 0.109 & 0.123 & Not applicable \\
Mean absolute GDP-parameter error & 0.124 & 0.137 & Not applicable \\
Mean valid metropolitan rows & 90.5 & 90.3 & Not applicable \\
Complete 100-row tables & 2/100 & 1/100 & Not applicable \\
\end{longtable}
}

The matched Blueprint-minus-Baseline changes in absolute error were +0.014 {[}\(-0.010\), 0.037{]} for Road and +0.013 {[}\(-0.015\), 0.041{]} for GDP; neither differed from zero after Holm correction (both \(p=0.527\)).

\paragraph{Prompt and scale sensitivity}\label{prompt-and-scale-sensitivity}

Early prompt-component experiments separately added an empirical-reference description, explicit theory information, or both. These experiments contained ten independent generations per condition and are retained as prompt sensitivity analyses. In the matched comparison of 98 Baseline--Blueprint pairs, Blueprint expanded Population coverage, altered rank--size structure and increased residual heterogeneity; the matched changes in Road and GDP exponent error are reported above.

\textbf{Supplementary Table 6 \textbar{} Urban Scaling sensitivity to early prompt components.}

{\def\LTcaptype{none} 
\begin{longtable}[]{@{}
  >{\raggedright\arraybackslash}p{(\linewidth - 12\tabcolsep) * \real{0.1111}}
  >{\raggedleft\arraybackslash}p{(\linewidth - 12\tabcolsep) * \real{0.1481}}
  >{\raggedleft\arraybackslash}p{(\linewidth - 12\tabcolsep) * \real{0.1481}}
  >{\raggedleft\arraybackslash}p{(\linewidth - 12\tabcolsep) * \real{0.1481}}
  >{\raggedleft\arraybackslash}p{(\linewidth - 12\tabcolsep) * \real{0.1481}}
  >{\raggedleft\arraybackslash}p{(\linewidth - 12\tabcolsep) * \real{0.1481}}
  >{\raggedleft\arraybackslash}p{(\linewidth - 12\tabcolsep) * \real{0.1481}}@{}}
\toprule\noalign{}
\begin{minipage}[b]{\linewidth}\raggedright
Early prompt condition
\end{minipage} & \begin{minipage}[b]{\linewidth}\raggedleft
Rank--size slope
\end{minipage} & \begin{minipage}[b]{\linewidth}\raggedleft
Rank--size \(R^2\)
\end{minipage} & \begin{minipage}[b]{\linewidth}\raggedleft
Road exponent
\end{minipage} & \begin{minipage}[b]{\linewidth}\raggedleft
Road \(R^2\)
\end{minipage} & \begin{minipage}[b]{\linewidth}\raggedleft
GDP exponent
\end{minipage} & \begin{minipage}[b]{\linewidth}\raggedleft
GDP \(R^2\)
\end{minipage} \\
\midrule\noalign{}
\endhead
\bottomrule\noalign{}
\endlastfoot
Empirical-reference description & \(-0.810\) & 0.956 & 0.516 & 0.586 & 0.816 & 0.579 \\
Theory information & \(-0.900\) & 1.000 & 0.974 & 0.949 & 1.019 & 0.985 \\
Both additions & \(-1.132\) & 0.924 & 0.862 & 0.804 & 0.795 & 0.886 \\
\end{longtable}
}

These ten-generation conditions predate the complete Blueprint design and are reported separately from the 100-generation comparison.

Across joint-generation sizes from 20 to 300 cities, the principal Road-sublinear and GDP-superlinear directions were retained. Parameter estimates at 200 and 300 cities used all valid returned rows because most requests did not return the requested length.

The fitted direction was also stable to the estimator and to the largest generated metropolitan area. Ordinary least squares and Theil--Sen agreed on the expected side of one in 0.960--0.980 of the 99 tables per condition and outcome; ordinary least squares before and after deleting the largest Population value agreed in 0.970--1.000. Mean slopes are reported below.

\textbf{Supplementary Table 7 \textbar{} Urban Scaling sensitivity to fitting method and the largest Population value.}

{\def\LTcaptype{none} 
\begin{longtable}[]{@{}
  >{\raggedright\arraybackslash}p{(\linewidth - 8\tabcolsep) * \real{0.1667}}
  >{\raggedright\arraybackslash}p{(\linewidth - 8\tabcolsep) * \real{0.1667}}
  >{\raggedleft\arraybackslash}p{(\linewidth - 8\tabcolsep) * \real{0.2222}}
  >{\raggedleft\arraybackslash}p{(\linewidth - 8\tabcolsep) * \real{0.2222}}
  >{\raggedleft\arraybackslash}p{(\linewidth - 8\tabcolsep) * \real{0.2222}}@{}}
\toprule\noalign{}
\begin{minipage}[b]{\linewidth}\raggedright
Condition
\end{minipage} & \begin{minipage}[b]{\linewidth}\raggedright
Outcome
\end{minipage} & \begin{minipage}[b]{\linewidth}\raggedleft
OLS mean \(\beta\) {[}95\% CI{]}
\end{minipage} & \begin{minipage}[b]{\linewidth}\raggedleft
Theil--Sen mean \(\beta\) {[}95\% CI{]}
\end{minipage} & \begin{minipage}[b]{\linewidth}\raggedleft
OLS without largest Population mean \(\beta\) {[}95\% CI{]}
\end{minipage} \\
\midrule\noalign{}
\endhead
\bottomrule\noalign{}
\endlastfoot
Baseline & Road & 0.887 {[}0.862, 0.910{]} & 0.896 {[}0.870, 0.920{]} & 0.889 {[}0.864, 0.914{]} \\
Blueprint & Road & 0.850 {[}0.820, 0.881{]} & 0.866 {[}0.833, 0.901{]} & 0.857 {[}0.826, 0.889{]} \\
Baseline & GDP & 1.044 {[}1.015, 1.072{]} & 1.045 {[}1.017, 1.074{]} & 1.046 {[}1.017, 1.074{]} \\
Blueprint & GDP & 1.013 {[}0.987, 1.039{]} & 1.022 {[}0.993, 1.051{]} & 1.018 {[}0.991, 1.045{]} \\
\end{longtable}
}

The OLS column repeats the primary intervals in Supplementary Table 5. Sensitivity intervals use 10,000 table-level bootstrap resamples of the 99 fit-valid tables in each condition.

\paragraph{Generated and empirical distributions in a separate generic-city cohort}\label{generated-and-empirical-distributions-in-a-separate-generic-city-cohort}

The distribution analysis used a separate prompt-component experiment with generic Population, total Road length and city GDP definitions. The empirical reference contains 3,525 positive complete records from the archived Chinese physical-city polygons. Sensitivity references requiring effective area above 5 or 20 contained 2,347 and 878 polygons. In all three references, Blueprint reduced the Wasserstein distance for the three variables jointly. A boundary-definition sensitivity used 676 positive complete municipal administrative units. The Blueprint--Baseline distance change was +0.136 {[}0.113, 0.159{]} for Population, \(-0.020\) {[}\(-0.040\), 0.002{]} for Road and 0.000 {[}\(-0.033\), 0.036{]} for GDP. These comparisons concern marginal distributions within this generic-city prompt family; the source-matched US metropolitan experiment remains the basis for the coefficients in Fig. 2a.

The main comparison used 91 complete Baseline tables, 58 complete Blueprint tables and 100 without-replacement empirical samples of 100 cities. Values were log-transformed within each table before the distributional distance was calculated.

\textbf{Supplementary Table 8 \textbar{} Generated--empirical distribution distances and numerical ranges in the generic-city Scaling cohort.}

{\def\LTcaptype{none} 
\begin{longtable}[]{@{}
  >{\raggedright\arraybackslash}p{(\linewidth - 12\tabcolsep) * \real{0.1111}}
  >{\raggedleft\arraybackslash}p{(\linewidth - 12\tabcolsep) * \real{0.1481}}
  >{\raggedleft\arraybackslash}p{(\linewidth - 12\tabcolsep) * \real{0.1481}}
  >{\raggedleft\arraybackslash}p{(\linewidth - 12\tabcolsep) * \real{0.1481}}
  >{\raggedleft\arraybackslash}p{(\linewidth - 12\tabcolsep) * \real{0.1481}}
  >{\raggedleft\arraybackslash}p{(\linewidth - 12\tabcolsep) * \real{0.1481}}
  >{\raggedleft\arraybackslash}p{(\linewidth - 12\tabcolsep) * \real{0.1481}}@{}}
\toprule\noalign{}
\begin{minipage}[b]{\linewidth}\raggedright
Variable
\end{minipage} & \begin{minipage}[b]{\linewidth}\raggedleft
Baseline \(W_1\)
\end{minipage} & \begin{minipage}[b]{\linewidth}\raggedleft
Blueprint \(W_1\)
\end{minipage} & \begin{minipage}[b]{\linewidth}\raggedleft
Empirical split-sample \(W_1\)
\end{minipage} & \begin{minipage}[b]{\linewidth}\raggedleft
Baseline log-range
\end{minipage} & \begin{minipage}[b]{\linewidth}\raggedleft
Blueprint log-range
\end{minipage} & \begin{minipage}[b]{\linewidth}\raggedleft
Empirical log-range
\end{minipage} \\
\midrule\noalign{}
\endhead
\bottomrule\noalign{}
\endlastfoot
Population & 0.626 {[}0.614, 0.636{]} & 0.193 {[}0.183, 0.206{]} & 0.180 {[}0.171, 0.191{]} & 0.862 {[}0.801, 0.924{]} & 2.665 {[}2.628, 2.712{]} & 3.858 {[}3.736, 3.985{]} \\
Road & 0.626 {[}0.617, 0.635{]} & 0.338 {[}0.316, 0.359{]} & 0.204 {[}0.193, 0.216{]} & 0.711 {[}0.658, 0.766{]} & 2.008 {[}1.910, 2.104{]} & 3.806 {[}3.667, 3.947{]} \\
GDP & 0.804 {[}0.791, 0.816{]} & 0.388 {[}0.357, 0.419{]} & 0.201 {[}0.191, 0.211{]} & 0.936 {[}0.857, 1.023{]} & 2.512 {[}2.411, 2.618{]} & 4.316 {[}4.234, 4.397{]} \\
\end{longtable}
}

\textbf{Supplementary Table 9 \textbar{} Additional distribution comparisons under Blueprint relative to Baseline.}

{\def\LTcaptype{none} 
\begin{longtable}[]{@{}
  >{\raggedright\arraybackslash}p{(\linewidth - 6\tabcolsep) * \real{0.2000}}
  >{\raggedleft\arraybackslash}p{(\linewidth - 6\tabcolsep) * \real{0.2667}}
  >{\raggedleft\arraybackslash}p{(\linewidth - 6\tabcolsep) * \real{0.2667}}
  >{\raggedleft\arraybackslash}p{(\linewidth - 6\tabcolsep) * \real{0.2667}}@{}}
\toprule\noalign{}
\begin{minipage}[b]{\linewidth}\raggedright
Blueprint minus Baseline
\end{minipage} & \begin{minipage}[b]{\linewidth}\raggedleft
Population
\end{minipage} & \begin{minipage}[b]{\linewidth}\raggedleft
Road
\end{minipage} & \begin{minipage}[b]{\linewidth}\raggedleft
GDP
\end{minipage} \\
\midrule\noalign{}
\endhead
\bottomrule\noalign{}
\endlastfoot
Wasserstein distance & \(-0.432\) {[}\(-0.447\), \(-0.416\){]} & \(-0.288\) {[}\(-0.312\), \(-0.265\){]} & \(-0.415\) {[}\(-0.449\), \(-0.382\){]} \\
Jensen--Shannon divergence & \(-0.323\) {[}\(-0.343\), \(-0.304\){]} & \(-0.333\) {[}\(-0.360\), \(-0.305\){]} & \(-0.320\) {[}\(-0.343\), \(-0.297\){]} \\
Distribution overlap & +0.362 {[}0.343, 0.380{]} & +0.324 {[}0.297, 0.351{]} & +0.324 {[}0.301, 0.346{]} \\
\end{longtable}
}

\paragraph{Population, Road and GDP interventions}\label{population-road-and-gdp-interventions}

Population interventions used three equivalent phrasings, repeated conditions and irrelevant-number controls. GDP remained mildly superlinear and Road sublinear under the three phrasings; repeats and irrelevant numbers produced no stable displacement.

Road and GDP were also varied across five levels in 100 fixed city settings. With Road as input, the mean within-city slope was 0.618 {[}0.561, 0.677{]} for Population and 0.606 {[}0.560, 0.649{]} for GDP; both response slopes were positive in 98 of 100 settings. With GDP as input, the corresponding slopes were 0.439 {[}0.388, 0.488{]} for Population and 0.351 {[}0.311, 0.390{]} for Road; both were positive in 96 of 100 settings.

\subsubsection{Distance Decay}\label{distance-decay}

\paragraph{Empirical parameters and prompt comparison}\label{empirical-parameters-and-prompt-comparison}

The empirical reference contains 28 Chinese and 22 US or European cities observed at three periods. Across the 150 city-period profiles, median inverse-S parameters were \(\alpha=3.210\) {[}5th--95th percentile: 1.797--5.126{]}, \(c=0.070\) {[}0--0.241{]} and \(D=15.750\) km {[}6.381--51.491 km{]}{[}6{]}.

The published 50-city collection supplies empirical parameter ranges, and a separate 2019 GISA benchmark for the same cities supplies curve-level detail. Each GISA profile used a GeoNames populated-place point as a common centre, 1-km distance bands, cosine-latitude area correction and the fraction of positive impervious-surface cells among all cells in each band; water cells remained in the denominator. The GISA profiles are used to compare value resolution, local curvature and outward reversals.

The median GISA 2019 inverse-S fit was \(R^2=0.981\), with \(\alpha=2.354\), \(c=0.048\) and \(D=22.449\) km. Median value resolution, normalized local curvature and outward reversals were 0.32, 0.0177 and 19, respectively. Generated curves were rounded to the same two-decimal precision before these local-detail metrics were compared.

Across matched Baseline--Blueprint generations, Blueprint increased \(\alpha\) by 0.261 {[}0.188, 0.334{]}, \(c\) by 0.009 {[}0.006, 0.013{]}, \(D\) by 15.308 km {[}13.300, 17.316{]}, outward reversals by 0.667 {[}0.211, 1.244{]}, and normalized local curvature by 0.00172 {[}0.00143, 0.00200{]}. The inverse-S selection rate was similar between conditions (difference \(-0.067\) {[}\(-0.200\), 0.067{]}).

\textbf{Supplementary Table 10 \textbar{} Distance Decay parameters and local-profile statistics.}

{\def\LTcaptype{none} 
\begin{longtable}[]{@{}
  >{\raggedright\arraybackslash}p{(\linewidth - 6\tabcolsep) * \real{0.2000}}
  >{\raggedleft\arraybackslash}p{(\linewidth - 6\tabcolsep) * \real{0.2667}}
  >{\raggedleft\arraybackslash}p{(\linewidth - 6\tabcolsep) * \real{0.2667}}
  >{\raggedleft\arraybackslash}p{(\linewidth - 6\tabcolsep) * \real{0.2667}}@{}}
\toprule\noalign{}
\begin{minipage}[b]{\linewidth}\raggedright
Quantity
\end{minipage} & \begin{minipage}[b]{\linewidth}\raggedleft
Baseline
\end{minipage} & \begin{minipage}[b]{\linewidth}\raggedleft
Complete Blueprint
\end{minipage} & \begin{minipage}[b]{\linewidth}\raggedleft
Published or GISA empirical reference
\end{minipage} \\
\midrule\noalign{}
\endhead
\bottomrule\noalign{}
\endlastfoot
Inverse-S median \(R^2\) & 0.997 & 0.995 & 0.981 (GISA 2019) \\
\(\alpha\) & 1.430 & 1.691 & 3.210 {[}1.797, 5.126{]} (published 150 profiles) \\
\(c\) & 0.010 & 0.016 & 0.070 {[}0, 0.241{]} (published 150 profiles) \\
\(D\) (km) & 22.892 & 37.114 & 15.750 {[}6.381, 51.491{]} (published 150 profiles) \\
Value resolution & 0.40 & 0.55 & 0.32 (GISA 2019) \\
Normalized local curvature & 0.0033 & 0.0048 & 0.0177 (GISA 2019) \\
Outward reversals & 0 & 0 & 19 (GISA 2019) \\
\end{longtable}
}

\paragraph{Model and distance-band sensitivity}\label{model-and-distance-band-sensitivity}

On the same 30 Blueprint cities, inverse-S was selected for 22 GPT-4o profiles and 19 profiles from the \texttt{gpt-5.6} gateway alias. The latter increased median value resolution from 0.560 to 0.975 and moved \(\alpha\) and \(c\) towards the empirical range.

Five synthesis designs varied the number of distance bands, band width and maximum distance. Both designs with 50 bands returned 100 complete curves and selected inverse-S for 92\%--95\% of the planned requests. Requests for 150 or 200 bands were frequently incomplete. Median \(D\) varied from 25.1 to 49.2 km across the designs, confirming that it describes decay scale relative to the specified physical support.

\textbf{Supplementary Table 11 \textbar{} Distance Decay sensitivity to the tested sets of synthesis distance bands.}

{\def\LTcaptype{none} 
\begin{longtable}[]{@{}
  >{\raggedright\arraybackslash}p{(\linewidth - 8\tabcolsep) * \real{0.1579}}
  >{\raggedleft\arraybackslash}p{(\linewidth - 8\tabcolsep) * \real{0.2105}}
  >{\raggedleft\arraybackslash}p{(\linewidth - 8\tabcolsep) * \real{0.2105}}
  >{\raggedleft\arraybackslash}p{(\linewidth - 8\tabcolsep) * \real{0.2105}}
  >{\raggedleft\arraybackslash}p{(\linewidth - 8\tabcolsep) * \real{0.2105}}@{}}
\toprule\noalign{}
\begin{minipage}[b]{\linewidth}\raggedright
Tested synthesis design
\end{minipage} & \begin{minipage}[b]{\linewidth}\raggedleft
Parseable profiles
\end{minipage} & \begin{minipage}[b]{\linewidth}\raggedleft
Inverse-S selected
\end{minipage} & \begin{minipage}[b]{\linewidth}\raggedleft
Median \(R^2\)
\end{minipage} & \begin{minipage}[b]{\linewidth}\raggedleft
Median \(D\) (km)
\end{minipage} \\
\midrule\noalign{}
\endhead
\bottomrule\noalign{}
\endlastfoot
50 bands \(\times\) 1 km; maximum 50 km & 100/100 & 95/100 & 0.997 & 25.1 \\
100 bands \(\times\) 1 km; maximum 100 km & 95/100 & 68/100 & 0.995 & 37.1 \\
150 bands \(\times\) 1 km; maximum 150 km & 23/100 & 11/100 & 0.992 & 49.2 \\
200 bands \(\times\) 0.5 km; maximum 100 km & 21/100 & 19/100 & 0.995 & 32.0 \\
50 bands \(\times\) 2 km; maximum 100 km & 100/100 & 92/100 & 0.997 & 43.2 \\
\end{longtable}
}

Six intervention designs produced near--far direction scores from 0.810 to 0.951, and all complete-profile mean slopes were negative. A paraphrase changed the primary near--far rate by 0.015 {[}\(-0.115\), 0.145{]}. At 15--16 km, repeated and irrelevant-number controls changed density by \(-0.019\) {[}\(-0.077\), 0.035{]} and \(-0.031\) {[}\(-0.104\), 0.035{]}, respectively.

\textbf{Supplementary Table 12 \textbar{} Distance Decay sensitivity to the tested intervention distance bands.}

{\def\LTcaptype{none} 
\begin{longtable}[]{@{}
  >{\raggedright\arraybackslash}p{(\linewidth - 6\tabcolsep) * \real{0.2000}}
  >{\raggedleft\arraybackslash}p{(\linewidth - 6\tabcolsep) * \real{0.2667}}
  >{\raggedleft\arraybackslash}p{(\linewidth - 6\tabcolsep) * \real{0.2667}}
  >{\raggedleft\arraybackslash}p{(\linewidth - 6\tabcolsep) * \real{0.2667}}@{}}
\toprule\noalign{}
\begin{minipage}[b]{\linewidth}\raggedright
Tested intervention design
\end{minipage} & \begin{minipage}[b]{\linewidth}\raggedleft
Complete city profiles
\end{minipage} & \begin{minipage}[b]{\linewidth}\raggedleft
Near--far direction score
\end{minipage} & \begin{minipage}[b]{\linewidth}\raggedleft
Mean slope per km among complete profiles
\end{minipage} \\
\midrule\noalign{}
\endhead
\bottomrule\noalign{}
\endlastfoot
3 tested distances; 1-km bands & 80/100 & 0.817 {[}0.760, 0.870{]} & \(-0.0113\) \\
5 tested distances; 1-km bands & 67/100 & 0.816 {[}0.772, 0.858{]} & \(-0.0101\) \\
7 tested distances; 1-km bands & 66/100 & 0.810 {[}0.779, 0.840{]} & \(-0.0098\) \\
Maximum distance 100 km & 92/100 & 0.922 {[}0.898, 0.945{]} & \(-0.0060\) \\
0.5-km bands & 98/100 & 0.951 {[}0.936, 0.964{]} & \(-0.0105\) \\
2-km bands & 96/100 & 0.875 {[}0.852, 0.896{]} & \(-0.0075\) \\
\end{longtable}
}

\subsubsection{Jacobs-informed Vitality}\label{jacobs-informed-vitality}

\paragraph{Variables and empirical sites}\label{variables-and-empirical-sites}

\textbf{Supplementary Table 13 \textbar{} Vitality variables and operational definitions.}

{\def\LTcaptype{none} 
\begin{longtable}[]{@{}
  >{\raggedright\arraybackslash}p{(\linewidth - 4\tabcolsep) * \real{0.3333}}
  >{\raggedright\arraybackslash}p{(\linewidth - 4\tabcolsep) * \real{0.3333}}
  >{\raggedright\arraybackslash}p{(\linewidth - 4\tabcolsep) * \real{0.3333}}@{}}
\toprule\noalign{}
\begin{minipage}[b]{\linewidth}\raggedright
Variable
\end{minipage} & \begin{minipage}[b]{\linewidth}\raggedright
Role
\end{minipage} & \begin{minipage}[b]{\linewidth}\raggedright
Operational definition
\end{minipage} \\
\midrule\noalign{}
\endhead
\bottomrule\noalign{}
\endlastfoot
Residential Density & Predictor & Dwellings per hectare within 400 m; analysed as \(z[\log(1+x)]\) \\
Functional Mix & Predictor & Normalized land-use entropy within 400 m; standardized for analysis \\
Short Blocks & Predictor & Negative log of area-weighted mean block area within 400 m \\
Aged Buildings & Predictor & Proportion of buildings at least 50 years old within 400 m \\
Tall Buildings & Predictor & Proportion of buildings at least 10 storeys high within 400 m \\
Pedestrian Activity & Outcome & Aggregate of median pedestrian counts across 11 hourly periods; analysed as \(\log(1+x)\) \\
\end{longtable}
}

The 52-site morphology variables use City of Melbourne 2024 CLUE building, dwelling and land-use data. Pedestrian Activity uses the Pedestrian Counting System from 2025 through June 2026. In the ranking task, each site was represented by an anonymous identifier and the five morphology variables; observed Pedestrian Activity defined the reference order used to score the model ranking.

Jacobs' concentration, mixed primary uses, short blocks and mixture of building ages were operationalized as Residential Density, Functional Mix, Short Blocks and Aged Buildings, respectively. The prespecified Jacobs-informed expectation was a positive association between each of these four indicators and Pedestrian Activity. Tall Buildings was an additional exploratory built-intensity indicator and did not carry a Jacobs-derived coefficient sign.

\paragraph{Generated relationships and empirical ranking}\label{generated-relationships-and-empirical-ranking}

Each generated table was analysed with a five-predictor standardized linear regression and heteroskedasticity-consistent standard errors. Across 100 Baseline tables, the median coefficients were 0.605 for Residential Density, 0.166 for Functional Mix, 0.132 for Short Blocks, \(-0.061\) for Aged Buildings and \(-0.026\) for Tall Buildings. The median adjusted \(R^2\) was 0.980 and median five-fold cross-validated \(R^2\) was 0.978.

The concurrent Baseline--Blueprint comparison used the mean coefficient across 100 independently generated complete tables in each condition. The empirical intervals in Supplementary Table 14 and Fig. 2c are HC3 95\% confidence intervals from the 52-site model. An eight-spatial-group bootstrap was retained as a dependence-sensitive uncertainty analysis.

\textbf{Supplementary Table 14 \textbar{} Vitality relationships under Baseline and Blueprint prompts and at 52 empirical sites.}

{\def\LTcaptype{none} 
\begin{longtable}[]{@{}
  >{\raggedright\arraybackslash}p{(\linewidth - 6\tabcolsep) * \real{0.2000}}
  >{\raggedleft\arraybackslash}p{(\linewidth - 6\tabcolsep) * \real{0.2667}}
  >{\raggedleft\arraybackslash}p{(\linewidth - 6\tabcolsep) * \real{0.2667}}
  >{\raggedleft\arraybackslash}p{(\linewidth - 6\tabcolsep) * \real{0.2667}}@{}}
\toprule\noalign{}
\begin{minipage}[b]{\linewidth}\raggedright
Quantity
\end{minipage} & \begin{minipage}[b]{\linewidth}\raggedleft
Baseline
\end{minipage} & \begin{minipage}[b]{\linewidth}\raggedleft
Complete Blueprint
\end{minipage} & \begin{minipage}[b]{\linewidth}\raggedleft
52-site empirical reference
\end{minipage} \\
\midrule\noalign{}
\endhead
\bottomrule\noalign{}
\endlastfoot
Residential Density coefficient & 0.599 {[}0.554, 0.647{]} & 0.699 {[}0.657, 0.741{]} & 0.287 {[}\(-0.020\), 0.594{]} \\
Functional Mix coefficient & 0.171 {[}0.137, 0.204{]} & 0.081 {[}0.055, 0.107{]} & 0.312 {[}0.068, 0.556{]} \\
Short Blocks coefficient & 0.136 {[}0.097, 0.176{]} & 0.157 {[}0.117, 0.197{]} & \(-0.841\) {[}\(-1.574\), \(-0.109\){]} \\
Aged Buildings coefficient & \(-0.048\) {[}\(-0.085\), \(-0.011\){]} & \(-0.075\) {[}\(-0.110\), \(-0.039\){]} & 1.189 {[}0.473, 1.905{]} \\
Tall Buildings coefficient & \(-0.045\) {[}\(-0.081\), \(-0.011\){]} & \(-0.072\) {[}\(-0.105\), \(-0.039\){]} & 0.080 {[}\(-0.145\), 0.305{]} \\
Adjusted \(R^2\) & 0.978 {[}0.976, 0.981{]} & 0.978 {[}0.976, 0.981{]} & 0.375 \\
Held-out \(R^2\) & 0.975 {[}0.972, 0.978{]}, random five-fold & 0.976 {[}0.973, 0.978{]}, random five-fold & \(-0.659\), spatial-group held out \\
\end{longtable}
}

The spatial-group bootstrap intervals were {[}\(-0.125\), 0.852{]} for Residential Density, {[}\(-0.062\), 0.642{]} for Functional Mix, {[}\(-1.188\), 0.362{]} for Short Blocks, {[}0.091, 1.953{]} for Aged Buildings and {[}\(-0.148\), 0.668{]} for Tall Buildings. Because the generated and empirical prediction analyses use different holdout units, their held-out \(R^2\) values are reported separately; coefficient comparisons use the standardized estimates in Supplementary Table 14.

The model also ranked the 52 empirical sites using only the five morphology variables. Pairwise accuracy was 0.603 {[}spatial-group bootstrap 95\% CI: 0.515, 0.681{]} for the primary prompt and 0.592 {[}0.512, 0.667{]} for a paraphrase. The two prompts agreed on 0.905 of site-pair directions.

\paragraph{Blueprint, intervention-factor and scale sensitivity}\label{blueprint-intervention-factor-and-scale-sensitivity}

The \(2\times2\times2\) prompt design comprised 800 generations; 795 returned valid content, and all 100 full-Blueprint generations returned 100 valid neighbourhoods. Baseline and Blueprint retained positive mean coefficients for Residential Density, Functional Mix and Short Blocks. Heterogeneity instructions increased the unique-value fraction for all six generated variables.

The full Blueprint generated positive coefficients in 100/100 tables for Residential Density, 79/100 for Functional Mix, 82/100 for Short Blocks, 29/100 for Aged Buildings and 37/100 for Tall Buildings. Relative to Baseline, the heterogeneity component increased the unique-value fraction by 0.096 for Residential Density, 0.044 for Functional Mix, 0.044 for Mean Block Area, 0.077 for Aged Buildings, 0.061 for Tall Buildings and 0.079 for Pedestrian Activity; all six bootstrap intervals excluded zero.

For each of the 52 observed morphology profiles, three otherwise identical variants were created. Functional Mix took a locally plausible low, middle or high value, defined as the 10th, 50th and 90th percentiles among the eight most similar observed sites; all other morphology variables were fixed. Eight pilot profiles were set aside before the main analysis. Of the remaining profiles, 38 had all three variants within the observed morphology space and entered the experiment. GPT-4o ranked the three anonymous variants by expected Pedestrian Activity. It returned the complete High\(>\)Mid\(>\)Low order in 31/38 profiles, or 0.816 {[}Wilson 95\% CI: 0.666, 0.908{]}. Pairwise rates were 32/38 for High\(>\)Mid, 32/38 for High\(>\)Low and 31/38 for Mid\(>\)Low. Six non-ranking responses remained in the denominator.

Residential Density, Short Blocks and Tall Buildings each produced a complete High--Mid--Low Pedestrian Activity order in 38 of 38 settings. Aged Buildings did so in 32 of 38 settings and was more sensitive to phrasing and order controls. Functional Mix controls produced complete orderings in 15--16 of 20 settings, with pairwise agreement of 0.979--1.000 among settings with complete paired outputs. The Tall Buildings intervention is reported as an exploratory response test outside the Jacobs-derived hypotheses.

\textbf{Supplementary Table 15 \textbar{} Vitality sensitivity to the intervened morphology indicator.}

{\def\LTcaptype{none} 
\begin{longtable}[]{@{}
  >{\raggedright\arraybackslash}p{(\linewidth - 6\tabcolsep) * \real{0.2143}}
  >{\raggedleft\arraybackslash}p{(\linewidth - 6\tabcolsep) * \real{0.2857}}
  >{\raggedleft\arraybackslash}p{(\linewidth - 6\tabcolsep) * \real{0.2857}}
  >{\raggedright\arraybackslash}p{(\linewidth - 6\tabcolsep) * \real{0.2143}}@{}}
\toprule\noalign{}
\begin{minipage}[b]{\linewidth}\raggedright
Intervened morphology indicator
\end{minipage} & \begin{minipage}[b]{\linewidth}\raggedleft
Complete High--Mid--Low order
\end{minipage} & \begin{minipage}[b]{\linewidth}\raggedleft
Wilson 95\% interval
\end{minipage} & \begin{minipage}[b]{\linewidth}\raggedright
Prompt-control result
\end{minipage} \\
\midrule\noalign{}
\endhead
\bottomrule\noalign{}
\endlastfoot
Functional Mix & 31/38, 0.816 & {[}0.666, 0.908{]} & 15--16/20 complete under paraphrase, alias/order, repeat and irrelevant-number controls; paired order agreement 0.979--1.000 \\
Residential Density & 38/38, 1.000 & {[}0.908, 1.000{]} & 16/16 consistent under each of the four controls \\
Short Blocks & 38/38, 1.000 & {[}0.908, 1.000{]} & 16/16 consistent under each control \\
Aged Buildings & 32/38, 0.842 & {[}0.696, 0.926{]} & 5/16 paraphrase, 11/16 order, 14/16 repeat and 8/16 irrelevant-number consistency \\
Tall Buildings & 38/38, 1.000 & {[}0.908, 1.000{]} & 16/16 consistent under each control \\
\end{longtable}
}

Aged Buildings produced the prespecified High--Mid--Low order in 32/38 settings, with lower consistency across the four prompt controls (5--14/16).

Across requested table sizes of 20--300 neighbourhoods, Residential Density, Functional Mix and Short Blocks remained positive and Aged Buildings remained negative. Tall Buildings became positive at 300 neighbourhoods, where the complete-return rate was 22\%; this scale-dependent result is retained in the sensitivity analysis.

\textbf{Supplementary Table 16 \textbar{} Vitality sensitivity to the requested number of neighbourhoods.}

{\def\LTcaptype{none} 
\begin{longtable}[]{@{}
  >{\raggedleft\arraybackslash}p{(\linewidth - 14\tabcolsep) * \real{0.1250}}
  >{\raggedleft\arraybackslash}p{(\linewidth - 14\tabcolsep) * \real{0.1250}}
  >{\raggedleft\arraybackslash}p{(\linewidth - 14\tabcolsep) * \real{0.1250}}
  >{\raggedleft\arraybackslash}p{(\linewidth - 14\tabcolsep) * \real{0.1250}}
  >{\raggedleft\arraybackslash}p{(\linewidth - 14\tabcolsep) * \real{0.1250}}
  >{\raggedleft\arraybackslash}p{(\linewidth - 14\tabcolsep) * \real{0.1250}}
  >{\raggedleft\arraybackslash}p{(\linewidth - 14\tabcolsep) * \real{0.1250}}
  >{\raggedleft\arraybackslash}p{(\linewidth - 14\tabcolsep) * \real{0.1250}}@{}}
\toprule\noalign{}
\begin{minipage}[b]{\linewidth}\raggedleft
Requested neighbourhoods
\end{minipage} & \begin{minipage}[b]{\linewidth}\raggedleft
Complete tables
\end{minipage} & \begin{minipage}[b]{\linewidth}\raggedleft
Mean valid rows
\end{minipage} & \begin{minipage}[b]{\linewidth}\raggedleft
Residential Density
\end{minipage} & \begin{minipage}[b]{\linewidth}\raggedleft
Functional Mix
\end{minipage} & \begin{minipage}[b]{\linewidth}\raggedleft
Short Blocks
\end{minipage} & \begin{minipage}[b]{\linewidth}\raggedleft
Aged Buildings
\end{minipage} & \begin{minipage}[b]{\linewidth}\raggedleft
Tall Buildings
\end{minipage} \\
\midrule\noalign{}
\endhead
\bottomrule\noalign{}
\endlastfoot
20 & 98/100 & 19.60 & 0.612 & 0.201 & 0.091 & \(-0.059\) & \(-0.046\) \\
50 & 98/100 & 49.00 & 0.641 & 0.200 & 0.130 & \(-0.075\) & \(-0.079\) \\
100 & 100/100 & 100.00 & 0.599 & 0.171 & 0.136 & \(-0.048\) & \(-0.045\) \\
200 & 88/100 & 188.21 & 0.608 & 0.156 & 0.134 & \(-0.046\) & \(-0.036\) \\
300 & 22/100 & 148.93 & 0.533 & 0.190 & 0.112 & \(-0.040\) & 0.047 \\
\end{longtable}
}

\subsubsection{Place Pulse Perception}\label{place-pulse-perception}

\paragraph{Human-judgement reference and request size}\label{human-judgement-reference-and-request-size}

Each of the six dimensions used 1,000 sampled Place Pulse comparison records and one human vote per record. Dimension-specific accuracy/\(\kappa\) was 0.572/0.245 for Safe, 0.594/0.272 for Beautiful, 0.579/0.251 for Wealthy, 0.528/0.158 for Lively, 0.501/0.156 for Boring and 0.523/0.187 for Depressing. Overall accuracy was 0.550 and Cohen's \(\kappa\) was 0.210. A same-pair plurality sensitivity excluded eight tied pluralities and gave \(\kappa=0.210\) for 5,992 tasks, compared with \(\kappa=0.210\) for the single-vote labels.

Judging one, two or three image pairs per request produced similar agreement. Accuracy/\(\kappa\) was 0.557/0.227, 0.560/0.221 and 0.560/0.233, respectively. Agreement declined at five pairs (0.533/0.188) and ten pairs (0.513/0.163), supporting independent or short-batch judging.

\paragraph{Image-generation prompt components}\label{image-generation-prompt-components}

Scene objects (O), co-occurrence rules (R) and regional/scene heterogeneity (H) formed eight prompt conditions. Of 800 planned images, 797 entered the analysis. O and H increased within-set CLIP distance by 0.0419 {[}0.0320, 0.0513{]} and 0.0179 {[}0.0081, 0.0272{]}, respectively. H reduced empirical-to-generated and generated-to-empirical nearest-neighbour similarity and increased centroid distance. R reduced within-set diversity by 0.0213 {[}\(-0.0312\), \(-0.0114\){]} and had small nearest-neighbour effects.

\textbf{Supplementary Table 17 \textbar{} Effects of Perception Blueprint components on image-representation statistics.}

{\def\LTcaptype{none} 
\begin{longtable}[]{@{}
  >{\raggedright\arraybackslash}p{(\linewidth - 8\tabcolsep) * \real{0.1579}}
  >{\raggedleft\arraybackslash}p{(\linewidth - 8\tabcolsep) * \real{0.2105}}
  >{\raggedleft\arraybackslash}p{(\linewidth - 8\tabcolsep) * \real{0.2105}}
  >{\raggedleft\arraybackslash}p{(\linewidth - 8\tabcolsep) * \real{0.2105}}
  >{\raggedleft\arraybackslash}p{(\linewidth - 8\tabcolsep) * \real{0.2105}}@{}}
\toprule\noalign{}
\begin{minipage}[b]{\linewidth}\raggedright
Blueprint component
\end{minipage} & \begin{minipage}[b]{\linewidth}\raggedleft
Within-generated CLIP distance
\end{minipage} & \begin{minipage}[b]{\linewidth}\raggedleft
Place Pulse \(\rightarrow\) generated similarity
\end{minipage} & \begin{minipage}[b]{\linewidth}\raggedleft
Generated \(\rightarrow\) Place Pulse similarity
\end{minipage} & \begin{minipage}[b]{\linewidth}\raggedleft
Centroid distance
\end{minipage} \\
\midrule\noalign{}
\endhead
\bottomrule\noalign{}
\endlastfoot
Scene objects (O) & +0.0419 {[}0.0320, 0.0513{]} & \(-0.0061\) {[}\(-0.0132\), \(-0.0008\){]} & \(-0.0280\) {[}\(-0.0357\), \(-0.0208\){]} & +0.0014 {[}\(-0.0050\), 0.0083{]} \\
Co-occurrence rules (R) & \(-0.0213\) {[}\(-0.0312\), \(-0.0114\){]} & +0.0006 {[}\(-0.0061\), 0.0066{]} & +0.0054 {[}\(-0.0011\), 0.0117{]} & +0.0073 {[}0.0005, 0.0141{]} \\
Regional/scene heterogeneity (H) & +0.0179 {[}0.0081, 0.0272{]} & \(-0.0192\) {[}\(-0.0260\), \(-0.0135\){]} & \(-0.0554\) {[}\(-0.0630\), \(-0.0482\){]} & +0.0232 {[}0.0168, 0.0302{]} \\
\end{longtable}
}

\paragraph{City-held-out image-representation comparison}\label{city-held-out-image-representation-comparison}

The empirical reference contains 3,840 distinct Place Pulse images from 16 held-out cities. These cities were reserved for this representation comparison and were not used to define the image prompts or select prompt components. Embedding metrics use the images only; Place Pulse votes enter the separate agreement analysis. Intervals jointly resample Place Pulse cities and generated images.

\textbf{Supplementary Table 18 \textbar{} CLIP comparison between generated streetscapes and held-out Place Pulse images.}

{\def\LTcaptype{none} 
\begin{longtable}[]{@{}
  >{\raggedright\arraybackslash}p{(\linewidth - 6\tabcolsep) * \real{0.2000}}
  >{\raggedleft\arraybackslash}p{(\linewidth - 6\tabcolsep) * \real{0.2667}}
  >{\raggedleft\arraybackslash}p{(\linewidth - 6\tabcolsep) * \real{0.2667}}
  >{\raggedleft\arraybackslash}p{(\linewidth - 6\tabcolsep) * \real{0.2667}}@{}}
\toprule\noalign{}
\begin{minipage}[b]{\linewidth}\raggedright
CLIP quantity
\end{minipage} & \begin{minipage}[b]{\linewidth}\raggedleft
Baseline (99 images)
\end{minipage} & \begin{minipage}[b]{\linewidth}\raggedleft
Complete Blueprint (100 images)
\end{minipage} & \begin{minipage}[b]{\linewidth}\raggedleft
Blueprint minus Baseline
\end{minipage} \\
\midrule\noalign{}
\endhead
\bottomrule\noalign{}
\endlastfoot
Place Pulse \(\rightarrow\) generated nearest-neighbour similarity & 0.7091 & 0.6844 & \(-0.0246\) {[}\(-0.0515\), \(-0.0065\){]} \\
Generated \(\rightarrow\) Place Pulse nearest-neighbour similarity & 0.7930 & 0.7119 & \(-0.0811\) {[}\(-0.0936\), \(-0.0463\){]} \\
Generated--empirical centroid distance & 0.2009 & 0.2356 & +0.0347 {[}0.0144, 0.0529{]} \\
\end{longtable}
}

In an equal-size 99-versus-99 comparison, the two nearest-neighbour differences were \(-0.0249\) {[}\(-0.0332\), \(-0.0173\){]} and \(-0.0705\) {[}\(-0.0899\), \(-0.0538\){]}, and the centroid-distance difference was +0.0348 {[}0.0275, 0.0424{]}. Two independent 99-image Place Pulse halves yielded approximately 0.876 similarity in both nearest-neighbour directions and a centroid distance of 0.0040.

\textbf{Supplementary Table 19 \textbar{} Sensitivity of the generated--Place Pulse comparison to image representation.}

{\def\LTcaptype{none} 
\begin{longtable}[]{@{}
  >{\raggedright\arraybackslash}p{(\linewidth - 4\tabcolsep) * \real{0.2727}}
  >{\raggedleft\arraybackslash}p{(\linewidth - 4\tabcolsep) * \real{0.3636}}
  >{\raggedleft\arraybackslash}p{(\linewidth - 4\tabcolsep) * \real{0.3636}}@{}}
\toprule\noalign{}
\begin{minipage}[b]{\linewidth}\raggedright
Blueprint minus Baseline
\end{minipage} & \begin{minipage}[b]{\linewidth}\raggedleft
DINOv2
\end{minipage} & \begin{minipage}[b]{\linewidth}\raggedleft
ResNet-50
\end{minipage} \\
\midrule\noalign{}
\endhead
\bottomrule\noalign{}
\endlastfoot
Within-generated mean distance & +0.0943 {[}0.0267, 0.1647{]} & +0.0151 {[}0.0019, 0.0277{]} \\
Place Pulse \(\rightarrow\) generated similarity & \(-0.0581\) {[}\(-0.1136\), \(-0.0041\){]} & +0.0007 {[}\(-0.0059\), 0.0085{]} \\
Generated \(\rightarrow\) Place Pulse similarity & \(-0.1204\) {[}\(-0.2000\), \(-0.0178\){]} & \(-0.0009\) {[}\(-0.0076\), 0.0053{]} \\
Centroid distance & +0.0608 {[}\(-0.0691\), 0.1786{]} & \(-0.0092\) {[}\(-0.0207\), 0.0036{]} \\
\end{longtable}
}

All three encoders showed greater within-generated dispersion. CLIP and DINOv2 also showed lower nearest-neighbour similarity to Place Pulse, whereas the ResNet-50 change was near zero. Image-to-empirical proximity therefore varied with the representation.

\paragraph{Theme intervention}\label{theme-intervention}

Twelve scenes each contained a baseline image, Natural, Traffic and Built theme edits, and one same-scene editing control. All 48 pairwise quality checks for the theme edits and editing controls passed identity-preservation and manipulation checks. Sixty images were scored one at a time by GPT-4o without condition labels, scene identifiers or study hypotheses. The estimand was the theme score minus the editing-control score in the same scene. Scene bootstrap intervals and exact paired sign-flip tests used the scene as the inferential unit. Holm correction was applied to the three prespecified relationships and Benjamini--Hochberg correction to the full \(3\times6\) matrix.

\textbf{Supplementary Table 20 \textbar{} Editing-control-adjusted Perception effects for the complete theme-by-dimension matrix.}

{\def\LTcaptype{none} 
\begin{longtable}[]{@{}
  >{\raggedright\arraybackslash}p{(\linewidth - 8\tabcolsep) * \real{0.1667}}
  >{\raggedright\arraybackslash}p{(\linewidth - 8\tabcolsep) * \real{0.1667}}
  >{\raggedleft\arraybackslash}p{(\linewidth - 8\tabcolsep) * \real{0.2222}}
  >{\raggedleft\arraybackslash}p{(\linewidth - 8\tabcolsep) * \real{0.2222}}
  >{\raggedleft\arraybackslash}p{(\linewidth - 8\tabcolsep) * \real{0.2222}}@{}}
\toprule\noalign{}
\begin{minipage}[b]{\linewidth}\raggedright
Theme
\end{minipage} & \begin{minipage}[b]{\linewidth}\raggedright
Dimension
\end{minipage} & \begin{minipage}[b]{\linewidth}\raggedleft
Effect {[}95\% CI{]}
\end{minipage} & \begin{minipage}[b]{\linewidth}\raggedleft
Exact sign-flip \(p\)
\end{minipage} & \begin{minipage}[b]{\linewidth}\raggedleft
BH \(q\) across 18 cells
\end{minipage} \\
\midrule\noalign{}
\endhead
\bottomrule\noalign{}
\endlastfoot
Natural & Beautiful & 2.233 {[}1.650, 2.792{]} & 0.000488 & 0.002930 \\
Natural & Depressing & \(-2.775\) {[}\(-3.567\), \(-2.025\){]} & 0.000488 & 0.002930 \\
Natural & Lively & 1.192 {[}0.733, 1.658{]} & 0.001953 & 0.004395 \\
Natural & Safe & 1.092 {[}0.717, 1.433{]} & 0.001953 & 0.004395 \\
Natural & Wealthy & 1.350 {[}0.917, 1.850{]} & 0.000977 & 0.003516 \\
Natural & Boring & \(-1.192\) {[}\(-1.500\), \(-0.917\){]} & 0.000488 & 0.002930 \\
Traffic & Beautiful & \(-0.767\) {[}\(-1.250\), \(-0.342\){]} & 0.003906 & 0.007813 \\
Traffic & Depressing & 0.083 {[}\(-0.300\), 0.475{]} & 0.757813 & 0.796875 \\
Traffic & Lively & 1.458 {[}0.958, 2.008{]} & 0.000977 & 0.003516 \\
Traffic & Safe & \(-0.408\) {[}\(-0.867\), 0.050{]} & 0.148438 & 0.222656 \\
Traffic & Wealthy & 0.242 {[}\(-0.050\), 0.508{]} & 0.171875 & 0.231027 \\
Traffic & Boring & \(-0.375\) {[}\(-0.858\), 0.083{]} & 0.195313 & 0.234375 \\
Built & Beautiful & 0.483 {[}\(-0.042\), 0.983{]} & 0.128906 & 0.210938 \\
Built & Depressing & \(-1.067\) {[}\(-1.775\), \(-0.417\){]} & 0.015625 & 0.028125 \\
Built & Lively & 0.150 {[}\(-0.058\), 0.375{]} & 0.312500 & 0.351563 \\
Built & Safe & 0.342 {[}\(-0.117\), 0.725{]} & 0.179688 & 0.231027 \\
Built & Wealthy & 0.975 {[}0.642, 1.292{]} & 0.001953 & 0.004395 \\
Built & Boring & 0.058 {[}\(-0.233\), 0.350{]} & 0.796875 & 0.796875 \\
\end{longtable}
}

The full 18-cell matrix is reported in Supplementary Table 20. Natural--Beautiful, Natural--Depressing and Traffic--Lively were prespecified; each had Holm-adjusted \(p=0.001465\).

\paragraph{Single-cue edits}\label{single-cue-edits}

Separate edits increased tree canopy, added visible pedestrian activity or repaired façades, with an editing control for each scene. Editing-control-adjusted changes were 1.208 {[}0.666, 1.833{]} for Tree canopy--Beautiful, \(-1.525\) {[}\(-2.358\), \(-0.667\){]} for Tree canopy--Depressing, 1.042 {[}0.542, 1.542{]} for Visible activity--Lively, 0.958 {[}0.750, 1.167{]} for Façade repair--Safe and 0.942 {[}0.675, 1.250{]} for Façade repair--Wealthy.

\subsubsection{Open-model analyses of selected output regularities}\label{open-model-analyses-of-selected-output-regularities}

Because GPT-4o does not expose internal activations, the selected tasks were repeated with Qwen2.5-VL-72B. The analysis began by giving both models the exact same task prompts and comparing their answers. Behavioural controls then tested whether city names or one fixed formula accounted for the response. Linear readouts measured whether selected city attributes and relation types were distinguishable in intermediate activations. Finally, activation edits tested whether adding, removing or replacing selected states shifted the answer. In the numerical tasks, the edits shifted short-answer logits while the prespecified statistics calculated from newly generated tables remained unchanged.

Linear probes follow work showing that spatial and temporal attributes can be decoded from language-model activations{[}23{]}. Separate geospatial work uses spatial autocorrelation and sparse autoencoders to map geographic structure in model activations{[}24{]}. Task-difference directions and activation edits follow function-vector and model-editing studies that test whether internal-state changes alter outputs{[}25--26{]}. Recent medical-domain interpretability work likewise combines activation mapping, lesioning and activation patching{[}27{]}.

\paragraph{Behavioural correspondence before hidden-state analysis}\label{behavioural-correspondence-before-hidden-state-analysis}

For each numerical case, we prepared six task descriptions with different fictional backgrounds. Each description asked the model to fill one table containing all tested Population, distance or Functional Mix values. GPT-4o and Qwen2.5-VL-72B-Instruct-AWQ received exactly the same descriptions, values and table templates. We then paired the corresponding Road, GDP, land-density or activity values in their tables. Figure 5a shows a mean response curve: values at the same table position were averaged across the six pairs of tables. In Article equation (16), \(r_t^{\mathrm{entry}}\) denotes the entry-by-entry correlation and \(r_t^{\mathrm{curve}}\) the mean-response-curve correlation, where \(t\) indexes the task. These output comparisons provide the behavioural basis for interpreting the Qwen internal-state analyses.

\textbf{Supplementary Table 21 \textbar{} Behavioural correspondence and held-out open-model task results.}

{\def\LTcaptype{none} 
\begin{longtable}[]{@{}
  >{\raggedright\arraybackslash}p{(\linewidth - 6\tabcolsep) * \real{0.2143}}
  >{\raggedleft\arraybackslash}p{(\linewidth - 6\tabcolsep) * \real{0.2857}}
  >{\raggedleft\arraybackslash}p{(\linewidth - 6\tabcolsep) * \real{0.2857}}
  >{\raggedright\arraybackslash}p{(\linewidth - 6\tabcolsep) * \real{0.2143}}@{}}
\toprule\noalign{}
\begin{minipage}[b]{\linewidth}\raggedright
Case
\end{minipage} & \begin{minipage}[b]{\linewidth}\raggedleft
Corresponding-value correlation
\end{minipage} & \begin{minipage}[b]{\linewidth}\raggedleft
Mean-response-curve correlation in Fig. 5a
\end{minipage} & \begin{minipage}[b]{\linewidth}\raggedright
Open-model task result
\end{minipage} \\
\midrule\noalign{}
\endhead
\bottomrule\noalign{}
\endlastfoot
Urban Scaling & \(r_t^{\mathrm{entry}}=0.932\) {[}0.900, 0.963{]}; 132 paired quantities \(=6\) examples \(\times11\) changes from the smallest of 12 Population inputs \(\times2\) outputs, Road and GDP & \(r_t^{\mathrm{curve}}=0.975\); 22 mean quantities \(=11\) Population changes \(\times2\) outputs & In held-out joint tables, \(\beta_{\mathrm{GDP}}-\beta_{\mathrm{Road}}=0.550\) {[}0.373, 0.668{]} \\
Distance Decay & \(r_t^{\mathrm{entry}}=0.977\) {[}0.970, 0.991{]}; 60 paired densities \(=6\) fictional city examples \(\times10\) distances & \(r_t^{\mathrm{curve}}=0.997\); 10 mean land-density values, one per distance & All 6 fictional city examples declined; mean standardized decay strength 0.862 {[}0.808, 0.910{]} \\
Vitality & \(r_t^{\mathrm{entry}}=0.998\) {[}0.998, 0.998{]}; 48 paired quantities \(=6\) neighbourhood examples \(\times8\) changes from the 0.10 reference among 9 Functional Mix inputs & \(r_t^{\mathrm{curve}}=0.998\); 8 mean log-activity changes & Endpoint log-activity contrast 1.294; little residual variation among the six examples \\
Perception & Not applicable & Complete pattern of 18 theme-by-dimension effects, \(r=0.929\) {[}0.862, 0.940{]} & Natural, Traffic and Built effects were estimated for all six dimensions \\
\end{longtable}
}

These correlations are calculated within each task. A paired quantity is a corresponding Road or GDP change, land-density value or log-activity change derived from matching positions in the two models' tables. A point on a mean response curve is that quantity averaged across the six pairs of tables.

\paragraph{Testing city-name and fixed-rule explanations}\label{testing-city-name-and-fixed-rule-explanations}

Fifty cities received the same six Population values under named and anonymous conditions. GDP and Road response slopes were similar across conditions. A separate set of 60 fictional cities received city-specific rules that deliberately contradicted standard Scaling directions; outputs followed the current rule, whereas placing the same numbers in irrelevant identifiers did not produce the response. City names retained some real-attribute ranking information, but city-name facts and repetition of one fixed formula did not account for the full behaviour.

\paragraph{Readable city attributes and relation types}\label{readable-city-attributes-and-relation-types}

Linear probes were trained on hidden activations elicited by city names and evaluated on spatially blocked held-out administrative units. Test \(R^2\) was 0.894 for Population, 0.821 for GDP and 0.379 for Road. The Population readout also distinguished Population or Residents changes from irrelevant-number changes in synthetic tasks.

Matched synthetic prompts explicitly specified superlinear, linear or sublinear relationships. A selected hidden state distinguished superlinear from sublinear prompts with test AUC 0.975 {[}0.967, 0.983{]}. This result establishes linear decodability of the task relation in the tested activation; the activation edits below provide the separate output test.

For the common numerical-task screen, relation-state AUC was 0.944 for Scaling, 0.750 for Decay and 1.000 for Vitality. Under the prespecified 0.90 rule, state editing during table generation was performed for Scaling and Vitality. Perception used paired image-state replacement in the full theme task.

\paragraph{Activation editing}\label{activation-editing}

The full relation-state replacement produced consistent changes, whereas the one-dimensional Population probe direction did not. Replacing the selected activation with the full activation from the opposite Scaling relation changed GDP and Road answer preference by 4.575 {[}4.208, 4.912{]} and 3.992 {[}3.665, 4.294{]} logit units. A separately selected relation direction was then tested in 12 new fictional city examples, yielding 7,296 intervention measurements. Its dose slope was 0.113 {[}0.099, 0.127{]} for GDP and 0.245 {[}0.227, 0.261{]} for Road, exceeding Population, irrelevant-number and equal-norm random directions.

Scaling and Vitality relations were recognized in hidden states (AUC 0.944 and 1.000) and remained present in held-out generated tables (\(\beta_{\mathrm{GDP}}-\beta_{\mathrm{Road}}=0.550\) {[}0.373, 0.668{]}; 6/6 Vitality tables positive). In separate short-answer experiments, activation editing shifted answer preferences. When the common-screen directions were applied during table generation, the symmetric effects \(E_{b,t}\) were zero for both tasks. Under the prespecified 0.90 rule, this table-generation test was performed for Scaling and Vitality only.

\paragraph{Perception state replacement}\label{perception-state-replacement}

The Qwen2.5-VL-72B \(3\times6\) theme-effect pattern correlated with the GPT-4o comparison pattern at \(r=0.929\) {[}0.862, 0.940{]}. Replacing theme-image token states with paired same-scene editing-control states reduced the Qwen pattern by 3.712 {[}3.084, 4.312{]} logit units when each Qwen change was weighted by the direction and magnitude of the corresponding GPT-4o difference. This reduction exceeded an equal-norm random perturbation at the same positions. Within image tokens, high-change positions produced greater reduction than low-change positions and random controls. The matched text-token perturbation produced the largest reduction, so the prespecified modality-specificity gate was not met.

The 7B checkpoint preserved the overall theme pattern (\(r=0.847\) {[}0.760, 0.875{]}) and the three central effect directions. Theme-to-editing-control replacement reduced the GPT-4o-weighted pattern by 0.272 {[}0.167, 0.367{]}; the matched random perturbation produced the larger absolute change in every scene (specificity \(-0.594\) {[}\(-0.767\), \(-0.430\){]}). The 72B and 7B checkpoints therefore shared directional behaviour, while editing-control state-replacement specificity was higher for the 72B checkpoint in this within-family comparison.

\subsection{Reproducibility resources and analysis units}\label{reproducibility-resources-and-analysis-units}

Supplementary Data 1 contains the prompt templates and inference settings, Supplementary Data 2 describes the empirical sources and spatial definitions, Supplementary Data 3 contains parsed outputs, and Source Data provides the values plotted in the figures. Numerical synthesis uses an independently generated table or fictional example as the resampling unit. Vitality empirical coefficients use HC3 intervals, with an eight-spatial-group bootstrap as a dependence-sensitive analysis; Perception intervention inference uses the scene. Requested and valid sample sizes are reported separately when responses were missing or incomplete.

\subsubsection{Analysis inclusion and technical checks}\label{analysis-inclusion-and-technical-checks}

Analyses used the first usable response for each request; calls that produced no output followed the predefined retry policy. In an initial technical Vitality ranking run, an output-schema error caused alias placeholders to be repeated in 116/118 responses. The schema was corrected before ranking outcomes were inspected, and the same inputs and estimand were rerun; only the corrected run is analysed here.

The Perception Blueprint experiment planned 800 images and analysed 797; three requests produced no usable image. All analysed images were 1672$\times$941 pixels. The theme experiment comprised 36 theme edits, 12 reused Baseline images and 12 editing controls; all 48 edited--control pairs met the identity-preservation and manipulation criteria, and all 60 images were scored blind. The single-cue experiment comprised 12 Baseline images and 48 predefined edits; all 48 edited--control pairs met the same criteria, and all 60 images were scored blind.

Completeness used the originally requested row or band count. Responses with fewer valid rows contributed only to analyses defined for all valid rows; completeness and model-selection rates retained the planned requests as the denominator.

\subsection{Supplementary Figures}\label{supplementary-figures}

\begin{figure}[H]
\centering
\includegraphics[width=\textwidth,height=0.78\textheight,keepaspectratio]{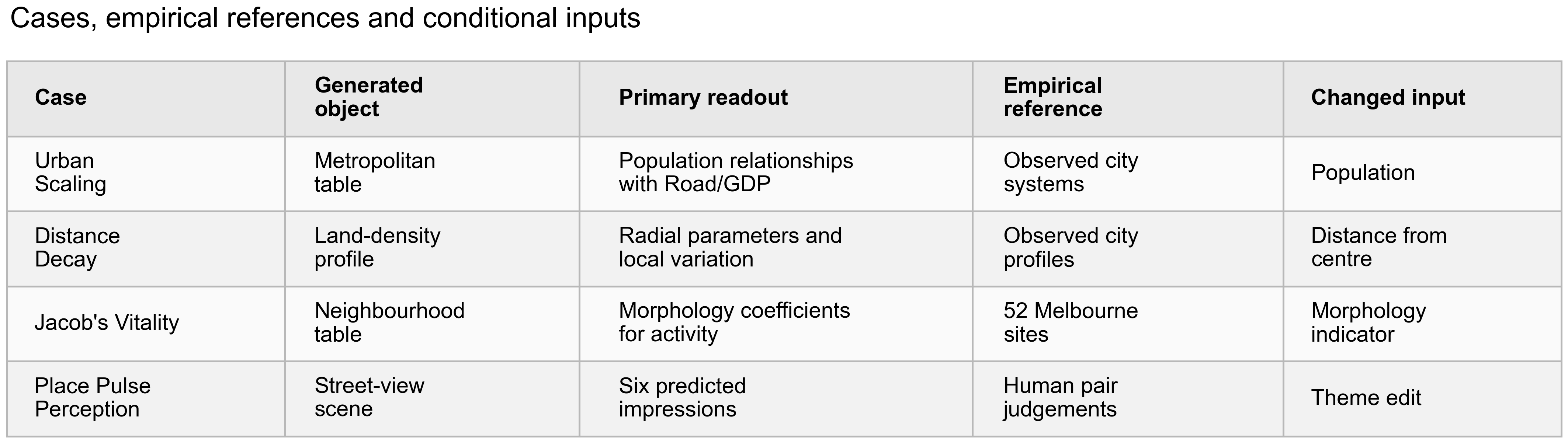}
\caption{\textbf{Case definitions, empirical references and conditional inputs.} The rows identify the generated object, primary relationship or outcome, empirical reference and changed input for Urban Scaling, Distance Decay, Jacob's Vitality and Place Pulse Perception.}
\label{fig:s1}
\end{figure}

\begin{figure}[H]
\centering
\includegraphics[width=\textwidth,height=0.78\textheight,keepaspectratio]{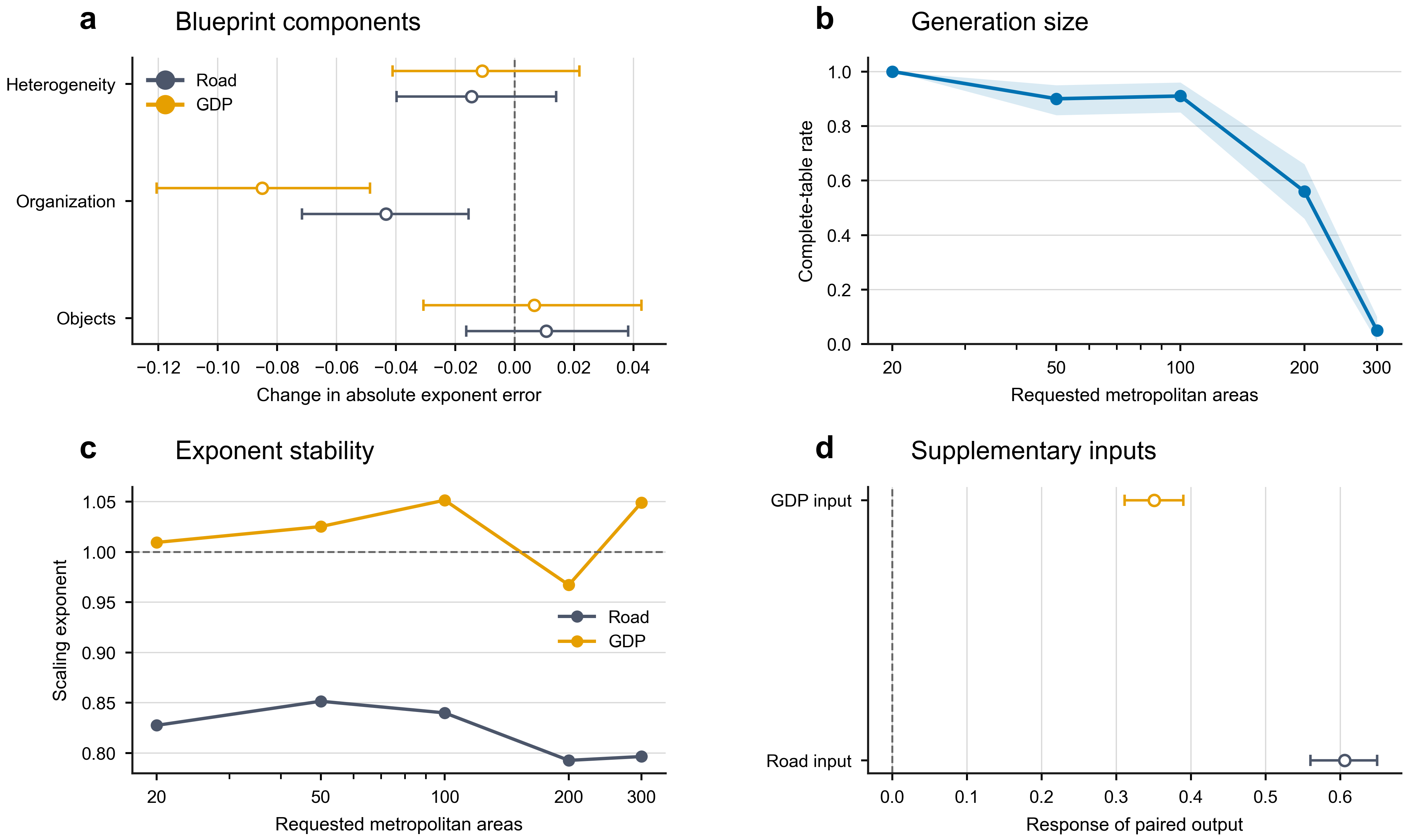}
\caption{\textbf{Urban Scaling prompt components, generation size and supplementary inputs.} \textbf{a,} Marginal effects of explicit objects, system organization and heterogeneity on absolute Road and GDP exponent error; error bars are percentile 95\% confidence intervals from 10,000 resamples of matched generated-table replicate indices. \textbf{b,} Complete-table rate from 20 to 300 requested metropolitan areas; shading denotes percentile 95\% confidence intervals from 10,000 generation-level bootstrap resamples. \textbf{c,} Mean Road and GDP exponents across generation sizes; points are condition means and no uncertainty intervals are drawn. \textbf{d,} Paired-output responses when Road or GDP is changed while Population is fixed; bars denote setting-bootstrap percentile 95\% confidence intervals.}
\label{fig:s2}
\end{figure}

\begin{figure}[H]
\centering
\includegraphics[width=\textwidth,height=0.78\textheight,keepaspectratio]{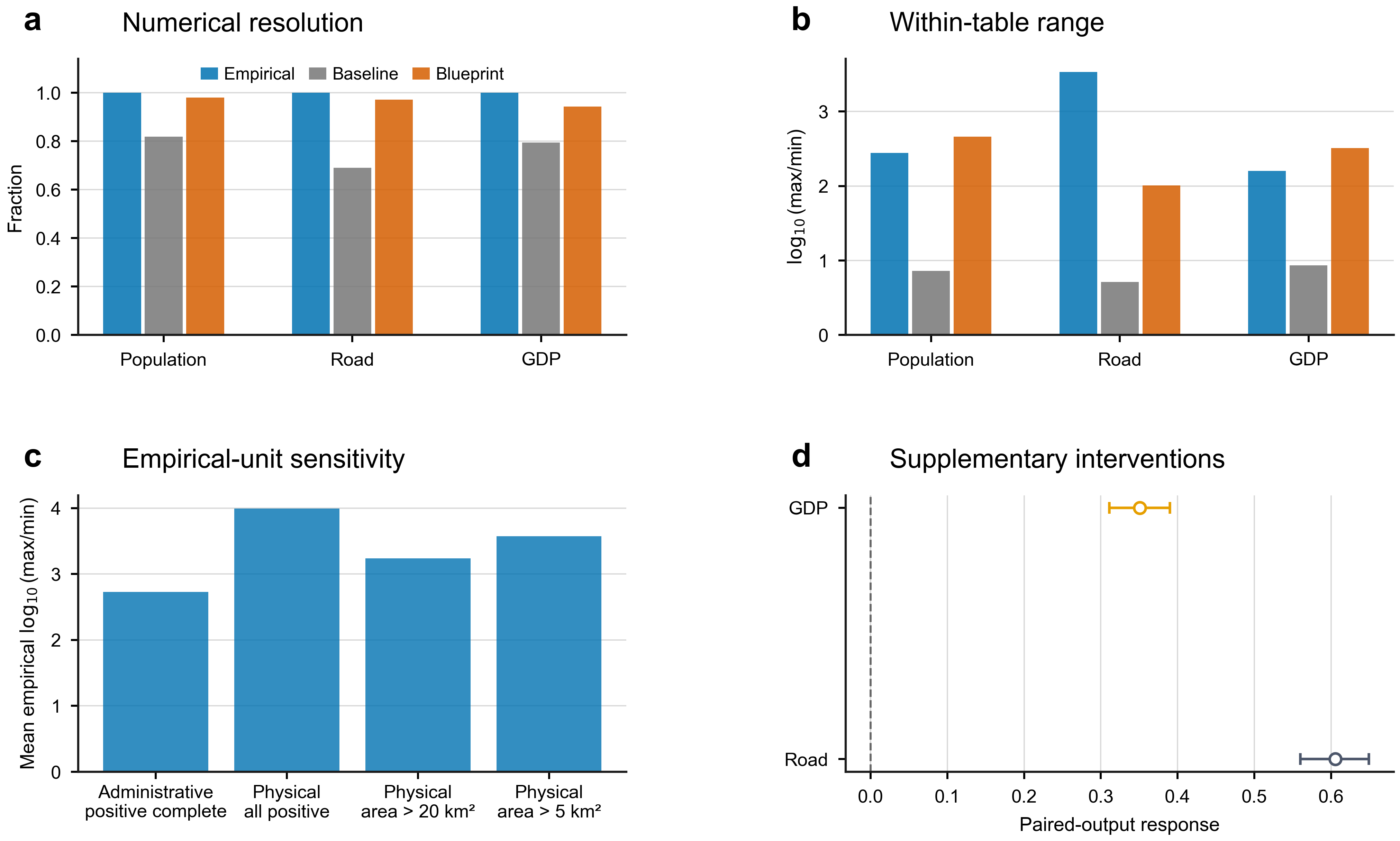}
\caption{\textbf{Urban Scaling distributions, empirical-unit sensitivity and supplementary interventions.} \textbf{a,b,} Unique-value fractions and within-table \(\log_{10}\)(maximum/minimum) ranges for Baseline, Blueprint and empirical 100-city resamples. \textbf{c,} Mean empirical range under the administrative-unit reference and three inclusion thresholds within the archived physical-city polygons. \textbf{d,} Paired-output responses when Road or GDP is changed; bars denote percentile 95\% confidence intervals.}
\label{fig:s3}
\end{figure}

\begin{figure}[H]
\centering
\includegraphics[width=\textwidth,height=0.78\textheight,keepaspectratio]{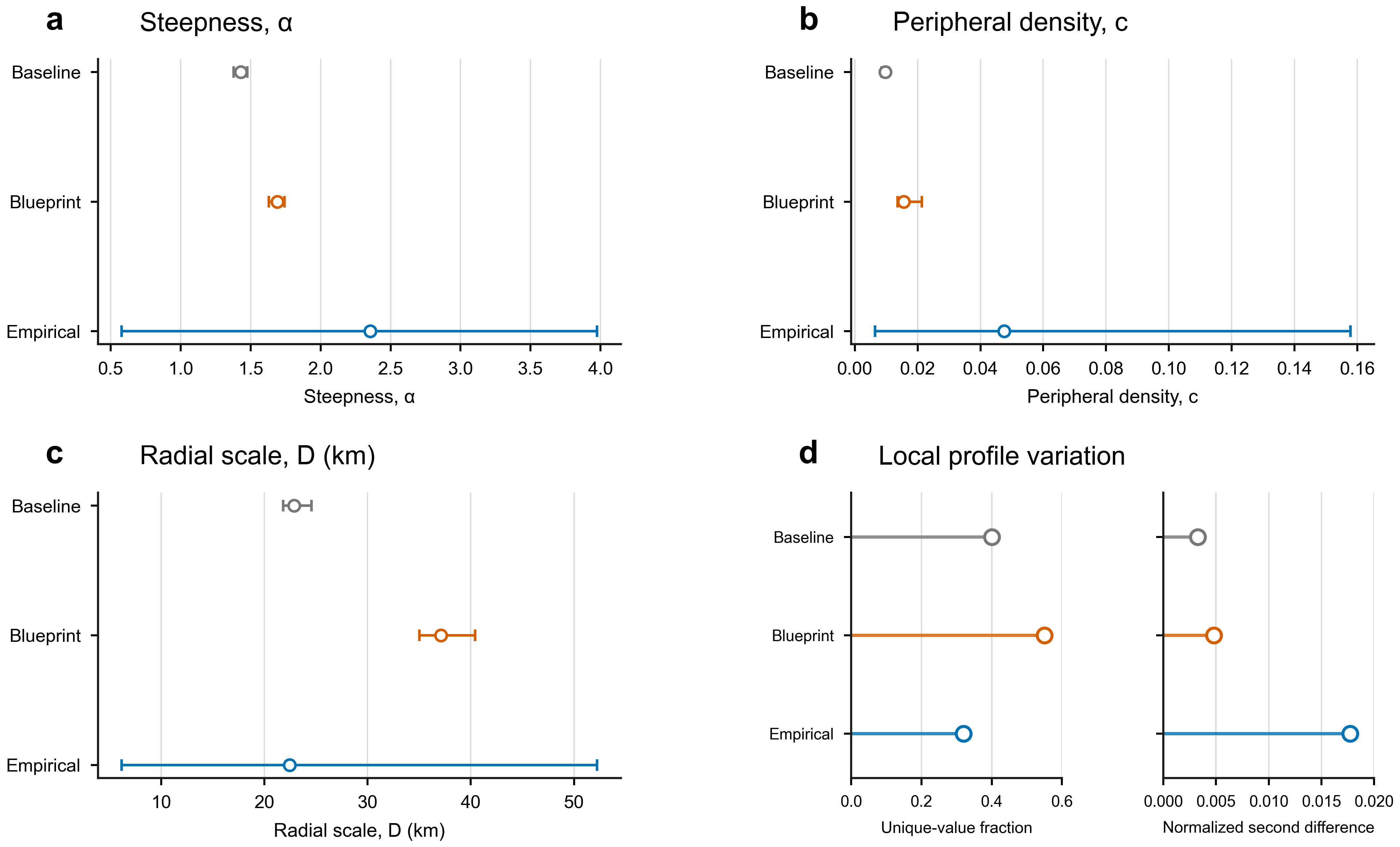}
\caption{\textbf{Distance Decay parameters and local profile variation.} \textbf{a--c,} Median steepness \(\alpha\), peripheral density \(c\) and radial scale \(D\) for Baseline and Blueprint profiles and the 50-city empirical reference. Generated intervals are percentile 95\% confidence intervals; empirical intervals span the 2.5th--97.5th percentiles. \textbf{d,} Median unique-value fraction and normalized second difference for the same three sources.}
\label{fig:s4}
\end{figure}

\begin{figure}[H]
\centering
\includegraphics[width=\textwidth,height=0.78\textheight,keepaspectratio]{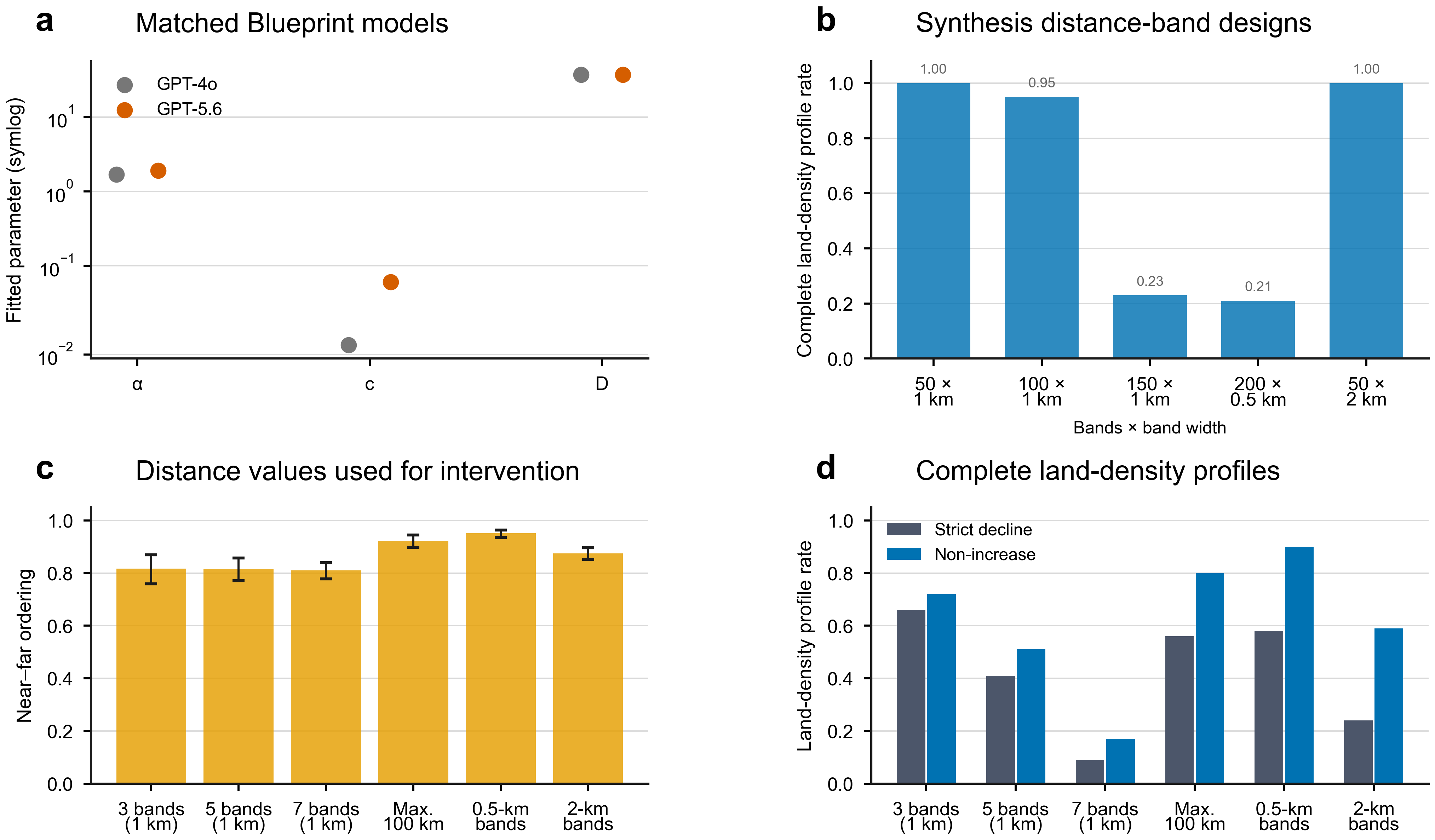}
\caption{\textbf{Distance Decay model and distance-band sensitivity.} \textbf{a,} Parameters fitted to matched GPT-4o and \texttt{gpt-5.6} gateway-alias Blueprint profiles; the symlog axis accommodates \(\alpha\), \(c\) and \(D\) on one scale. \textbf{b,} Complete-profile rate as the requested number and width of synthesis distance bands change. \textbf{c,} Near--far ordering across the six tested intervention designs, retaining the planned number of city examples as the denominator; bars denote percentile 95\% confidence intervals. \textbf{d,} Strict-decline and non-increase rates among the tested city examples.}
\label{fig:s5}
\end{figure}

\begin{figure}[H]
\centering
\includegraphics[width=\textwidth,height=0.78\textheight,keepaspectratio]{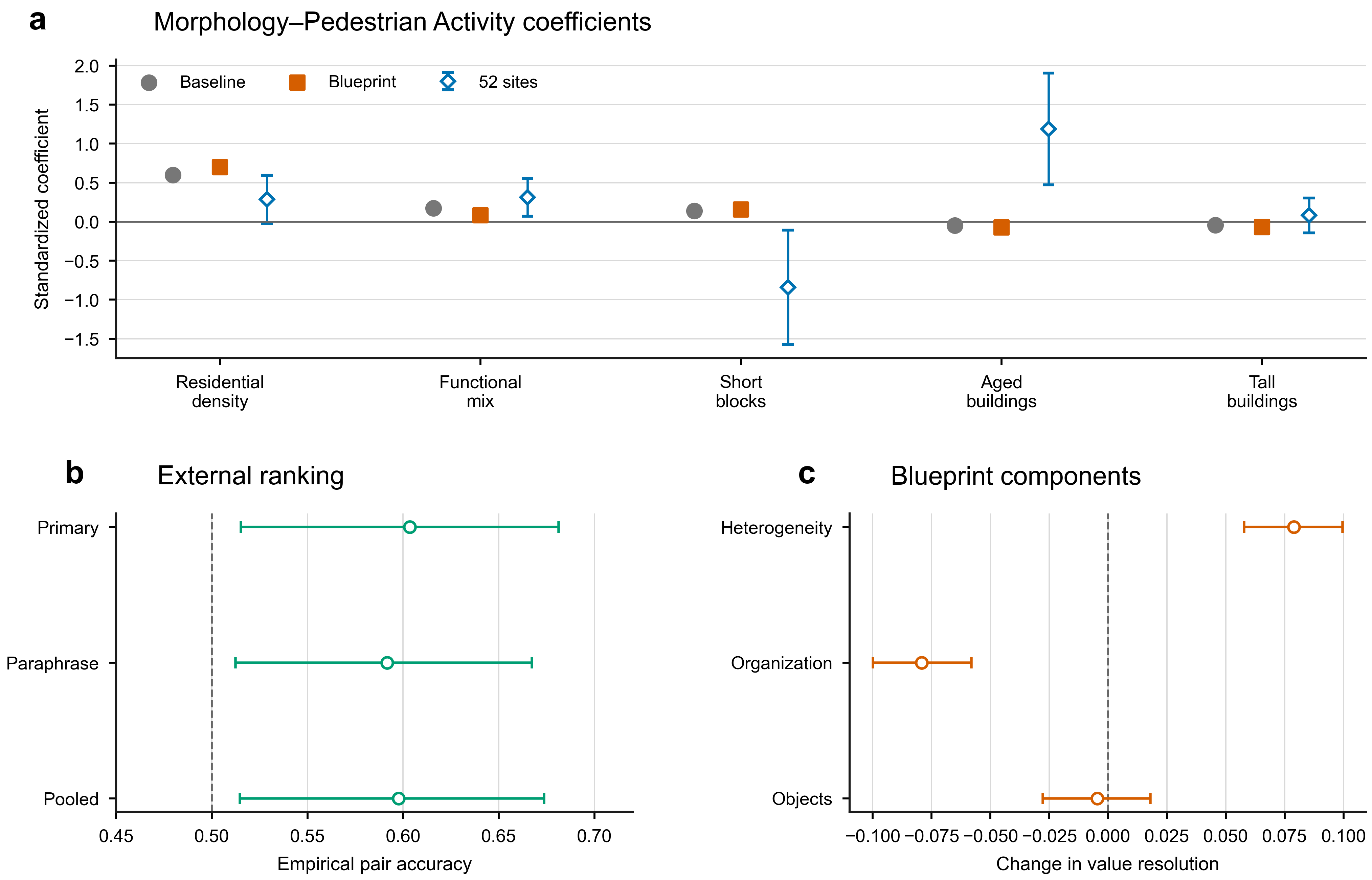}
\caption{\textbf{Jacob's Vitality synthesis and empirical validation.} \textbf{a,} Standardized morphology coefficients in Baseline and Blueprint tables and at 52 Melbourne sites. Baseline and Blueprint points are means; empirical error bars are HC3 95\% confidence intervals. \textbf{b,} Pedestrian Activity pair-ranking accuracy from morphology indicators alone for the primary prompt, its paraphrase and their pooled results; bars are spatial-group-bootstrap percentile 95\% confidence intervals. \textbf{c,} Marginal effects of the three Blueprint components on Pedestrian Activity value resolution; error bars are percentile 95\% confidence intervals from 10,000 resamples of matched generated-table replicate indices.}
\label{fig:s6}
\end{figure}

\begin{figure}[H]
\centering
\includegraphics[width=\textwidth,height=0.78\textheight,keepaspectratio]{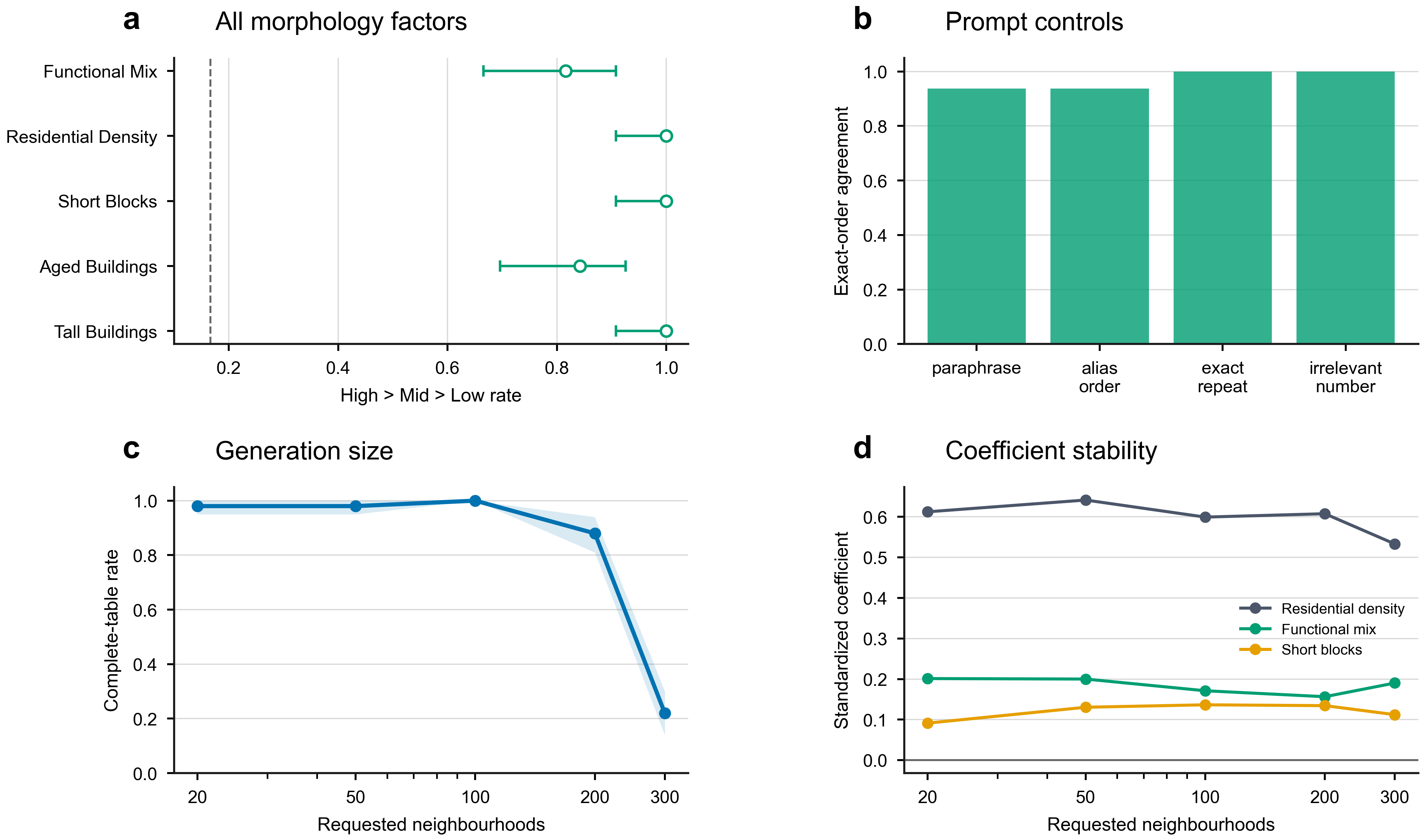}
\caption{\textbf{Jacob's Vitality intervention-factor and generation-size sensitivity.} \textbf{a,} Complete High--Mid--Low Pedestrian Activity ordering when each morphology factor is changed in 38 neighbourhood settings; for Functional Mix, these are locally low, middle and high levels (\(q_{10}\), \(q_{50}\) and \(q_{90}\) among similar observed sites). Bars are Wilson 95\% confidence intervals and the dashed line denotes the \(1/6\) random-order reference. \textbf{b,} Exact-order agreement under prompt controls; point estimates are shown without intervals. \textbf{c,} Complete-table rate from 20 to 300 requested neighbourhoods; shading denotes percentile 95\% confidence intervals from 10,000 generation-level bootstrap resamples. \textbf{d,} Standardized coefficients for three consistently positive predictors across generation sizes; points are condition means and no uncertainty intervals are drawn.}
\label{fig:s7}
\end{figure}

\begin{figure}[H]
\centering
\includegraphics[width=\textwidth,height=0.78\textheight,keepaspectratio]{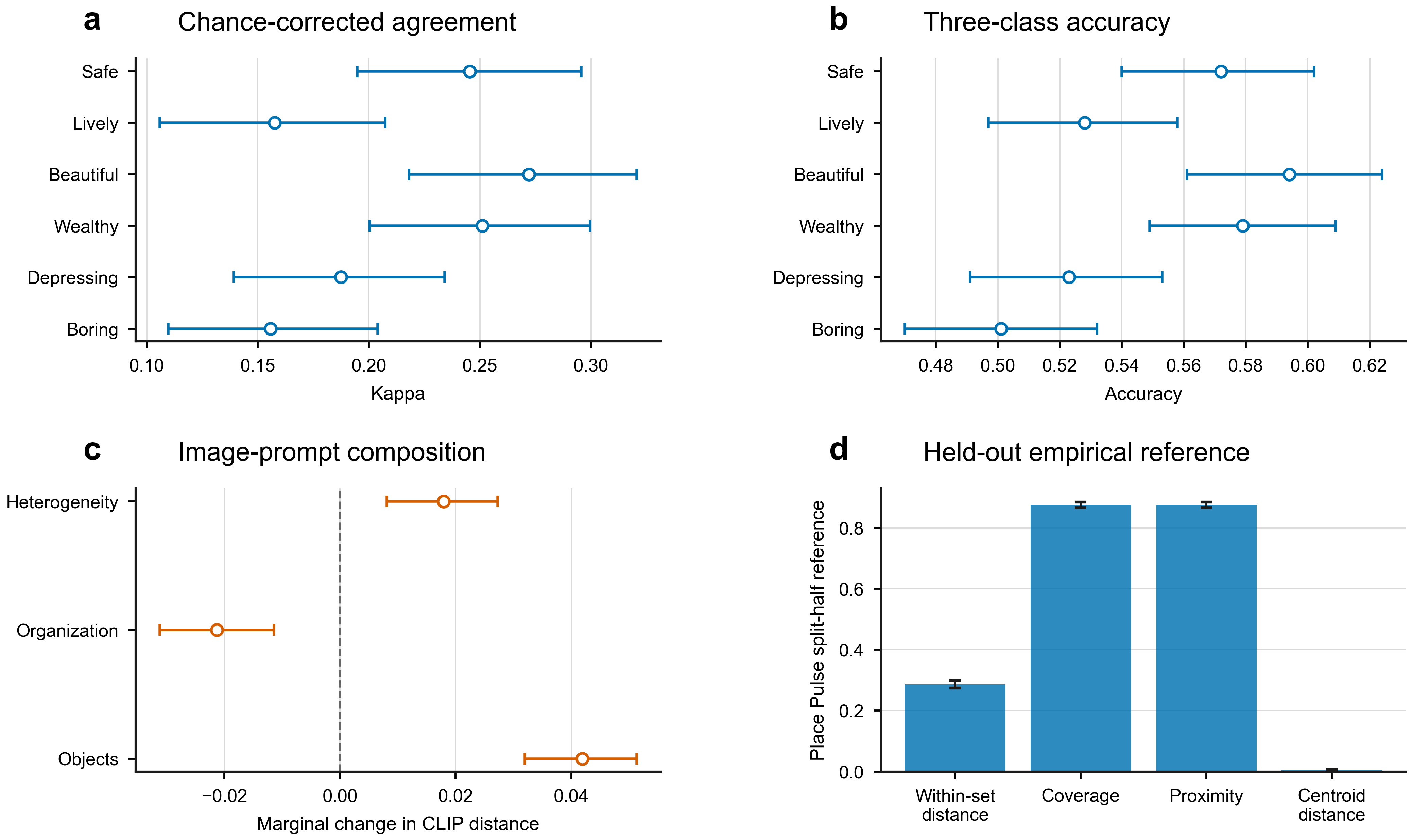}
\caption{\textbf{Place Pulse human agreement, image-prompt composition and empirical split-half references.} \textbf{a,b,} Three-class Cohen's \(\kappa\) and accuracy between GPT-4o choices and one sampled human vote for 1,000 pairs per dimension; bars are task-bootstrap percentile 95\% confidence intervals. \textbf{c,} Marginal effects of explicit objects, spatial organization and heterogeneity on within-generated CLIP distance; error bars are pointwise, unadjusted percentile 95\% confidence intervals from 2,000 image-resampling replicates. \textbf{d,} Place Pulse split-half values for within-set distance, bidirectional nearest-neighbour similarity and centroid distance; bars span the 2.5th--97.5th percentiles across 2,000 repeated disjoint, size-matched Place Pulse splits.}
\label{fig:s8}
\end{figure}

\begin{figure}[H]
\centering
\includegraphics[width=\textwidth,height=0.78\textheight,keepaspectratio]{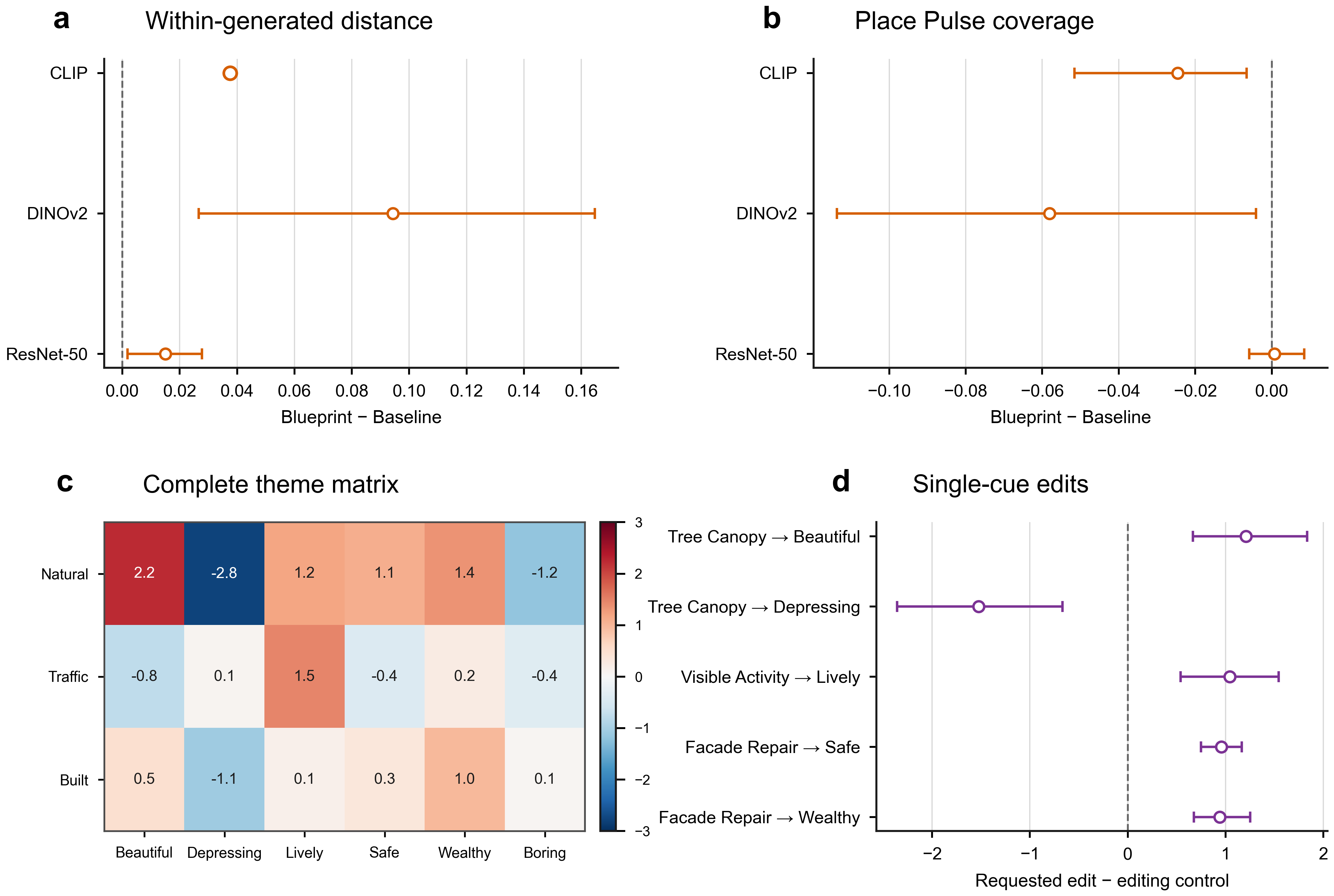}
\caption{\textbf{Perception representation and intervention sensitivity.} \textbf{a,b,} Blueprint-minus-Baseline changes in within-generated distance and Place Pulse-to-generated nearest-neighbour similarity under CLIP, DINOv2 and ResNet-50. For DINOv2 and ResNet-50 in \textbf{a} and for all three encoders in \textbf{b}, error bars are percentile 95\% confidence intervals from 2,000 resamples of generated images, with Place Pulse cities also resampled for the metric in \textbf{b}. No interval was computed for the CLIP within-generated estimate in \textbf{a}, which is shown as a point only. \textbf{c,} Complete \(3\times6\) editing-control-adjusted GPT-4o perception matrix; entries show point estimates, with scene-bootstrap confidence intervals and adjusted \(q\) values in Supplementary Table 20. \textbf{d,} Five prespecified single-cue effects; bars are scene-bootstrap percentile 95\% confidence intervals.}
\label{fig:s9}
\end{figure}

\begin{figure}[H]
\centering
\includegraphics[width=\textwidth,height=0.78\textheight,keepaspectratio]{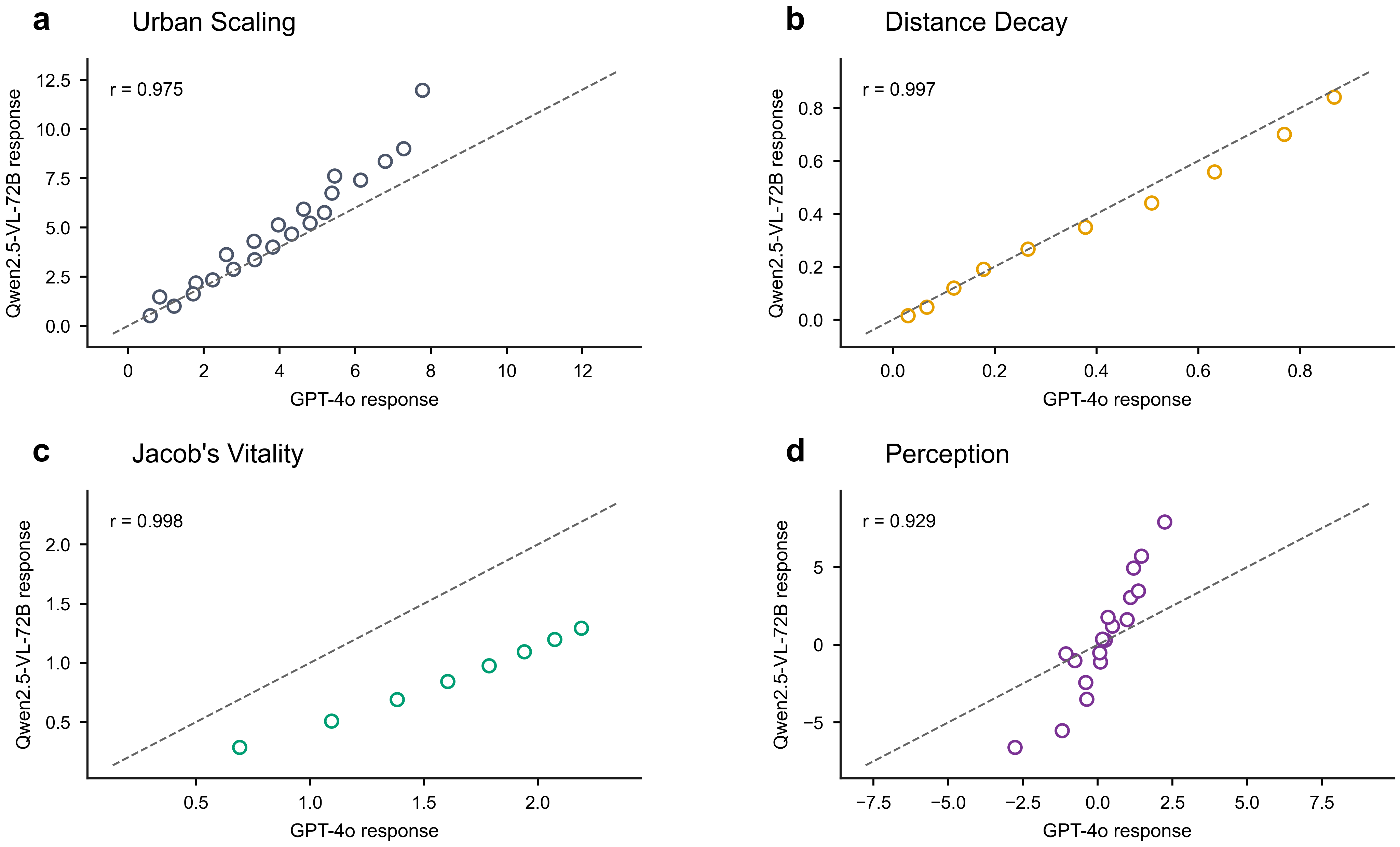}
\caption{\textbf{GPT-4o and open-model behavioural correspondence.} \textbf{a--d,} Mean response curves from GPT-4o and Qwen2.5-VL-72B for Urban Scaling, Distance Decay and Jacob's Vitality, together with the complete Perception theme matrix. For the numerical tasks, corresponding table values at the same tested value were averaged across fictional examples. Dashed lines indicate equality. The plotted Pearson correlations summarize these mean curves; primary corresponding-value correlations are reported in Supplementary Table 21. Absolute response units are not compared across tasks.}
\label{fig:s10}
\end{figure}

\begin{figure}[H]
\centering
\includegraphics[width=\textwidth,height=0.78\textheight,keepaspectratio]{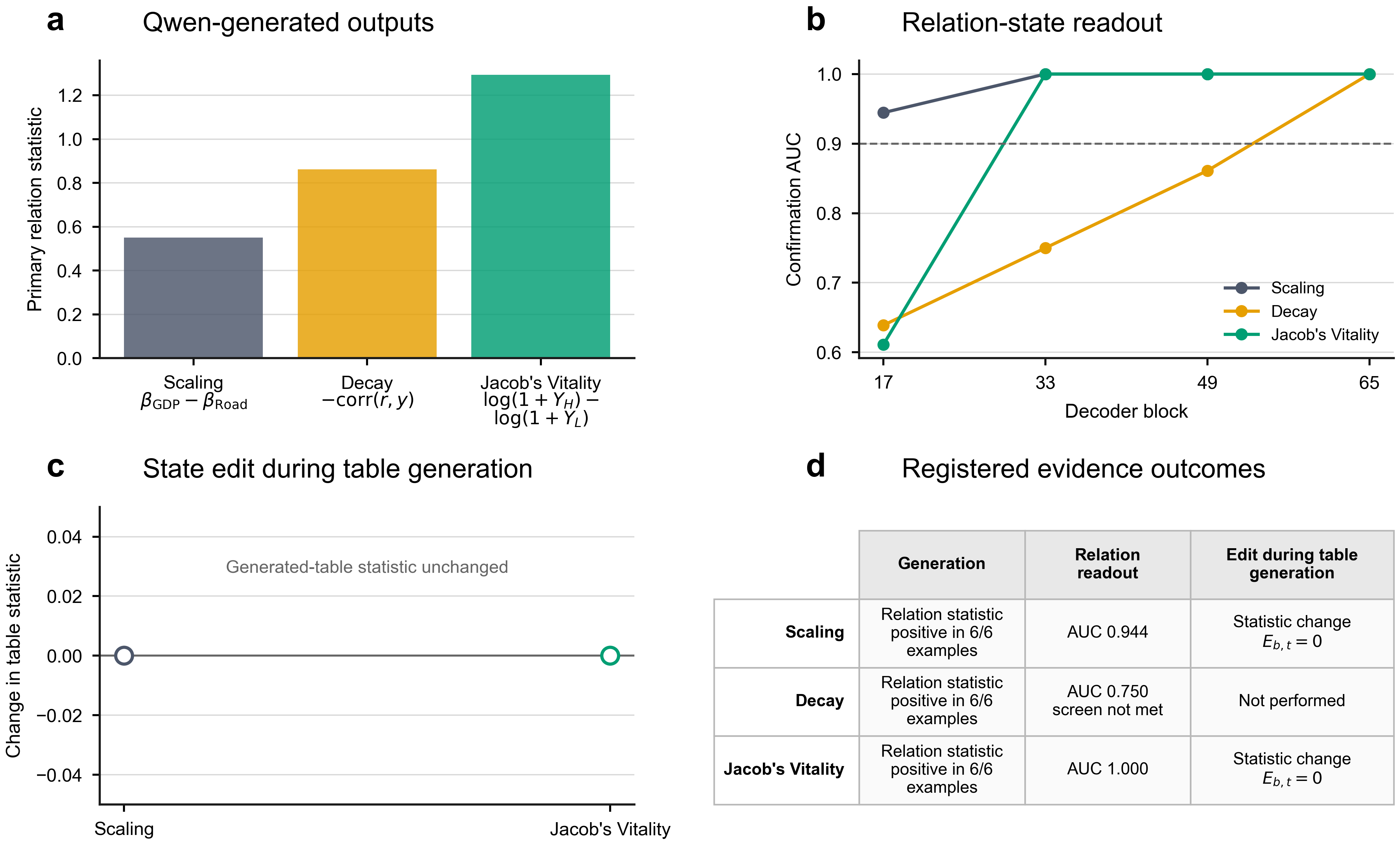}
\caption{\textbf{Numerical relation generation, hidden-state readout and state edits during table generation.} \textbf{a,} Open-model relationship statistics in held-out Scaling, Distance Decay and Jacob's Vitality generations. \textbf{b,} Confirmation AUC across candidate decoder blocks; the dashed line marks the prespecified 0.90 criterion. \textbf{c,} Changes \(E_{b,t}\) in the prespecified Scaling and Jacob's Vitality statistics when the selected direction was applied during table generation; both estimates and confidence intervals were zero. \textbf{d,} Results of the generation, relation-readout and table-generation sequence. Under the prespecified 0.90 rule, the state edit during table generation was performed for Scaling and Jacob's Vitality only.}
\label{fig:s11}
\end{figure}

\begin{figure}[H]
\centering
\includegraphics[width=\textwidth,height=0.78\textheight,keepaspectratio]{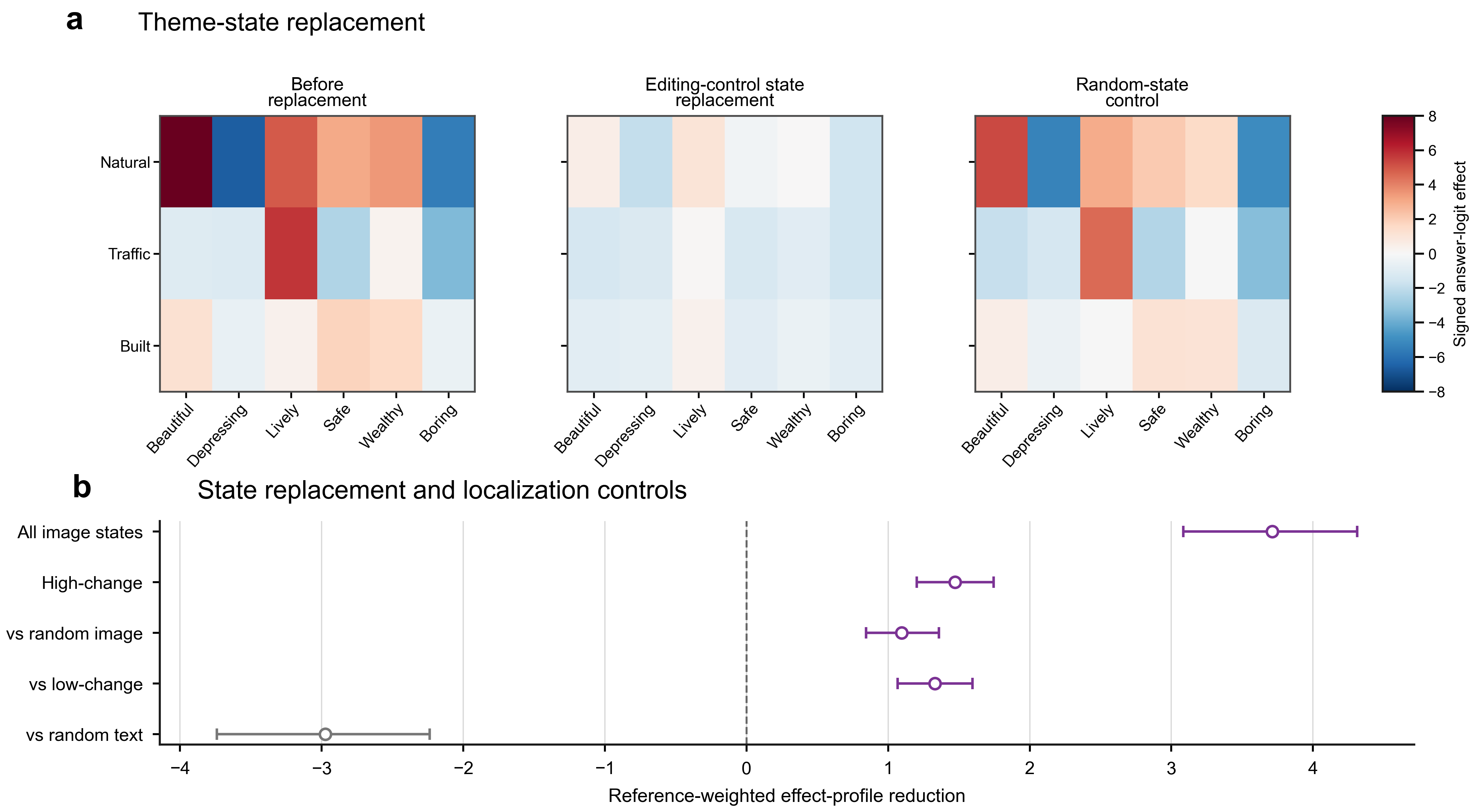}
\caption{\textbf{Perception image-state replacement and localization.} \textbf{a,} Signed theme-answer effects before state replacement, after copying paired same-scene editing-control states into the theme-image positions, and after equal-norm random replacement. \textbf{b,} Reduction in the Qwen effect pattern, weighted by the corresponding GPT-4o theme differences, for all image states, high-change image positions and matched image- and text-stream controls; bars are scene-bootstrap 95\% confidence intervals. Within image tokens, high-change positions produced greater reduction than low-change positions; matched text-token perturbations produced the largest reduction.}
\label{fig:s12}
\end{figure}

\subsection{Supplementary Data}\label{supplementary-data}

The three files organize model inputs, empirical sources and analysis-ready model outputs. Source Data contains the values plotted in Figs. 2--5 and Supplementary Figs. 1--12; Supplementary Data 3 contains the parsed unit-level outputs used to calculate these summaries.

\textbf{Supplementary Data 1 \textbar{} Prompt templates and inference settings (.xlsx).} The workbook contains prompt templates and fully rendered prompts for the reported analyses across all four cases, organized by task role and condition, together with model identifiers, decoding settings, output schemas and the configuration of the within-family 7B Perception follow-up. Its README defines the template notation and maps each analysis to its prompt.

\textbf{Supplementary Data 2 \textbar{} Empirical sources and spatial definitions (.xlsx).} The workbook identifies the empirical source and analysis sample for each case, the 50 GeoNames centres used to construct the 1-km Distance Decay bands, the 16 held-out Place Pulse cities and the operational definitions of the Vitality variables. It also identifies third-party or locally archived raw inputs that are not redistributed.

\textbf{Supplementary Data 3 \textbar{} Parsed model outputs (.zip).} The archive contains distributable row-, request-, setting- and scene-level model-output tables, a data dictionary and tables linking each figure panel and the within-family 7B Perception analyses to the corresponding observations and denominators. It includes the deidentified 6,000-task Place Pulse sample and the 7B behaviour and state-replacement matrices; source images, location identifiers, raw voter-level records and identifiers, and raw hidden-state arrays are not included.

\subsection{Supplementary References}\label{supplementary-references}

\begin{enumerate}
\def\labelenumi{\arabic{enumi}.}
\tightlist
\item
  Tu, T., Zhang, E. \& Long, Y. Profile and theoretical advances in urban big data studies: a systematic review of 57 representative journals (2013--2023). \emph{Environ. Plan. B Urban Anal. City Sci.} \textbf{53}, 673--688 (2026). \url{https://doi.org/10.1177/23998083251346582}
\item
  Bettencourt, L. M. A. The origins of scaling in cities. \emph{Science} \textbf{340}, 1438--1441 (2013). \url{https://doi.org/10.1126/science.1235823}
\item
  Arcaute, E. et al.~Constructing cities, deconstructing scaling laws. \emph{J. R. Soc. Interface} \textbf{12}, 20140745 (2015). \url{https://doi.org/10.1098/rsif.2014.0745}
\item
  Cabrera-Arnau, C. \& Bishop, S. R. The effect of dragon-kings on the estimation of scaling law parameters. \emph{Sci. Rep.} \textbf{10}, 20226 (2020). \url{https://doi.org/10.1038/s41598-020-77232-6}
\item
  Jiao, L. Urban land density function: a new method to characterize urban expansion. \emph{Landsc. Urban Plan.} \textbf{139}, 26--39 (2015). \url{https://doi.org/10.1016/j.landurbplan.2015.02.017}
\item
  Jiao, L. \& Dong, T. Inverse S-shape rule of urban land density distribution and its applications. \emph{J. Geomatics} \textbf{43}, 8--16 (2018). \url{https://doi.org/10.14188/j.2095-6045.2018217}
\item
  Jacobs, J. \emph{The Death and Life of Great American Cities}. Random House (1961).
\item
  Huang, J. et al.~Re-examining Jane Jacobs' doctrine using new urban data in Hong Kong. \emph{Environ. Plan. B Urban Anal. City Sci.} \textbf{50}, 76--93 (2023). \url{https://doi.org/10.1177/23998083221106186}
\item
  Dubey, A., Naik, N., Parikh, D., Raskar, R. \& Hidalgo, C. A. Deep learning the city: quantifying urban perception at a global scale. In \emph{European Conference on Computer Vision 2016} 196--212 (Springer, 2016). \url{https://doi.org/10.1007/978-3-319-46448-0_12}
\item
  Chen, J. et al.~Global 1 km × 1 km gridded revised real gross domestic product and electricity consumption during 1992--2019 based on calibrated nighttime light data. \emph{Sci. Data} \textbf{9}, 202 (2022). \url{https://doi.org/10.1038/s41597-022-01322-5}
\item
  Chen, Y., Xu, C., Ge, Y., Zhang, X. \& Zhou, Y. A 100 m gridded population dataset of China's seventh census using ensemble learning and big geospatial data. \emph{Earth Syst. Sci. Data} \textbf{16}, 3705--3718 (2024). \url{https://doi.org/10.5194/essd-16-3705-2024}
\item
  Huang, X., Li, J., Yang, J., Zhang, Z., Li, D. \& Liu, X. 30 m global impervious surface area dynamics and urban expansion pattern observed by Landsat satellites: from 1972 to 2019. \emph{Sci. China Earth Sci.} \textbf{64}, 1922--1933 (2021). \url{https://doi.org/10.1007/s11430-020-9797-9}
\item
  GeoNames. \emph{GeoNames geographical database}. \url{https://www.geonames.org/} (accessed 20 August 2026).
\item
  City of Melbourne. \emph{Pedestrian Counting System (counts per hour)}. \url{https://data.melbourne.vic.gov.au/explore/dataset/pedestrian-counting-system-monthly-counts-per-hour/} (accessed 15 August 2026).
\item
  City of Melbourne. \emph{Pedestrian Counting System - Sensor Locations}. \url{https://data.melbourne.vic.gov.au/explore/dataset/pedestrian-counting-system-sensor-locations/} (accessed 15 August 2026).
\item
  City of Melbourne. \emph{Blocks for Census of Land Use and Employment (CLUE)}. \url{https://data.melbourne.vic.gov.au/explore/dataset/blocks-for-census-of-land-use-and-employment-clue/} (accessed 15 August 2026).
\item
  Salesses, P. \& Hidalgo, C. A. \emph{Place Pulse}. figshare Dataset (2020). \url{https://doi.org/10.6084/m9.figshare.11859993.v1}
\item
  White, H. A heteroskedasticity-consistent covariance matrix estimator and a direct test for heteroskedasticity. \emph{Econometrica} \textbf{48}, 817--838 (1980). \url{https://doi.org/10.2307/1912934}
\item
  Sen, P. K. Estimates of the regression coefficient based on Kendall's tau. \emph{J. Am. Stat. Assoc.} \textbf{63}, 1379--1389 (1968). \url{https://doi.org/10.1080/01621459.1968.10480934}
\item
  Radford, A. et al.~Learning transferable visual models from natural language supervision. In \emph{Proceedings of the 38th International Conference on Machine Learning}, \emph{PMLR} \textbf{139}, 8748--8763 (2021). \url{https://proceedings.mlr.press/v139/radford21a.html}
\item
  Oquab, M. et al.~DINOv2: learning robust visual features without supervision. \emph{Trans. Mach. Learn. Res.} (2024). \url{https://openreview.net/forum?id=a68SUt6zFt}
\item
  He, K., Zhang, X., Ren, S. \& Sun, J. Deep residual learning for image recognition. In \emph{Proceedings of the IEEE Conference on Computer Vision and Pattern Recognition} 770--778 (2016). \url{https://doi.org/10.1109/CVPR.2016.90}
\item
  Gurnee, W. \& Tegmark, M. Language models represent space and time. In \emph{The Twelfth International Conference on Learning Representations} (2024). \url{https://openreview.net/forum?id=jE8xbmvFin}
\item
  De Sabbata, S., Mizzaro, S. \& Roitero, K. Geospatial mechanistic interpretability of large language models. In \emph{Geography According to Foundation Models}, \emph{Frontiers in Artificial Intelligence and Applications} \textbf{422}, 135--150 (IOS Press, 2026). \url{https://doi.org/10.3233/FAIA260476}
\item
  Todd, E., Li, M. L., Sharma, A. S., Mueller, A., Wallace, B. C. \& Bau, D. Function vectors in large language models. In \emph{The Twelfth International Conference on Learning Representations} (2024). \url{https://openreview.net/forum?id=AwyxtyMwaG}
\item
  Meng, K., Bau, D., Andonian, A. \& Belinkov, Y. Locating and editing factual associations in GPT. \emph{Adv. Neural Inf. Process. Syst.} \textbf{35}, 17359--17372 (2022). \url{https://doi.org/10.52202/068431-1262}
\item
  Marinescu, R., Gruber, V.-E. \& Fajardo Vargas, D. Medical interpretability and knowledge maps of large language models. In \emph{The Fourteenth International Conference on Learning Representations} (2026). \url{https://openreview.net/forum?id=BhqFWlYKUi}
\end{enumerate}
\end{document}